\documentclass[twocolumn]{aastex631}
\usepackage{amsmath,amsfonts,amssymb,graphicx,chngcntr,multirow, float,booktabs,afterpage}
\usepackage{rotating}
\usepackage[]{hyperref}
\hypersetup{colorlinks=true}

\newcommand{\XMM}{{\it XMM-Newton}}

\received{Nov. 26, 2025}
\revised{Feb. 3, 2026}
\accepted{Feb. 18, 2026}

\shorttitle{TDE X-ray Spectral-Timing}

\begin{document}

\title{X-ray Spectral-Timing Properties of Tidal Disruption Events}

\author[0000-0002-0568-6000]{Joheen Chakraborty}
\thanks{Both authors contributed equally to this work}
\email{joheen@mit.edu}
\affiliation{Department of Physics \& Kavli Institute for Astrophysics and Space Research, Massachusetts Institute of Technology, Cambridge, MA 02139, USA}

\author[0000-0003-4127-0739]{Megan Masterson}
\thanks{Both authors contributed equally to this work}
\email{mmasters@mit.edu}
\affiliation{Department of Physics \& Kavli Institute for Astrophysics and Space Research, Massachusetts Institute of Technology, Cambridge, MA 02139, USA}

\author{Andrew Mummery}
\affiliation{School of Natural Sciences, Institute for Advanced Study, 1 Einstein Drive, Princeton, NJ 08540, USA}

\author[0000-0003-0172-0854]{Erin Kara}
\affiliation{Department of Physics \& Kavli Institute for Astrophysics and Space Research, Massachusetts Institute of Technology, Cambridge, MA 02139, USA}

\author[0009-0001-9034-6261]{Christos Panagiotou}
\affiliation{Department of Physics \& Kavli Institute for Astrophysics and Space Research, Massachusetts Institute of Technology, Cambridge, MA 02139, USA}

\author[0000-0003-4054-7978]{Riccardo Arcodia}
\affiliation{Department of Physics \& Kavli Institute for Astrophysics and Space Research, Massachusetts Institute of Technology, Cambridge, MA 02139, USA}

\author[0000-0002-7303-8144]{Vera Berger}
\affiliation{Department of Physics \& Kavli Institute for Astrophysics and Space Research, Massachusetts Institute of Technology, Cambridge, MA 02139, USA}

\begin{abstract}
We perform the first systematic study of the minute-to-hours-timescale stochastic variability observed in the X-ray luminosity of tidal disruption events (TDEs) using \XMM\ data and Fourier analysis methods. We measure the spectral properties, power spectral densities (PSDs), fractional variability amplitudes, and energy dependence of the variability for 18 TDEs spanning 54 observations, of which 27 occur in thermal disk-dominated states and 27 show a nonthermal hard X-ray corona. Compared to pure thermal sources, we find TDEs with coronae are more X-ray variable and show steeper PSDs indicating longer correlation timescales. This state-transition behavior is qualitatively similar to X-ray binaries, which show higher fractional variability in the hard state than in the soft state. However, newborn TDE coronae show systematically flatter PSDs and softer energy spectra than their long-lived AGN counterparts. We also show that the variability amplitude of thermal TDEs increases with photon energy, consistent with variations sourced by local temperature fluctuations and exponentially enhanced in the Wien tail. Our work demonstrates that combining spectral and timing properties of X-ray TDEs can probe the microphysics of newly formed accretion flows around supermassive black holes, and that the coronae formed in TDEs fundamentally differ from those in AGN.
\end{abstract}


\section{Introduction} \label{sec:intro}

Accretion onto compact objects powers some of the most energetic phenomena in the universe, ranging from stellar-mass black holes in X-ray binaries (XRBs) to supermassive black holes (SMBHs) in active galaxy nuclei (AGN). Despite differing in mass scale by 6-8 orders of magnitude, these systems share the common property of stochastic variability arising from turbulent accretion flows, whereby magnetohydrodynamic instabilities drive fluctuations that modulate the observed electromagnetic emission. In principle, the statistical properties of this variability encode information about interesting quantities in the accretion flow, such as the origin of the turbulent fluctuations, coherence length scales, viscous time/length scales, and disk-corona coupling. 

Many decades of effort have thus been invested into meticulous study of the time variability of accreting systems, resulting in a vast literature of techniques and results. For instance, early studies found that the X-ray emission from SMBHs is more variable than optical/UV \citep{Barr86}, can be characterized by red noise power spectra (e.g. \citealt{Uttley02,Markowitz03,Gonzales12}), and sometimes shows a discrete break in variability amplitude whose frequency scaling with mass and accretion rate \citep{McHardy06,Ponti12}. Combining spectral and timing information such as the variability/RMS spectrum \citep{Vaughan03} has further enabled deconvolving the contributions from the disk, corona, and soft excess, showing emission from the corona is typically the most variable on short timescales while reprocessed disk emission is smeared out. Similarly, reverberation lags have been used to probe the geometry of the accreting black holes, in particular the strength and vertical extent of the corona and its relation to spectral state transitions and jets \citep{Uttley14,Kara16,Wang22}. Though these results are far from exhaustive, they showcase that much has been learned about the inner accretion flows of black holes via spectral-timing studies.

Tidal disruption events (TDEs) are a relatively newer addition to the observational census of accreting compact objects. TDEs result from stellar orbits scattered within the tidal radii ($R_t$) of a SMBH, where the differential tidal force across the body of a star exceeds its self-gravity. Upon passing within $R_t$, the star is torn apart, and the resulting stellar debris eventually loses enough orbital energy and angular momentum to circularize around the SMBH in a compact accretion disk. Though the basic features were predicted decades ago \citep{Rees88,Phinney89}, many open questions remain related to the origin of the electromagnetic emission, the nature of the hosts, and the efficiency of the circularization process following initial disruption, which has direct implications for subsequent accretion disk formation. 

Observationally, TDEs are broadband, multiwavelength phenomena, with quickly growing samples enabled by wide-area time-domain surveys \citep[see e.g.][for a recent review]{Gezari21}. Of all the multiwavelength emission features, the most compelling evidence that TDEs are indeed able to form compact accretion disks is their soft, thermal-like X-ray emission, which has been documented in many studies spanning several instruments and surveys \citep{Esquej06,Auchettl17,Saxton20,Sazonov21,Guolo24,Grotova25}. Recent advances in detailed spectral energy distribution (SED) modeling have shown rigorously that the accretion disks formed in TDEs are indeed compatible with theoretical expectations of $\sim 0.5M_\odot$ of unbound stellar mass viscously expanding from an initial radius $\sim R_t$; the success of these models paves the way for using TDEs to probe fundamental SMBH properties and accretion physics \citep{Wen20,Mummery23,Guolo25b}. Nevertheless, many of the X-ray observables of TDEs have proven to be heterogeneous and unexpected, with a rich phenomenology of both photometric and spectroscopic properties. Studying these properties will improve our understanding of low-mass, otherwise quiescent SMBHs, super-Eddington accretion, and newborn, rapidly-evolving accretion disks.

While the X-ray spectral properties of TDEs have been studied systematically \citep{Auchettl17,Sazonov21,Guolo24,Grotova25}, the variability characteristics---which have proven valuable for understanding accretion physics in AGN and XRBs---have not yet been examined across a large sample. In this paper, we carry out the first systematic analysis of the rapid X-ray variability properties of TDEs, on minutes-to-hours (1-10 ks) timescales using archival \XMM\ data. We use Fourier timing analysis methods developed in the AGN, X-ray binary, and cataclysmic variable literatures. The rapid disk evolution and spectral state transitions seen in TDEs offer a unique window into how their variability properties may change as accretion disks and coronae form and evolve on human-observable timescales, which is otherwise impossible to do for SMBHs.

In Section~\ref{sec:methods}, we give an overview of the Fourier timing methods used to characterize the time series variability and describe our accompanying spectral analysis. In Section~\ref{sec:results}, we present the main results of applying these methods to the archival \XMM\ data. In Section~\ref{sec:discussion}, we discuss the implications of our findings at the population level and draw comparisons to AGN and XRB results. In Appendix~\ref{appendix:figs} we present plots and data for all of our results. We make concluding remarks and highlight areas for future study in Section~\ref{sec:conclusion}.

\section{Methods} \label{sec:methods}

\begin{figure}
    \centering
    \includegraphics[width=\linewidth]{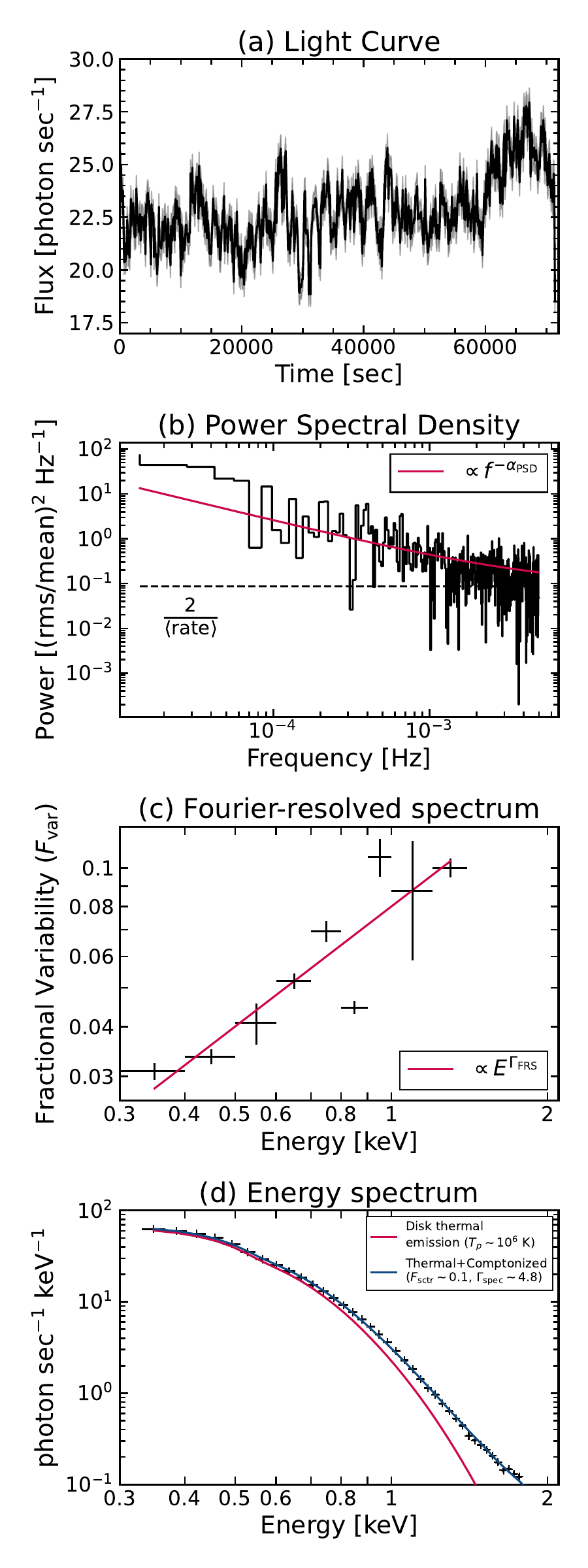}
    \caption{(a) time-series light curve with 100 second bins, (b) power spectral density, (c) Fourier-resolved spectrum, and (d) energy spectrum of the TDE AT2022lri.}
    \label{fig:lri_example}
\end{figure}

\subsection{Time-series variability metrics} \label{subsec:var_metrics}

Accreting black holes exhibit stochastic, aperiodic variability in their X-ray emission across a range of timescales. For an example, see Fig.~\ref{fig:lri_example}a, which depicts the light curve of the TDE AT2022lri observed by \XMM. In principle, the underlying processes generating this variability are shaped by the complex, multi-scale physics of the accretion process. The \textit{modus operandi} of timing studies is that by carefully characterizing the underlying stochastic process, one can understand something about the nature of the accretion flow which generated it. The regular sampling cadence and multi-timescale variability properties of X-ray data lend itself particularly well to Fourier techniques, meaning the starting-point for variability studies is the Power Spectral Density (PSD). 

The PSD of a time series quantifies the distribution of variability power as a function of temporal frequency. The power $P(f)$ represents the contribution to the total variance from variations occurring at frequency $f$. For a discrete, evenly-sampled time series consisting of $N$ flux measurements $x_i$ at times $t_i$ with sampling cadence $\Delta T$ and mean count rate $\bar{x}$, the PSD is estimated via the periodogram, which is the modulus-squared of the Discrete Fourier Transform:
\begin{equation}
P(f_j) = \frac{2\Delta T}{N\bar{x}^2} \left| \sum_{i=1}^{N} x_i e^{-2\pi i f_j t_i} \right|^2
\label{eq:psd}
\end{equation}
Note that we have normalized the PSD such that it has units of (rms/mean)$^2$ Hz$^{-1}$; the advantage of this will soon become clear.

In Fig.~\ref{fig:lri_example}b, we show the PSD of the \XMM\ data of AT2022lri. The PSD is typically evaluated at frequencies $f_j = j/(N\Delta T)$ for $j = 1, 2, \ldots, N/2$, where $f_{N/2} = 1/(2\Delta T)$ is the Nyquist frequency, the highest frequency at which the signal can be reconstructed for a given sampling rate. Some culturally significant PSDs are those having $P(f)\propto f^0 \propto\mathrm{const.}$, which is called ``white noise'' and has equal power at all temporal frequencies; $P(f)\propto f^{-\alpha}$ with $\alpha>1$, called ``red noise'' due to its stronger variability power concentrated at lower frequencies; and $P(f)\propto f^{-1}$, aptly named ``pink noise'' or flicker noise. The slope of the PSD is fundamentally a statement about the correlation timescales present in the underlying process. Steeper PSDs indicate stochastic processes dominated by long-range correlations which occur over large timescales. Flat (white) PSDs indicate processes with \textit{zero} correlation time, as the values are randomly chosen at every time with no ``memory'' of the preceding variability. Measuring a correlation timescale directly from the PSD or light curve is not straightforward in general; we will return to this point in Section~\ref{subsec:physical_interp}.

The PSDs of AGN typically show red noise with $\alpha\sim2$, ranging from timescales of minutes to days and sometimes showing a discrete low-frequency turnover to a shallower slope at a ``break frequency'' corresponding to $\sim$days \citep{Uttley02,Markowitz03,McHardy06,Gonzales12}. In our case, however, most TDE X-ray observations are not long enough to measure such low frequencies. To remain consistent across the population, we limit our analysis to timescales of 1-10 kiloseconds to ensure fair comparison between differences in system properties and data quality.

In real observational data, an additional noise component arises from measurement errors, appearing as a ``white noise floor'' at constant power $P_{\rm noise}$ in our fractional normalization (shown as the dashed line in Fig.~\ref{fig:lri_example}b). This noise floor arises from the Poisson nature of photon counting: since the variance of a Poisson process equals its mean ($\sigma^2=\bar{x}$), the fractional variance is $\sigma^2/\bar{x}^2=1/\bar{x}$. The normalization convention used in X-ray astronomy introduces a factor of 2 (Eq.~\ref{eq:psd}), yielding a Poisson noise level of $P_{\rm noise}=2/\bar{x}$. Only power exceeding this level is driven by intrinsic source variability, so we subtract $P_{\rm noise}$ from the measured PSDs when quoting variability metrics.

Integrating the PSD between two frequencies yields the contribution to the total variance from variations occurring within that frequency range:
\begin{equation}
 F_{\rm var} = \sqrt{\int_{f_1}^{f_2} [P(f) - P_{\rm noise}] \, df}
\label{eq:fvar}
\end{equation}
where $F_{\rm var}$ (known as the ``fractional variability RMS amplitude'') directly follows as a result of using the RMS/mean normalization (Eq.~\ref{eq:psd}). In our toy example (Fig.~\ref{fig:lri_example}), integrating the PSD between different frequency ranges yields $F_{\rm var,0.4-1\;ks}\approx 2\%$ (as the PSD approaches the Poisson noise floor at these frequencies); $F_{\rm var,1-10\;ks}\approx3.6\%$; and $F_{\rm var,10-50\;ks}\approx 4\%$, demonstrating the red-noise character of the light curve which is visible by-eye. For comparison, directly computing from the time series gives $\sigma_F/\mu_F \approx 7\%$ but loses the information of variability \textit{timescale}. We digress here to briefly mention that $F_{\rm var}$ is commonly measured directly from the time domain using the formalism of \cite{Vaughan03} rather than from the PSD. This is because time-domain methods do not require evenly spaced or continuous data; however, they also lose some control over specific variability timescales and possess no unbiased estimator, unlike the PSD. We use Fourier methods in this paper to ensure that we compute variability statistics on consistent timescales across many sources, but the results should be qualitatively comparable to time-domain estimates. This difference should nevertheless be kept in mind when directly comparing numbers to other estimates in the literature.

One can carry out the same PSD-based estimation above with energy-resolved light curves to compute $F_{\rm var}$ separately for distinct energy bands. The result is the ``Fourier-resolved spectrum'' (FRS; e.g. \citealt{Revnivtsev99,Papadakis07,Panagiotou20}). The FRS is closely related to RMS spectrum, which is the time-domain analog commonly used to assess thee energy-dependent variability for AGN \citep{Parker20} and XRBs \citep{Konig24}. The costs and benefits of the FRS versus the RMS spectrum are exactly the same as those involved in computing $F_{\rm var}$ from the PSD as opposed to from the light curve. In Fig.~\ref{fig:lri_example}c, we show the FRS of AT2022lri, computed for frequencies of $10^{-4}-10^{-3}$ Hz (1-10 kiloseconds) in a series of narrow-energy bands. One clearly observes a trend of increasing fractional variability at higher energies, which can be quantified by a power-law fit to the FRS having a positive index ($\Gamma_{\rm FRS}>0$). By contrast, AGN show a striking diversity of behaviors in their RMS spectra (e.g. \citealt{Hu22}); we will discuss the physical implications of this in Section~\ref{sec:discussion}.

\subsection{Spectral diagnostics} \label{subsec:spectra}

The energy spectrum provides complementary information to the aforementioned quantities derived from the light curve and PSD. We show the X-ray spectrum of AT2022lri in Fig.~\ref{fig:lri_example}d. As is usually the case in TDEs, the dominant component is a thermal blackbody peaking in soft X-rays. AT2022lri is among the handful of optically-selected TDEs which also show an additional, non-thermal emission component indicative of a Comptonizing corona.

All spectra were fit with the same model of \texttt{tbabs}$\times$\texttt{zashift}$\times$\texttt{simpl}$\times$\texttt{tdediscspec}. The thermal blackbody continuum is represented by \texttt{tdediscspec} \citep{Mummery21b}, a model for the X-ray spectrum resulting from a cool ($k_B T\lesssim0.3$ keV) accretion disk which accounts for the aspherical emission surface, the temperature gradient present in a disk (as opposed to a single-temperature blackbody prescription), and disk opacity effects which are typically encapsulated in color-correction factors. The model contains three free parameters: $T_p$, the hottest temperature in the accretion disk; $R_p$, the radius at which this region resides; and $\gamma$, a nuisance factor dependent on the disk inclination angle and inner boundary condition. For a more detailed discussion, see \cite{Mummery21b}. We also allow for a hard X-ray power law component with \texttt{simpl} \citep{Steiner09}, an empirical Comptonization model in which a fraction ($F_{\rm sctr}$) of the photons from an arbitrary input seed spectrum are inverse Compton-scattered into a higher-energy power-law. For sources where clear absorption/emission residuals from the base model were visible, we added the \texttt{gabs} multiplicative gaussian absorption model, with the normalization forced to be $<0$ in cases of emission. We detected absorption troughs in ASASSN-14li, AT2020ksf, AT2020adgm, and GSN 069, all of which have been previously reported in the literature \citep{Miniutti13,Kara18,Wevers24,Kosec23,Pasham24,Kosec25}. We recovered an emission feature near $\sim 1$ keV in AT2022lri, which was reported and discussed in \cite{Yao24}.

\subsection{Observations and data reduction} \label{subsec:data}

We surveyed the \XMM\ archive for pointed observations of TDEs reported in the literature, and retained only those which resulted in an X-ray detection with a minimum continuous duration of $\geq$10 kiloseconds after filtering for background flaring. We give the basic source properties in Table~\ref{tab:sources}. All data were retrieved from the \XMM\ Science Archive (\href{https://nxsa.esac.esa.int/nxsa-web/}{https://nxsa.esac.esa.int/nxsa-web/}) and were reduced using \XMM\ SASv21.0.0 and HEASoft v6.33. Our sample contains 18 sources with 54 observations, whose light curves, energy spectra, FRS, PSDs, and all derived measurements are shown in the Figures and Table of Appendix~\ref{appendix:figs}. We note that most of our sources are optically selected TDEs, with only 3/18 selected from X-ray archives/surveys (3XMM J2150, GSN 069, eRASSt J0456) due to the generally sparser data coverage of those TDEs. This may be a source of unknown selection effects, but it is unlikely to severely bias our results due to the general consistency between optical and X-ray TDE populations  \citep{Guolo24,Yao23}.

We applied a uniform data reduction procedure for every observation ID (OBSID).  Source products were extracted from a circular region of 33'' radius, while the background was extracted from source-free circular region falling on the same detector with a 60'' radius. We retained events with PATTERN$\leq$4 (single and double events only) and discarded time intervals with a 10--12 keV count rate $\geq 1$ counts s$^{-1}$. For both light curves and spectra, we used only the data from the EPIC-pn detector due to its superior effective area compared to the EPIC-MOS detectors. For subsequent timing analyses, we retained only the longest continuous segment after filtering for background flares. Light curves were extracted with the \texttt{evselect}. We used the \texttt{epiclccorr} command to perform background-subtraction and absolute corrections for the detector efficiency, PSF, bad pixels, vignetting, and non-uniform exposure in each time series bin. We used 20-second time binning for all soft-band light curves ($\leq2$ keV)---with the exception of 3XMM J2150-05, which only passed our minimum duration cutoff after using 200-second bins due to frequent drop-outs resulting from high background---and 200-second binning for hard-band light curves ($>2$ keV) due to the generally smaller count rates at those energies.

We extracted spectra with the \texttt{evselect} command using a binning factor of 5, then grouped spectral channels using the \texttt{specgroup} command with a minimum group width of 1/3 the resolution FWHM at each energy (\texttt{oversample} $=3$). All spectral fitting was performed using the Cash statistic \citep{Cash1979} and \href{https://heasarc.gsfc.nasa.gov/docs/xanadu/xspec/python/html/index.html}{\texttt{PyXspec}}, the \texttt{Python} interface to the \texttt{XSPEC} spectral fitting program \citep{Arnaud96}. For each OBSID, we retained data between 0.3 keV up to the energy where the source spectrum becomes fainter than the background.

Several \XMM\ observations of GSN 069 include quasi-periodic eruptions \citep{Miniutti19}; we excluded them from our analysis, as we were only interested in the accretion disk variability.

\subsection{Fitting Procedures and Error Estimation} \label{subsec:psd}

To obtain the PSD slopes, we fit the unbinned PSD with a model $P(f) = N_0 f^{-\alpha} + C$, with two components accounting for the Poisson noise (the constant, $C$) and intrinsic source variability (the power law, $N_0 f^{-\alpha}$). Each point in an unbinned PSD follows a $\chi^2$ distribution with 2 degrees of freedom, which arises from summing two Gaussian-distributed variables (the real and imaginary parts of the Fourier transform) in quadrature. To properly account for these statistics, we use the Whittle likelihood function, given by
\begin{equation}
    \ln \mathcal{L} = - \sum_j \left(\frac{I_j}{S_j} + \ln S_j\right),
\end{equation}
where $I_j$ ($S_j$) is the observed (model) power at a frequency $f_j$. For more details, we refer the reader to \cite{Vaughan2005} and \cite{Vaughan2010}. Given the data quality in our sample, we fix the constant to the theoretical Poisson noise value, 2/$\bar{x}$, as detailed in Section \ref{subsec:var_metrics}. We then fit each PSD for the value of $\alpha_{\rm PSD}$ and $N_0$ using a Markov Chain Monte Carlo (MCMC) sampler with the \texttt{emcee} package \citep{Foreman-Mackey2013}. Not all sources show significant intrinsic source variability. We keep only those sources which have $\alpha_{\rm PSD}/\Delta\alpha_{\rm psd}>2$, i.e. excluding sources in which the only contribution is the instrumental Poisson noise constant. The remaining sources are not included in comparisons of the PSD slopes. For sources with a detected intrinsic PSD component, we measured the fractional variability from Eq.~\ref{eq:fvar}. The error on $F_{\rm var}$ was estimated via an MCMC chain with 1000 samples. Fourier-resolved spectra were fit and error estimations were also performed via MCMC; further details of the energy binning used for the FRS are given in Appendix~\ref{appendix:figs}. Spectral fit errors were estimated via running MCMC chains within \texttt{PyXSPEC} using the Goodman-Weare algorithm, then quoting the 16th and 84th percentiles of the posterior.

\section{Results} \label{sec:results}

\begin{table*}
\centering
\caption{Properties of TDEs in the sample. Galactic neutral absorption ($N_H$) is estimated from the MW dust maps of \cite{HI4PI}. SMBH masses are drawn from late-time SED modeling of the plateau emission where available and host galaxy scaling relations otherwise. In cases of asymmetric error bars, we quote their average.}
\label{tab:sources}
\hspace*{-2.5cm}
\resizebox{1.15\textwidth}{!}{%
\begin{tabular}{l|cccccccc}
\hline\hline
Source & $\log(M_{\rm BH}/M_\odot)$ & $z$ & RA & Dec & $N_{\rm H}$ & Peak Date & \XMM\ OBSIDs \\
 & & & (deg) & (deg) & ($10^{20}$ cm$^{-2}$) & (MJD) & \\
\hline
3XMM J2150-05 & $4.39 \pm 0.13$$^{1}$ & 0.055 & 327.5937 & -5.8525 & 2.81 & 53680 & 0404190101, 0603590101 \\
ASASSN-14ko & $7.86 \pm 0.36$$^{2}$ & 0.0425 & 81.3255 & -46.0056 & 3.49 & --- & 0900410101, 0900410201, 0900410301 \\
ASASSN-14li & $6.44 \pm 0.19$$^{1}$ & 0.0206 & 192.0625 & 17.7740 & 1.88 & 56983 & 0694651201, 0694651401, 0694651501 \\
 & & & & & & & 0722480201, 0770980101, 0770980601 \\
ASASSN-15oi & $6.38 \pm 0.17$$^{1}$ & 0.0484 & 309.7875 & -30.7556 & 4.87 & 57259 & 0722160701 \\
AT2018fyk & $7.69 \pm 0.39$$^{3}$ & 0.0600 & 342.5666 & -44.8648 & 1.15 & 58389 & 0831790201, 0853980201, 0911790601 \\
AT2019azh & $6.52 \pm 0.13$$^{1}$ & 0.0222 & 123.3206 & 22.6483 & 4.16 & 58558 & 0822041101, 0823810201, 0823810301 \\
 & & & & & & & 0823810401, 0842592601 \\
AT2019teq & $6.52 \pm 0.8$$^{4}$ & 0.0870 & 284.7729 & 47.5182 & 4.54 & 58794 & 0902760901, 0913992001, 0935190101 \\
AT2020adgm & $7.5\pm 0.5$$^{5}$ & 0.0560 & 63.2602 & -53.0727 & 1.2 & 59225 & 0852600301, 0891803701, 0891803801 \\
AT2020afhd & $6.7 \pm 0.5^{6}$ & 0.0270 & 48.3987 & -2.1518 & 4.72 & 60352 & 0953010301 \\
AT2020ddv & $6.28 \pm 0.78$$^{7}$ & 0.1600 & 149.6390 & 46.9112 & 1.35 & 58915 & 0842592501 \\
AT2020ksf & $5.97 \pm 0.17$$^{1}$ & 0.0920 & 323.8636 & -18.2765 & 3.58 & 58976 & 0882591201 \\
AT2020ocn & $6.65\pm 0.63$$^{7}$ & 0.0700 & 208.4744 & 53.9973 & 1.32 & 58972 & 0863650101, 0863650201, 0872392901 \\
 & & & & & & & 0902760701 \\
AT2021ehb & $6.66 \pm 0.29$$^{1}$ & 0.0180 & 46.9492 & 40.3113 & 9.97 & 59315 & 0840140201, 0840140301, 0840140401 \\
 & & & & & & & 0840140901, 0840141001, 0882590901 \\
AT2022lri & $5.79\pm 0.08$$^{1}$ & 0.0328 & 35.0334 & -22.7209 & 1.6 & 59682 & 0882591901, 0915390201, 0932390701 \\
AT2023mhs & --- & 0.0482 & 205.8153 & 19.2503 & 1.57 & 60134 & 0935191201 \\
GSN 069 & $6.45 \pm 0.12$$^{1}$ & 0.0180 & 19.7851 & -34.1919 & 2.29 & 55391 & 0740960101, 0864330201, 0864330301 \\
 & & & & & & & 0864330401, 0884970101, 0884970201 \\
OGLE16aaa & $7.54 \pm 0.34$$^{4}$ & 0.1655 & 16.8375 & -64.2725 & 2.71 & 57403 & 0793183201 \\
eRASSt J0456-20 & $7.0 \pm 0.4$$^{8}$ & 0.0770 & 74.2096 & -20.6306 & 2.77 & --- & 0884960601, 0891801101, 0891801701 \\
 & & & & & & & 0931791501 \\
\hline
\end{tabular}
}
\tablenotetext{}{References:
$^{1}$ \citealt{Guolo25b};
$^{2}$ \citealt{Payne21};
$^{3}$ \citealt{Wevers20};
$^{4}$ \citealt{Mummery24a};
$^{5}$ \citealt{Pasham24};
$^{6}$ \citealt{Liu25};
$^{7}$ \citealt{Hammerstein23};
$^{8}$ \citealt{Liu23}
}
\end{table*}

We split the observations into two classes based on their spectral continuum properties. Observations with $F_{\rm sctr} \leq 0.01$ were classified as purely thermal, while observations with $F_{\rm sctr} > 0.01$ were classified as having a corona.\footnote{Throughout the rest of this work, we will refer to TDEs with $F_{\rm scat} < 0.01$ as thermal TDEs and those with $F_{\rm sctr} > 0.01$ as TDEs with a corona or corona TDEs. It is important to note that our sample \textit{does not contain any jetted TDEs}. Even TDEs with a corona are often dominated by thermal disk emission.} While a scattering fraction of 1\% is modest, even those sources with the smallest nonzero $F_{\rm sctr}$ show clear nonthermal components in their spectra which are visible by-eye (Appendix~\ref{appendix:figs}). In any case, the vast majority have $F_{\rm sctr}\gg 0.01$. For TDEs with coronae, the underlying disk parameters $T_p$ and $R_p$ cannot be uniquely recovered beyond a particular corona strength (also reported in \citealt{Guolo24}, where they found a cutoff of $F_{\rm sctr}\sim0.2$). Rather than further partition the population with a corona, we simply do not use $R_p$ or $T_p$ for TDEs with a corona in our subsequent population comparisons. 

In total, our definitions resulted in 27 thermal observations and 27 with a corona. These observations consist of 12 unique thermal TDEs and 9 TDEs with a corona. The total sums to more than $>18$ because some sources transition between states within our data and are thus double-counted, namely AT2019azh \citep{Wevers20}, AT2020ocn \citep{Cao24}, and eRASSt J0456-20 \citep{Liu23}. We note there are also other sources which show soft-to-hard transitions reported in the literature, such as AT2018fyk \citep{Wevers21}, AT2021ehb \citep{Yao22}, and AT2019teq \citep{Berger26}; however, in those cases, only the corona-dominated states passed our data quality thresholds, so the sources are not double-counted. Furthermore, as described in Section~\ref{subsec:psd}, some sources (particularly in faint observations) do not show detectable variability in addition to the Poisson noise. We quote only $F_{\rm var}$ and $\alpha_{\rm PSD}$ upper limits for those sources and retain only their spectral information. This resulted in reduced sample sizes of 14 thermal/24 corona observations having timing information. Of those, an even smaller fraction have sufficiently high count rates to measure the FRS in at least two bands, resulting in 12 thermal/23 corona observations with FRS fits.

\begin{figure*}
    \centering
    \includegraphics[width=\linewidth]{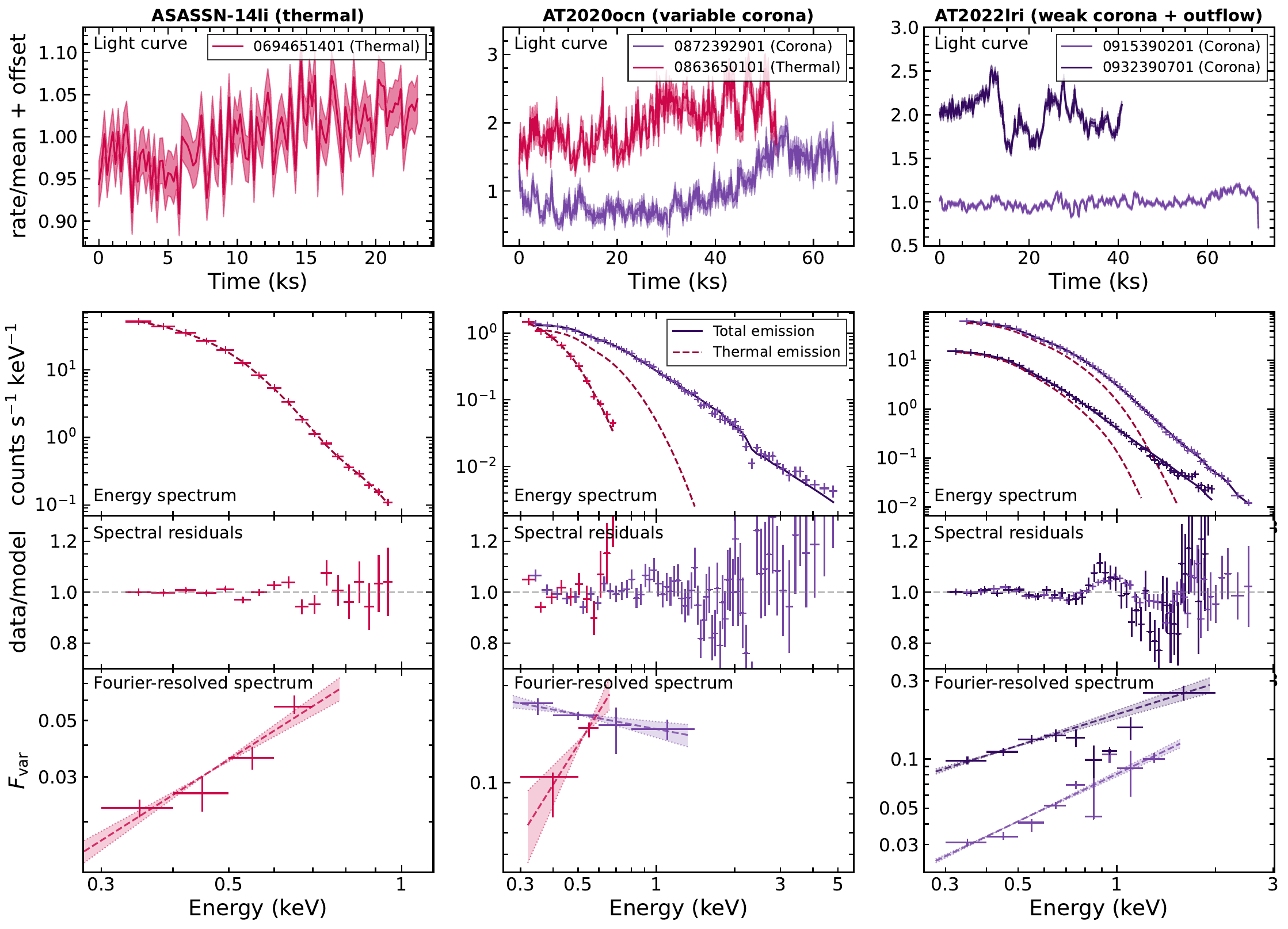}
    \caption{Normalized light curves (with offsets of 0 or 1), energy spectra, spectral residuals, and Fourier-resolved spectra of some representative TDEs. ASASSN-14li depicts the rising FRS typical of thermal TDEs. AT2020ocn, which showed a state transition from a pure thermal spectrum to one with a corona, showed a corresponding change in the slope of its FRS. For AT2022lri only, we have removed the gaussian emission component near 1 keV for visualization purposes only, such that the line is visible in the spectral residuals panel. We observe a deficit in the FRS coinciding with the spectral emission component, indicating a physically distinct emission and variability process than the disk or corona continuum.}
    \label{fig:fig2}
\end{figure*}

In Fig.~\ref{fig:fig2}, we show the normalized light curves, energy spectra, spectral residuals, and FRS of some representative TDEs from the various spectral classes. ASASSN-14li is a purely thermal TDE, which shows a rising FRS, as is common for the population. AT2020ocn underwent a state transition whereby it developed a Comptonizing corona of $F_{\rm sctr}\sim0.34$ at late times; the sign of $\Gamma_{\rm FRS}$ is observed to switch during the state transition, i.e. the emergence of a corona changes the energy-dependence of the variability. AT2022lri shows a weak corona ($0.03<F_{\rm sctr}<0.26$) at all epochs, and also has an outflow seen near $\sim1$ keV. Remarkably, the FRS decreases exactly at the location of the outflow emission feature, indicative of a different emission process than disk turbulence (see Section~\ref{sec:discussion}). Similar features have been associated with AGN spectral lines: for example, several Type I AGN exhibit a local decrease in variability near the Fe K$\alpha$ emission line \citep{Markowitz03}, while IRAS 13224--2809 shows a localized variability increase in the ultrafast outflow-dominated absorption band \citep{Parker20}. We believe this is the first instance in a TDE. Further study is thus warranted, though AT2022lri is the only TDE in our sample with high enough SNR for this detection.

\subsection{Population features} \label{subsec:population}

\begin{figure*}
    \centering
    \includegraphics[width=\linewidth]{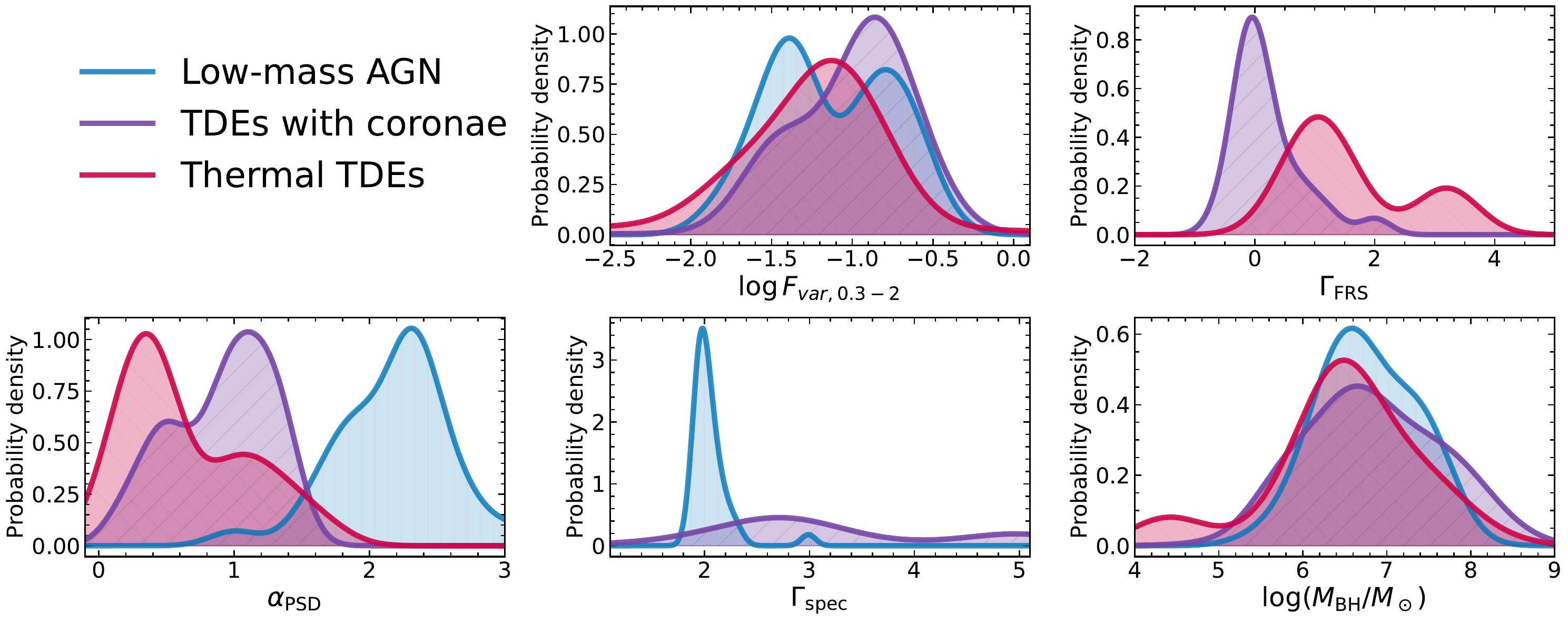}
    \caption{Probability densities of 0.3-2 keV fractional variability RMS amplitude ($F_{\rm var}$), power-law index of the Fourier-resolved spectrum ($\Gamma_{\rm FRS}$), power-law index of the PSD ($\alpha_{\rm PSD}$), power-law index of the X-ray spectrum in sources with a corona ($\Gamma_{\rm spec}$), and $\log(M_{\rm BH}/M_\odot)$, for samples of thermal TDEs (red), TDEs with coronae (purple), and low-mass AGN (blue). We use Gaussian kernel density estimation with a bandwidth $\sigma N^{-1/4}$, where $\sigma$ is the standard deviation of a parameter within a given sample and $N$ is the sample size. AGN $M_{\rm BH}$ uncertainties uniformly were assumed to be 0.3 dex. AGN PSDs were drawn from \cite{Gonzales12} and photon indices from \cite{Liu16}; for both samples, we retained only sources with $\log(M_{\rm BH}/M_\odot)<7.5$ to construct a comparable-mass sample to the TDEs.}
    \label{fig:agn_comparison}
\end{figure*}

We then extended the above analysis to the entire population. Results from our timing and spectral analyses are reported in  Fig.~\ref{fig:agn_comparison} and Table~\ref{tab:population_stats}. We measured the distributions of 0.3-2 keV $F_{\rm var}$, $\alpha_{\rm PSD}$, $\Gamma_{\rm FRS}$, disk $T_P$, disk $R_p$, $F_{\rm sctr}$, spectral photon index ($\Gamma_{\rm spec}$), SMBH mass, and 0.3-10 keV luminosity separately for the thermal and corona populations. We also compared our results to Type 1 AGN with comparable SMBH mass, keeping only sources with $\log (M_{\rm BH}/M_\odot)\leq10^{7.5}$. We drew 0.2-2 keV PSDs from \cite{Gonzales12} (32 total, excluding Seyfert 2 AGN) and spectral fit parameters from \cite{Liu16} (64 total). \cite{Gonzales12} obtained their PSD fits over a broadband frequency range of $>3$ decades (from a lower bound of $\sim 10^{-5}$ Hz in the longest observations to an upper bound of $\sim 10^{-2}$ Hz due to their 100 second time binning). To ensure consistency with our analysis, we integrated their PSD model fits only between $10^{-4}-10^{-3}$ Hz to estimate $F_{\rm var}$ on the same timescales. \cite{Liu16} fit the background-subtracted 2-10 keV spectra of a large sample of AGN using the models \texttt{BNTORUS}, \texttt{PEXMON}, and an unabsorbed power-law, to encapsulate the  intrinsic power law (with absorption and scattering effects), disk reflection of the intrinsic continuum, and the soft excess, respectively. They report the resulting photon indices in their Table 2, which we draw from in our comparison. We note that although we used only their fits to Type 1 AGN, the $\Gamma_{\rm spec}$ distribution is similar for obscured sources as well (see their Fig. 8).

To quantitatively compare the three distributions, we perform pair-wise two sample Kolmogorov-Smirnov (KS) tests, which measures the maximum absolute difference between the empirical distribution of a measured sample and the cumulative distribution function of another reference population. The corresponding $p$-value quantifies the likelihood that the null hypothesis (that both populations are drawn from the same distribution) is true. The results of the KS tests are given in Table~\ref{tab:ks_tests}. We find several statistically significant differences between the thermal TDE, corona TDE, and low-mass AGN populations, which now discuss in detail for each variable.

\begin{table*}
\centering
\caption{X-ray variability and spectral properties of TDEs and AGN. Values are reported as 16th/50th/84th percentiles; dashes indicate parameters not measured for that population. AGN PSDs and spectral parameters are drawn from \cite{Gonzales12} and \cite{Liu16}, respectively. Numbers in parentheses indicate sample size. * indicates the sample sizes used for timing  measurements ($\Gamma_{\rm FRS}$, $\alpha_{\rm PSD}$, and $F_{\rm var}$), which could not be made in all observations (Section~\ref{subsec:data}).}
\begin{tabular}{l|ccc}
\hline\hline
Parameter & Thermal TDEs & Corona TDEs & AGN ($<10^{7.5}~M_\odot$) \\
& (27/14*) & (27/24*) & (64 spectra/34 PSDs) \\
\hline
0.3-2 keV variability ($F_{\rm var}$) & $0.08^{+0.05}_{-0.04}$ & $0.13^{+0.05}_{-0.09}$ & $0.09^{+0.08}_{-0.07}$ \\
PSD slope ($\alpha_{\rm PSD}$) & $0.49^{+0.59}_{-0.12}$ & $1.01^{+0.29}_{-0.48}$ & $2.25^{+0.22}_{-0.45}$ \\
FRS slope ($\Gamma_{\rm FRS}$) & $1.19^{+1.95}_{-0.31}$ & $0.04^{+0.65}_{-0.26}$ & --- \\
Disk $T_p$ ($10^5$ K) & $5.70^{+3.02}_{-1.20}$ & --- & --- \\
Disk $R_p$ ($10^{12}$ cm) & $0.54^{+0.68}_{-0.29}$ & --- & --- \\
Corona scattering fraction ($F_{\rm sctr}$) & --- & $0.25^{+0.24}_{-0.20}$ & --- \\
Corona spectral slope ($\Gamma_{\rm spec}$) & --- & $2.84^{+2.16}_{-0.62}$ & $2.00^{+0.15}_{-0.06}$ \\
$\log_{10}(M_{\rm BH}/M_\odot)$ & $6.48^{+0.64}_{-0.28}$ & $6.66^{+0.95}_{-0.14}$ & --- \\
$\log_{10}(L_X / {\rm erg\,s}^{-1})$ & $42.3^{+1.09}_{-0.56}$ & $43.0^{+0.31}_{-0.45}$ & --- \\
\hline
\label{tab:population_stats}
\end{tabular}
\end{table*}

\begin{table*}
\centering
\caption{Two-sample Kolmogorov-Smirnov test $p$-values comparing TDE and AGN populations. Bold values indicate statistically significant differences ($p < 0.05$). Dashes indicate comparisons not performed.}
\begin{tabular}{lccc}
\hline
Parameter & Thermal vs Corona & AGN vs Thermal & AGN vs Corona \\
\hline\hline
0.3-2 keV variability ($F_{\rm var}$) & $\mathbf{0.03}$ & $0.20$ & $\mathbf{0.02}$ \\
FRS slope ($\Gamma_{\rm FRS}$) & $\mathbf{3.6 \times 10^{-5}}$ & --- & --- \\
PSD slope ($\alpha_{\rm PSD}$) & $\mathbf{0.01}$ & $\mathbf{5.9 \times 10^{-9}}$ & $\mathbf{1.1 \times 10^{-14}}$ \\
Corona spectral slope ($\Gamma_{\rm spec}$) & --- & --- & $\mathbf{4.9 \times 10^{-12}}$ \\
$\log_{10}(M_{\rm BH}/M_\odot)$ & $0.11$ & $0.27$ & $0.78$ \\
$\log_{10}(L_X / {\rm erg\,s}^{-1})$ & $\mathbf{8.9 \times 10^{-3}}$ & --- & --- \\
$\Delta t$ (days) & $0.28$ & --- & --- \\
\hline
\end{tabular}
\label{tab:ks_tests}
\end{table*}

\textbf{Fractional variability amplitude ($F_{\rm var}$):} We find that thermal TDEs exhibit systematically lower 0.3-2 keV variability amplitudes compared to TDEs with coronae. The thermal population shows $F_{\rm var,thermal}=0.08^{+0.05}_{-0.04}$, while TDEs with coronae are more variable with $F_{\rm var,corona}=0.13^{+0.05}_{-0.09}$. A KS test comparing the thermal and corona TDE populations yields $p=0.03$, providing statistically significant evidence that the populations are distinct, and thus that the emergence of a corona enhances soft X-ray variability in TDEs.

When comparing TDEs to low-mass AGN ($M_{\rm BH} < 10^{7.5}~M_\odot$), which show $F_{\rm var,AGN}=0.09^{+0.08}_{-0.07}$ \citep{Gonzales12}, we find both thermal and corona TDEs show $F_{\rm var}$ distributions consistent with AGN ($p=0.10/0.76$ for thermal/corona TDEs, respectively). The AGN distribution appears to show a hint of bimodality, with one mode near the peak of thermal TDEs and another nearer to TDEs with coronae (Fig.~\ref{fig:agn_comparison}). This is almost certainly a mass-related effect, as AGN show a well-known anticorrelation between SMBH mass and X-ray variability amplitude \citep[e.g.][]{Ponti12}. Thus, constructing an AGN sample extending to higher/lower SMBH mass will result in a smaller/larger distribution of $F_{\rm var}$, respectively. This is qualitatively different from TDEs; not only do we find no significant scaling between $M_{\rm BH}$ and $F_{\rm var}$ (Section~\ref{subsec:scalings}), but TDEs are not in long-term steady-state configurations, unlike AGN. Thus, the disk properties---$T_p$, $R_p$, spectral state, and consequently, $\alpha_{\rm PSD}$, $F_{\rm var}$, and other variability metrics---can evolve on human-observable timescales even within a single system, smearing out any putative $M_{\rm BH}$ scaling. On the other hand, AGN show statistically stationary variability properties, enabling short-term variability metrics to serve as robust probes of the underlying system parameters.

\textbf{Energy-dependence of the variability ($\Gamma_{\rm FRS}$):} Of the 12 thermal TDEs with sufficient data quality to measure their Fourier-resolved spectra (FRS), all of them have a positive power-law index ($\Gamma_{\rm FRS}>0$) with characteristic values $\Gamma_{\rm FRS}=1.19^{+1.95}_{-0.31}$, meaning that variability amplitude increases with energy. In contrast, 12/23 corona TDE observations have positive $\Gamma_{\rm FRS}$, and 11/23 have negative $\Gamma_{\rm FRS}$, with a distribution centered on $\Gamma_{\rm FRS}=0.04^{+0.65}_{-0.26}$. A two-sample KS test confirms the difference is significant ($p=3.6\times 10^{-5}$), i.e. the energy-dependence of X-ray variability differs between TDEs with and without coronae.

The FRS or RMS spectra of AGN show a qualitatively diverse range of behaviors, consistent also with the much wider variety of energy spectra observed in AGN between, for example, Compton-thick/thin sources, super-soft AGN, narrow-line Seyfert 1s, and reflection-dominated sources. Figure 2 of \cite{Hu22} gives an illustrative compendium of AGN RMS spectra (note the difference from FRS, and their use of 1-100~ks timescales), showing at least five different shapes including concave-up, concave-down, monotonically increasing, monotonically decreasing, and flat. The FRS of TDEs are simpler and more uniform by comparison, showing broadly monotonic trends with the primary difference imparted by corona emergence. The hard state of XRBs tend to show similarly flat RMS spectra to what we find in the TDEs with coronae, potentially resulting from variability in the normalization \citep{Gierlinski05}. 

\textbf{PSD slopes ($\alpha_{\rm PSD}$):} Thermal TDEs show the flattest PSDs with $\alpha_{\rm thermal}=0.49^{+0.59}_{-0.12}$, closest to white noise. TDEs with coronae occupy an intermediate regime with $\alpha_{\rm corona}=1.01^{+0.29}_{-0.48}$---roughly flicker noise/pink noise---while AGN display the steepest PSDs with $\alpha_{\rm AGN}=2.25^{+0.22}_{-0.45}$, indicating a typical red noise process. We find significant differences across all three populations (Table~\ref{tab:ks_tests}). The progression from flat to steep PSDs across these populations suggests an approximately continuous evolution in the variability properties, potentially related to the development of coronae with increasing strength and/or the transition from super-Eddington to sub-Eddington accretion regimes. As the PSD shape is the most direct quantitative measurement of the underlying stochastic process generating the light curve, the difference in $\alpha_{\rm PSD}$ is the clearest indication that the variability processes between TDEs with and without coronae are fundamentally distinct. As the PSD slope is a rough proxy for the correlation timescale, the emergence of a corona in TDEs increases the correlation timescale of the underlying stochastic variability process causing the soft X-ray fluctuations. At the same time, TDE coronae have shorter coherence times than those of similar-mass AGN. The origins of these changes are uncertain, but we discuss some possibilities in Section~\ref{sec:discussion}, and they are a worthwhile area for future work.

\textbf{Spectral shape ($\Gamma_{\rm spec})$:} The X-ray spectral photon index ($\Gamma_{\rm spec}$) measures the spectral hardness of a corona and reflects the physical properties of the hot Comptonizing plasma (e.g. electron temperature, optical depth). By definition, our thermal TDE sample does not require a corona component in spectral fitting and thus photon indices are not comparable. For the TDE with coronae population, we measure $\Gamma_{\rm spec, corona}=2.84^{+2.16}_{-0.62}$, indicating relatively soft spectra compared to AGN and hard-state XRBs. In comparison, the low-mass AGN sample from \cite{Liu16} are a significantly harder and lower-variance population, with $\Gamma_{\rm spec, AGN}=2.00^{+0.15}_{-0.06}$. Previous studies have found the softer photon indices in TDEs with coronae reflect higher optical depths and cooler electron temperatures than AGN \citep{Cao23,Cao24,Liu23,Saxton2025}, which we discuss in Section~\ref{subsec:corona_agn_xrb}.

\subsection{Scaling relations (and lack thereof)} \label{subsec:scalings}

\begin{figure*}
    \centering
    \includegraphics[width=\linewidth]{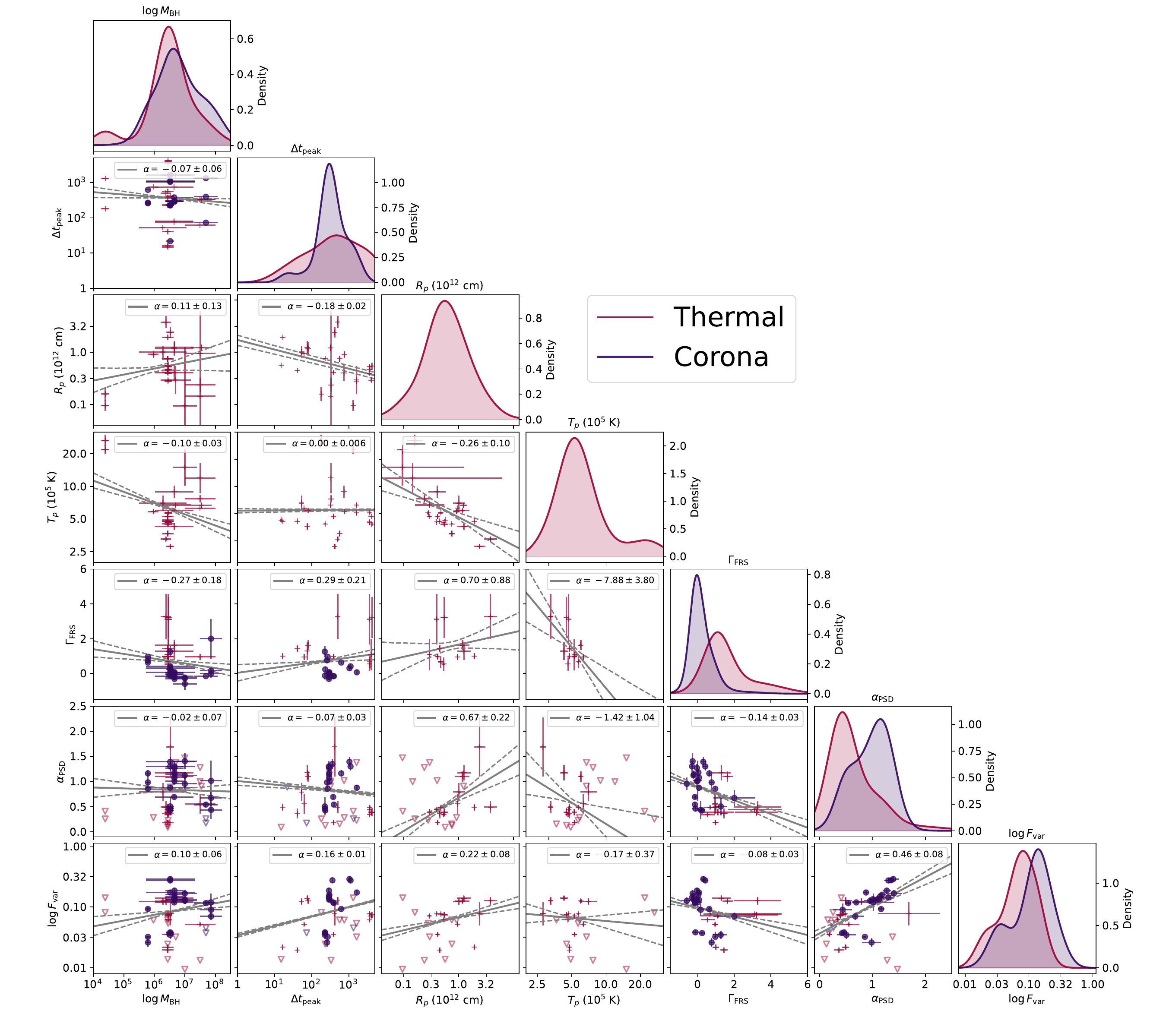}
    \caption{Corner plot showing the distributions and covariances of disk peak temperature/radius (measured from the spectrum) and the 0.3-2 keV fractional variability amplitude (from the light curve). Diagonal panels show probability densities estimated via Gaussian kernel density estimation with adaptive bandwidth $\sigma N^{-1/5}$ (same definitions as Fig.~\ref{fig:agn_comparison}).}
    \label{fig:rp_tp_fvar_rms_psd}
\end{figure*}

\begin{table*}
\centering
\caption{Best-fit linear models of the form $x\propto ay$, where $x$ is the row variable, $y$ is the column variable, and $a$ is the reported value. The fits and errors are determined via bootstrap resampling. Bold values indicate correlations with $>3\sigma$ significance. $T_p$ and $R_p$ use thermal TDE measurements only; $\Gamma_{\rm spec}$ and $F_{\rm sctr}$ use TDEs with coronae only; other parameters use all valid observations. $L_X$ is measured from $0.3-10$ keV.}
\hspace*{-3.3cm}
\resizebox{1.2\textwidth}{!}{%
    \begin{tabular}{l||c|c|c|c|c|c|c|c|c}
    Parameter & $\log F_{\rm var}$ & $\alpha_{\rm PSD}$ & $\Gamma_{\rm FRS}$ & $\log (L_X/\rm erg\;s^{-1})$ & $\log (T_p/10^5\;\rm K)$ & $\log(R_p/10^{12}\;\rm cm)$ & $\Gamma_{\rm spec}$ & $F_{\rm sctr}$ & $\log (\Delta t/\rm days)$ \\
    \hline\hline
    $\alpha_{\rm PSD}$ & $\mathbf{+0.46 \pm 0.08}$ & & & & & & & & \\[0.3em]
    \hline
    $\Gamma_{\rm FRS}$ & $-0.08 \pm 0.03$ & $\mathbf{-0.14 \pm 0.03}$ &  & & & & & & \\[0.3em]
    \hline
    $\log (L_X/\rm erg\;s^{-1})$ & $\mathbf{-0.10 \pm 0.02}$ & $+0.05 \pm 0.06$ & $-0.56 \pm 0.22$ &  & & & & & \\[0.3em]
    \hline
    $\log (T_p/10^5\;\rm K)$ & $-0.17 \pm 0.37$ & $-1.42 \pm 1.04$ & $-7.88 \pm 3.80$ & $-0.23 \pm 0.14$ &  & & & & \\[0.3em]
    \hline
    $\log(R_p/10^{12}\;\rm cm)$ & $+0.22 \pm 0.08$ & $\mathbf{+0.67 \pm 0.22}$ & $+0.70 \pm 0.88$ & $+0.52 \pm 0.22$ & $-0.26 \pm 0.10$ &  & & & \\[0.3em]
    \hline
    $\Gamma_{\rm spec}$ & $\mathbf{-0.23 \pm 0.01}$ & $\mathbf{-0.11 \pm 0.03}$ & $+0.16 \pm 0.07$ & $\mathbf{+0.17 \pm 0.02}$ & --- & --- &  & & \\[0.3em]
    \hline
    $F_{\rm sctr}$ & $\mathbf{+0.58 \pm 0.04}$ & $\mathbf{+0.33 \pm 0.08}$ & $-0.45 \pm 0.16$ & $\mathbf{+0.38 \pm 0.08}$ & --- & --- & $\mathbf{-1.19 \pm 0.13}$ &  & \\[0.3em]
    \hline
    $\log (\Delta t/\rm days)$ & $\mathbf{+0.16 \pm 0.01}$ & $-0.07 \pm 0.03$ & $+0.29 \pm 0.21$ & $\mathbf{-0.48 \pm 0.002}$ & $+0.00 \pm 0.006$ & $\mathbf{-0.18 \pm 0.02}$ & $\mathbf{-1.44 \pm 0.13}$ & $\mathbf{+0.36 \pm 0.03}$ &  \\[0.3em]
    \hline
    $\log (M_{\rm BH}/M_\odot)$ & $+0.10 \pm 0.06$ & $-0.02 \pm 0.07$ & $-0.27 \pm 0.18$ & $-0.05 \pm 0.06$ & $\mathbf{-0.10 \pm 0.03}$ & $+0.11 \pm 0.13$ & $\mathbf{-0.68 \pm 0.16}$ & $+0.02 \pm 0.06$ & --- \\[0.3em]
\end{tabular}%
}
\label{tab:corr}
\end{table*}

To assess the presence of scaling relations between our measurements, we performed linear regression with bootstrapping between each pair of variables. We generated 1000 random realizations of the data points centered on the measured values, and perturbed their values assuming their measurement uncertainties represent independent $1\sigma$ gaussian errors. For each realization we performed a linear fit, and report the resulting mean and standard deviation from the 1000 regressions. We used this procedure for each pair-wise combination using the combined thermal+corona populations (unless otherwise noted) of fractional RMS variability amplitude ($F_{\rm var}$), FRS power-law index ($\Gamma_{\rm FRS}$), PSD power-law index ($\alpha_{\rm PSD}$), peak disk temperature ($T_p$; thermal TDEs only), radius of $T_p$ ($R_p$; thermal TDEs only), spectral photon index ($\Gamma_{\rm spec}$; corona TDEs only), scattering fraction ($F_{\rm sctr}$; corona TDEs only), SMBH mass ($\log M_{\rm BH}/M_\odot$), X-ray luminosity ($\log L_{0.3-10\rm keV}$), and time since peak luminosity ($\Delta t_{\rm peak}$; we used the optical peak for optically-selected TDEs, X-ray peak for X-ray selected TDEs, and excluded repeating TDEs from this measurement). We report the resulting scalings in Table~\ref{tab:corr}, and plot a subset of then in Fig.~\ref{fig:rp_tp_fvar_rms_psd}. We identified significant ($>3\sigma$) correlations between the following parameters:
\begin{itemize}    
    \item \textit{Positive correlation between $\alpha_\mathrm{PSD}$ with $F_{\rm var}$:} More variable sources show steeper PSDs, which is again driven partly of the two separate populations (Fig.~\ref{fig:rp_tp_fvar_rms_psd}). This also naturally arises from the measurement of $F_\mathrm{var}$ as the square root of the integral under the PSD; steeper PSDs will yield larger $F_\mathrm{var}$ (although normalization also impacts this measurement).

    \item \textit{Negative correlation between $\alpha_{\rm PSD}$ and $\Gamma_{\rm FRS}$:} Steeper PSDs are associated with a flatter FRS, which is again mostly an artifact of the two TDE populations (Fig.~\ref{fig:rp_tp_fvar_rms_psd}).

    \item \textit{Positive correlation between $F_{\rm sctr}$ and $F_{\rm var}$:} Stronger coronae show higher fractional variability amplitude. This fits with numerous literature findings that coronae are compact, turbulent, and rapidly variable (e.g. \citealt{Bambic24,Kara25}). It is therefore natural to find that as its contribution to the total emission increases, so too does its imprint on the overall variability.

    \item \textit{Negative correlation between $F_\mathrm{var}$ and $L_{0.3-10\,\mathrm{kev}}$:} Higher luminosity sources are less variable. Naively, this could be explained for thermal sources as arising from observing closer to the peak temperature in hotter disks, but such a correlation with disk temperature is not seen (see below). This could also be a selection effect, indicating that we can only measure low $F_\mathrm{var}$ in the most luminous sources.

    \item \textit{Positive correlation between $F_\mathrm{var}$ and $\Delta t_{\rm peak}$:} Sources are more variable at late times. This effect is likely dominated from the variable coronae forming at late times.

    \item \textit{Positive scalings of both $F_\mathrm{var}$ and $\alpha_\mathrm{PSD}$ with $R_p$:} Larger disks show steeper PSDs, which may be an effect of the longer characteristic timescales associated with the larger physical scale. The PSD integral also increases (the $F_{\rm var} \propto \alpha_{\rm PSD}$ relation), which could also explain the $F_{\rm var}$ scaling with $R_p$. 

    \item \textit{Scalings of both $F_{\rm sctr}$ and $\Gamma_\mathrm{spec}$ with $\Delta t_{\rm peak}$:} The scattering fraction and photon index evolve significantly with time, in opposite directions, indicating TDE coronae become generally stronger and harder over time. This is consistent with other studies of individual sources \citep[e.g.,][]{Saxton2025,Berger26} and spectroscopy-based sample works \citep[e.g.,][]{Wevers20}.

    \item \textit{Scalings of both $\Gamma_{\rm spec}$ and $F_{\rm sctr}$ with $F_{\rm var}$ and $\alpha_{\rm PSD}$:} We find that coronae which are stronger (higher $F_{\rm sctr}$) and harder (lower $\Gamma_{\rm spec}$) are associated with higher variability and redder noise, in accordance with a picture where the TDEs which host more AGN-like coronae also exhibit more AGN-like variability.

\end{itemize}

We also recover a number of correlations that have been previously observed. Notably, we find that the temperature of the disk scales inversely with radius, black hole mass, and luminosity, confirming that on average more massive black holes have cooler disks, smaller disks are hotter, and disks cool with time \citep[e.g.][]{Auchettl17,Mummery23,Guolo25b}. Additionally, we recover that larger disks are more luminous and that generally the X-ray luminosity of TDEs decreases with time, which are both intuitively expected. 

On the other hand, some scaling relations are \textit{absent} which would otherwise be expected a priori. They are:
\begin{itemize}
    \item \textit{No scaling between $T_p$ and $F_{\rm var}$:} One would expect that the disk temperature is anti-correlated with $F_\mathrm{var}$ as a result of the exponential enhancement of Wien tail variability argued in \cite{Mummery22} and indeed observed in the slopes of our Fourier-resolved spectra (Section~\ref{subsec:population}). Lower-$T_p$ disks are observed deeper in the Wien tail for a fixed observing energy band (0.3-2 keV), so they should be more variable than higher-$T_p$ disks due to the exponential dependence of Wien-tail flux on temperature, $F\propto\exp(-h\nu/kT)$. This effect is \textit{not} seen, and the implications are discussed in Section~\ref{subsec:physical_interp}. 
    
    \item \textit{No scaling between $\log M_{\rm BH}$ and $F_{\rm var}$:} AGN are known to show a tight, highly significant inverse scaling between $F_{\rm var}$/$\sigma^2_{\rm rms}$ and SMBH mass (e.g. \citealt{Ponti12}), with the X-ray variability even used in some cases as an estimator for the black hole mass. It is therefore noteworthy that such a scaling is \textit{not} observed in TDEs, even when we restrict our analysis to the 2-10 keV band for TDEs with coronae to allow a more fair comparison to AGN. We speculate this is because of the steady-state nature of AGN accretion flows, whose variability is believed to be generated from a statistically stationary process over sufficiently long timescales. In contrast, TDEs host newborn, rapidly evolving accretion disks, and the turbulent variability changes dramatically even within individual sources on human-observable timescales (see Section~\ref{subsec:population}). On the other hand, our observations probe only a small range of SMBH masses, and it is possible that a larger sample over a greater dynamic range could reveal some scaling.
\end{itemize}

\subsection{Evolution of individual sources} \label{subsec:evol}

\begin{figure*}
    \centering
    \includegraphics[width=\linewidth]{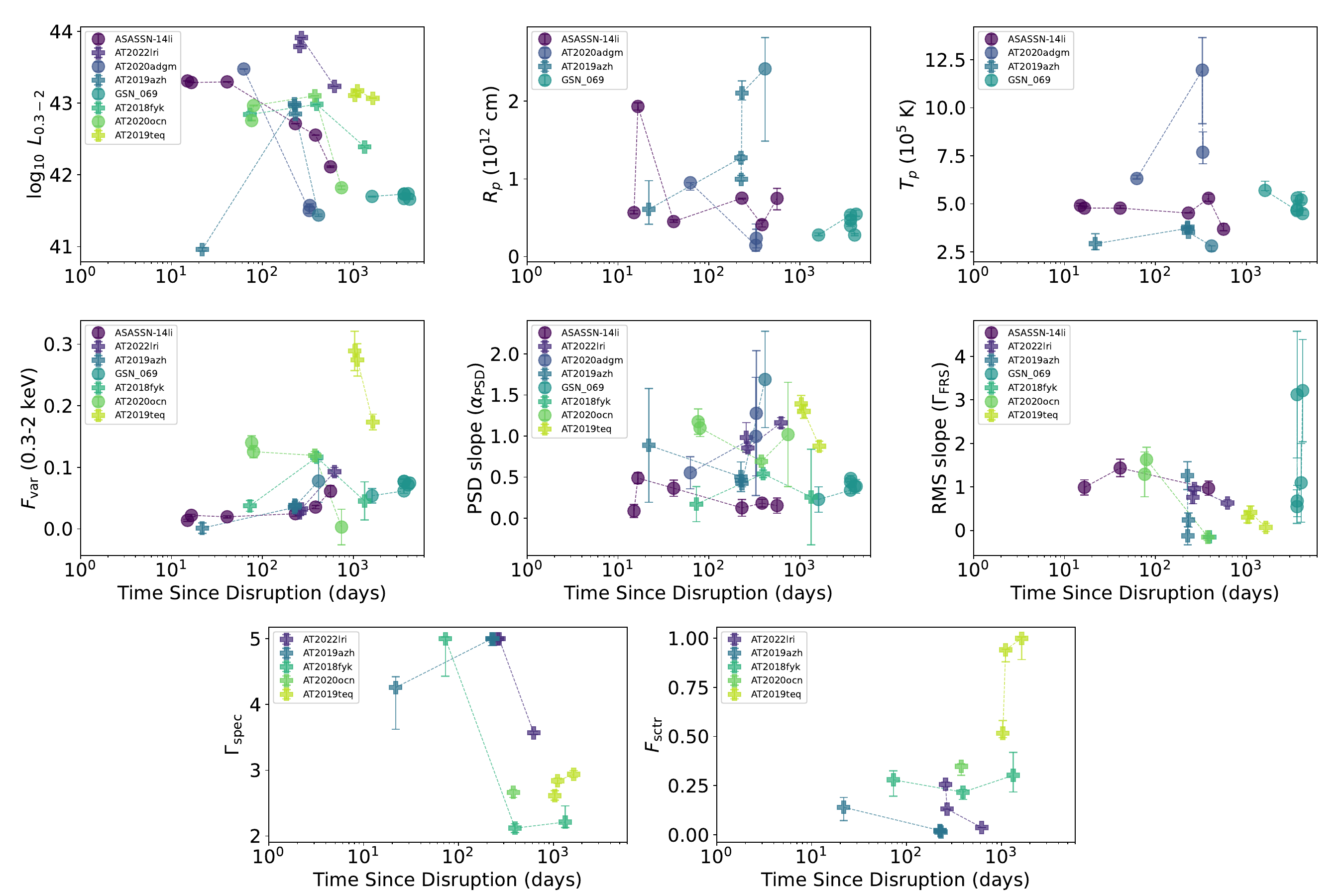}
    \caption{Evolution of spectral and variability properties in eight sources having multi-epoch \XMM\ observations. Thermal/corona epochs are denoted by circle/plus markers respectively.}
    \label{fig:evol}
\end{figure*}

In Fig.~\ref{fig:evol} we show the time evolution of 0.3-2 keV X-ray flux, $T_p$, $R_p$, $F_{\rm var}$, $\alpha_{\rm PSD}$, $\Gamma_{\rm FRS}$, $F_{\rm sctr}$, and $\Gamma_{\rm spec}$ for the 8 sources which have at least three epochs spaced by $>100$ days and sufficient data quality to measure the PSD. We mark thermal epochs with circles and corona epochs with crosses. Overall, the behaviors are unpredictable and diverse. In the most ``well-behaved'' source, ASASSN-14li, we see the luminosity progressively decrease while the disk cools, and the fractional variability increases alongside. The same is seen in GSN 069, albeit at much later times relative to first detection. Some other sources show a similar long-term luminosity decline (AT2020adgm, AT2022lri, AT2019azh), though it is not always accompanied by a corresponding increase in the variability. Some sources show relatively small flux evolution over the timescales probed in our sample, such as AT2019teq and GSN 069, potentially indicative of a longer viscous time in their accretion disks \citep{Guolo25a}. 

The development of an X-ray corona associated with a soft-to-hard state transition can also provide insight into microphysics of coronal heating and disk-corona coupling. Of the well-monitored systems, four have undergone clear and previously reported state transitions (AT2018fyk -- \citealt{Wevers21}, AT2021ehb -- \citealt{Yao22}, AT2020ocn -- \citealt{Guolo24}, AT2019teq -- \citealt{Berger26}). Of these sources, AT2018fyk shows a clear increase in variability from its early state ($\Delta t_{\rm peak} \lesssim 100$ days) to an observation at roughly a year post-disruption, indicative of a long-term increase in variability coinciding with the emergence of a corona in the spectrum. However, this trend is not universal; AT2020ocn shows an overall decreasing $F_{\rm var}$ with time, even as it forms a corona. Together with the scaling relations, these results suggest that while the presence of a corona generally increases variability, but other properties (e.g., $F_{\rm scat}, \Gamma$) can dictate the variability evolution in a given source.

Finally, we note that the $\Gamma_{\rm FRS}$ and $\alpha_{\rm PSD}$ measurements show heterogeneous behavior, often evolving non-monotonically in individual sources. As the uncertainties are relatively larger for these measurements, and there are also likely to be several competing effects setting those values---particularly in sources with both thermal and corona components---it is difficult to speculate on the underlying cause. Longer observations of TDEs with the next generation of X-ray observatories (e.g., AXIS, NewAthena) will be critical for better measuring these parameters and probing the detailed evolution.

\subsection{Outflow spectral properties} \label{subsec:outflow}

\begin{figure}
    \centering
    \includegraphics[width=\linewidth]{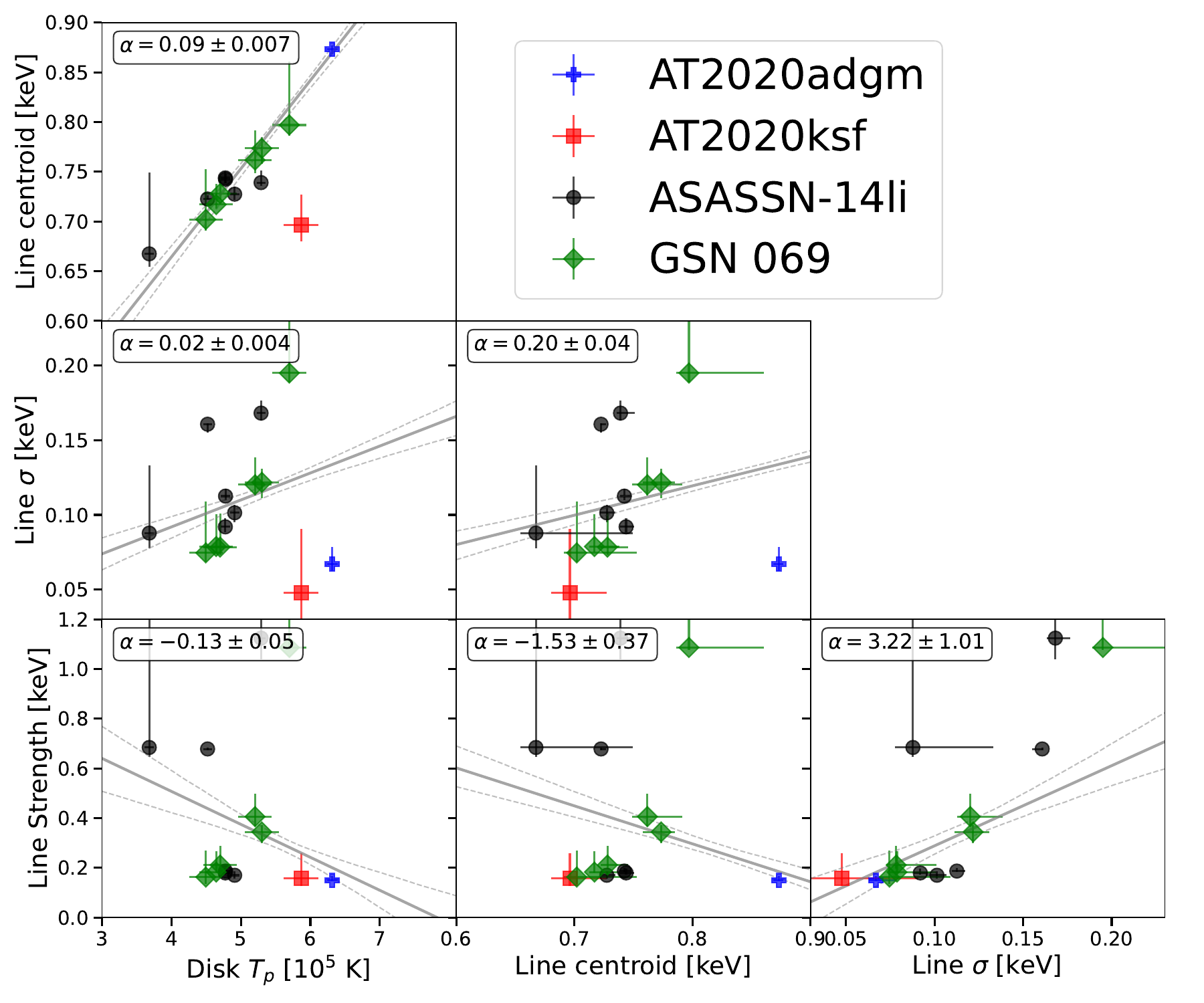}
    \caption{For sources showing a significant outflow absorption component (ASASSN-14li, AT2020adgm, AT2020ksf, GSN 069), we have modeled the absorption with a gaussian and shown each pair-wise combination of disk $T_p$, gaussian centroid, $\sigma$, and strength.}
    \label{fig:outflow}
\end{figure}

A serendipitous byproduct of our analysis is a population-wide survey of TDEs with significant absorption features in their EPIC-pn spectra. In Fig.~\ref{fig:outflow} we show pair-wise correlations of the absorption line properties (centroid $E_{\rm cent}$, width $\sigma$, strength $A$) and the disk temperature ($T_p$) along with linear fits; although these absorption features have all been previously reported \citep{Kara18,Kosec23,Wevers24,Pasham24,Kosec25}, to our knowledge this is the first systematic study of how such line properties correlate with the continuum. We observe several suggestive correlations between the line properties and continuum temperature, albeit with large scatter. While some tentative scalings may be seen between the line properties themselves, they are relatively higher-uncertainty fits and thus more difficult to characterize. 

It is constraining that the line and continuum properties show several correlations. A scaling between temperature and line energy has been observed in the quasi-periodic eruptions (QPEs) of ZTF19acnskyy, with a linear relation of approximately $(T/10^5\;\mathrm{K})\propto (10.8\pm 0.56)(E_{\rm cent}/\rm keV)$ \citep{Chakraborty25b}, remarkably similar to the scaling we find for TDEs. Given the different radiation and outflow properties predicted in shock-powered QPE models (e.g. \citealt{Vurm25}), it is surprising that the presumably radiation pressure-driven disk winds we observe in TDEs follow a similar relation. We speculate that this may be due, at least in part, to the change in ionizing continuum for different-temperature blackbodies. Due to the exponential dependence of the ionizing flux on the disk temperature at high energies, even modest changes in $T_p$ will result in a large difference in the number of ionizing photons for a given line species. The resulting line balance will vary sensitively with the continuum temperature, with all lower-energy transitions becoming fully stripped for even modest increases in the temperature. 

Another contributing effect is the dependence of outflow velocity on accretion rate, which itself will set the observed temperature and have a qualitatively similar effect to increasing the abundance of higher ionization photons. As we adopt only phenomenological gaussian modeling for our spectra---rather than full photoionization models which self-consistently fit for outflow column density, ionization state, and bulk velocity---there is a degeneracy between the dominant line species contributing to absorption (set by the ionization parameter) and the outflow velocity. Models for radiation pressure-driven winds in TDEs \citep{Dai18} and AGN \citep{Nardini15} indeed predict a correlation between $\dot{M}$ and outflow velocity, making it difficult to disentangle the effect of systematically varying ionization parameter from that of higher wind speeds. However, we note that the strong decline in the line strength with disk temperature seems to favor the ionization driven changes. As the accretion rate drops, we expect the disk to cool and the line strength to drop \citep{Giustini2019}, which is opposite to the correlation we observe. On the contrary, if the trends are driven by ionization state changes, then the hotter disks will lead to higher ionization state and on average fewer ionic species and weaker lines. We defer testing this hypothesis to a future analysis of high-resolution X-ray data that can fully resolve degeneracies between the ionization state and bulk velocity.

It is interesting to note that such evolution has not been seen in sources that show predominantly emission features; notably, both AT2022lri, a TDE in our sample with a corona, and 1ES 1927+654, a potential TDE in an AGN \citep{Trakhtenbrot2019,Ricci2020}, show features best modeled with emission at $\sim 1$keV \citep{Ricci2021,Masterson2022,Yao24}. The line properties in both of these sources seem to be independent of the underlying continuum, suggesting a different physical mechanism behind those lines seen primarily in emission versus those seen in absorption. \cite{Masterson2022} suggested that the 1 keV line in 1ES 1927+654 arises from reflection from a hot, geometrically thick disk illuminating itself and outflowing at $v \sim 0.3c$; face-on viewing angles required to see such emission are also where we would expect to see the maximum line-of-sight velocity, as these are often faster than the typical outflows seen in absorption. Again, however, future work and more high-quality data is needed to break degeneracies between ionization state and velocity shift in these models.

\subsection{Hard X-ray variability}

The majority of the TDEs in our sample are dominated by thermal emission in the soft X-ray band, even those with coronae. To disentangle the variability of the disk and corona, we also fit the 2-10 keV PSDs for most TDEs with coronae. We excluded ASASSN-14ko, as its hard X-ray flux is dominated by a companion dual AGN; and AT2019azh, as its hard X-ray count rate is too low. We show the resulting PSDs in Appendix Figure \ref{fig:psds_hard}.  The hard X-ray PSD slopes are still systematically lower than those of AGN, with a 16/50/84th percentile distribution of $\alpha_{\rm PSD}=0.94^{+0.77}_{-0.58}$. This confirms our result that corona-induced variability in TDEs is indeed fundamentally different from those in AGN, showing shallower PSDs with shorter coherence timescales. Notably, we find that the hard and soft PSD slopes are not strongly correlated; we suspect that this is driven by large uncertainties on the hard X-ray PSD slopes  ($\sim 3\times$ larger than the soft PSDs). 

\section{Discussion} \label{sec:discussion}

\subsection{TDE coronae: properties and comparison to AGN and X-ray binaries} \label{subsec:corona_agn_xrb}

\begin{figure*}
    \centering
    \includegraphics[width=\textwidth]{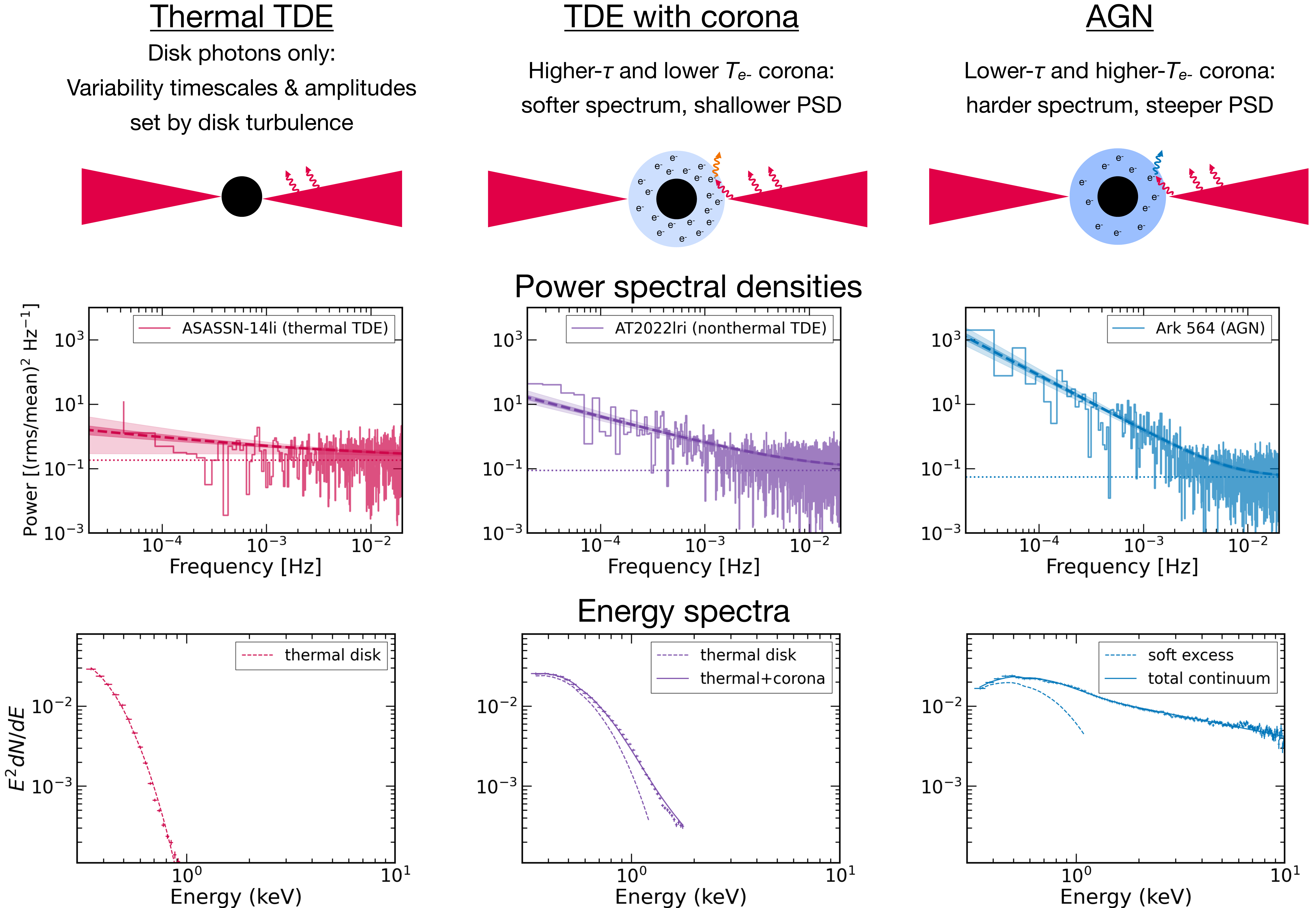}
    \caption{\textbf{Top row:} Schematic of our spectral-timing results for thermal and corona TDEs and similar-mass AGN. Purely thermal TDEs show the whitest noise (shortest $t_{\rm corr}$). TDEs which form a corona show redder noise with $\alpha\sim 1$ (longer $t_{\rm corr}$), but still with a steeper photon index and whiter noise than in AGN coronae. TDE coronae are typically associated with higher optical depth than those in AGN, which appears to result in both shorter $t_{\rm corr}$ and less energetic photons. In comparison, AGN show red noise ($\alpha\sim 2$) and harder photon indices. \textbf{Middle/bottom row:} PSDs/energy spectra for representative thermal and corona TDEs and AGN.}
    \label{fig:schematic}
\end{figure*}

Our main finding is that the spectral-timing properties of TDEs with and without coronae are distinct, and even TDEs with coronae differ significantly from AGN. We show a schematic representation of this in Fig.~\ref{fig:schematic}, along with representative examples of PSDs and energy spectra for each class. For comparison to AGN, we use data from the low-mass AGN Ark 564 \citep{McHardy2007,Kara2017}; this is a well-studied, highly variable, and unobscured AGN with $\log M_{\rm BH} \approx 6.3$ \citep{Lewin2022}, comparable to TDEs. Most notably, TDEs with a corona show: 1) larger 0.3-2 keV fractional variability on 1-10 kilosecond timescales, at the $\sim 13$\% level compared to $\sim 8$\% for thermal TDEs; 2) steeper PSD slopes akin to flicker/pink noise ($\alpha\sim 1$) as opposed to whiter noise ($\alpha\sim 0.5$) in thermal TDEs and red noise in AGN ($\alpha\sim 2$); and 3) a weaker energy-dependence than thermal sources, which all show increasing variability with energy (Fig.~\ref{fig:agn_comparison} top-right). Moreover, even within the corona TDE population we observe a scaling of $F_{\rm var}$ with $F_{\rm sctr}$ (Table~\ref{tab:corr}), indicating a continuous process whereby increasing the corona strength results in larger-amplitude variability. It is noteworthy that XRBs also show enhanced variability in the hard, corona-dominated state ($\gtrsim 20$\% fractional rms variability) compared to the soft thermal state ($\lesssim 5$\%, \citealt{Belloni16}), which is qualitatively similar to the state transitions in our TDE sample.

Our finding that corona formation is accompanied by a transition from white-noise-like PSDs to pink-noise PSDs indicates an increased correlation timescale ($t_{\rm corr}$), which can intuitively be understood as the length of ``memory'' of underlying stochastic process, i.e. the duration over which two points in the time series remain correlated. Precisely defining $t_{\rm corr}$ requires assuming a functional form for the correlation function, but qualitatively, any monotonically decreasing function is reasonable. One possible reason for the longer $t_{\rm corr}$ is that higher-energy photons escape from the corona after one diffusion time, which may be long compared to the characteristic timescale of disk turbulence (Section~\ref{subsec:physical_interp}). Moreover, the increase in $F_{\rm var}$ between thermal and corona TDEs may result from scattering of the already-variable seed disk photons by independently varying nonthermal heating/cooling of the corona (see e.g. Figure 7 of \citealt{Mummery21a}).

Another key result is that TDE coronae have unique PSD slopes, occupying intermediate values between the thermal TDE and AGN populations (Fig.~\ref{fig:agn_comparison}) and comparable to soft-state XRBs (e.g. \citealt{Arcodia21}). Previous spectral fitting studies have found individual TDE coronae also have higher optical depths and lower electron temperatures than those in AGN \citep{Cao23,Cao24,Liu23,Saxton2025}; our results indicate this is associated with shorter-timescale variability, though the mechanism is unclear. One possibility is that seed photons penetrate less deeply into higher-$\tau$ coronae, thereby encountering fewer $\gamma$-$e^-$ scatterings to yield a shorter correlation timescale.  Another possibility is that TDE coronae may be more compact or have a different corona temperature gradient as this can suppress short-timescale variability \citep{Arevalo2006}. In any case, these explanations are merely speculative, and our results motivate follow-up study of the relation between coronal properties and variability timescale.

\begin{figure}
    \centering
    \includegraphics[width=\linewidth]{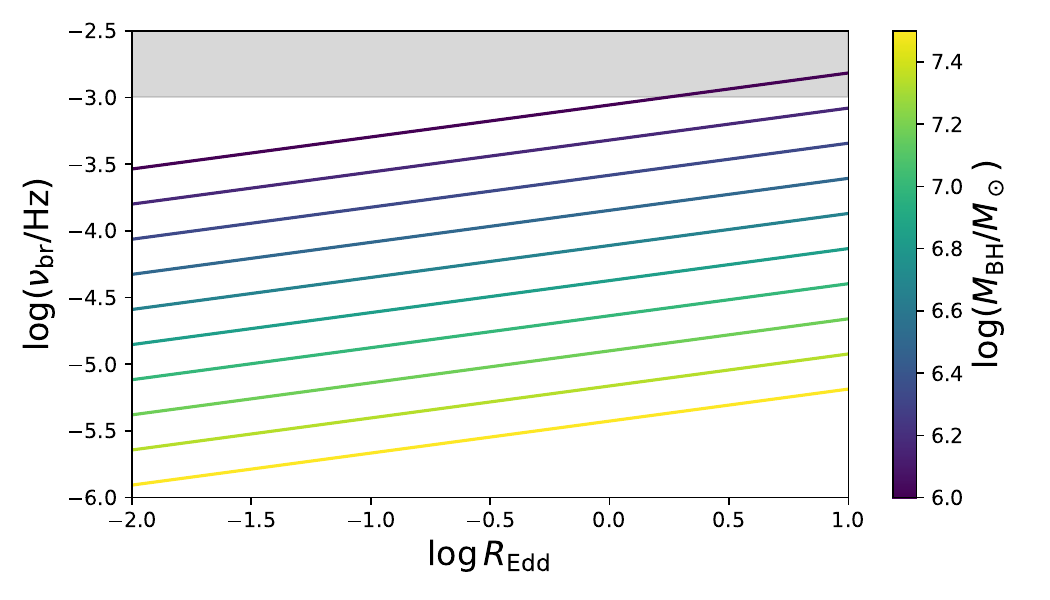}
    \caption{Predicted break frequency \citep{Gonzales12}, as a function of Eddington ratio ($R_{\rm Edd}$), for different SMBH masses in the range $10^6-10^{7.5} M_\odot$. The shaded region represents break frequencies above our sensitivity range, below which we would observe only the flattened ($\alpha_{\rm PSD}\sim 1$) portion of the PSD.}
    \label{fig:break_freqs}
\end{figure}

The PSDs of AGN and XRBs may also show a break in variability power, where the PSD power-law slope changes discretely to a different value. For instance, a few dozen AGN are known which show a PSD break at frequencies of order $10^{-6}-10^{-3}$ Hz, below which one observes $P(f)\propto f^{-1}$ and above which we see $\alpha_{\rm PSD}\sim2$ red noise. These break frequencies ($\nu_{\rm br}$) scale inversely with $M_{\rm BH}$ and proportionally with Eddington ratio/bolometric luminosity across AGN and XRBs, spanning $\sim 8$ orders of magnitude \citep{McHardy06,Gonzales12}. The short durations of TDE observations mean that we have only a few, poorly constrained low-frequency data points with which to constrain such a break. Likewise, the low count rates means that we are only sensitive to PSD breaks up to $\sim 10^{-3}$ Hz.  To assess the possibility that we are observing the low-frequency $\alpha_{\rm PSD} \sim 1$ PSD regime in rapidly accreting, low-mass SMBHs in TDEs, we plot the predicted $\nu_{\rm br}$ as a function of Eddington ratio for SMBH masses in the range of $10^6-10^{7.5}M_\odot$ in Fig.~\ref{fig:break_freqs}. For all but the most extreme cases of small $M_{\rm BH}$ and high $R_{\rm Edd}$, $\nu_{\rm br}$ lies below $10^{-3}$ Hz, meaning it is unlikely that the observed difference in PSD slopes is driven by $\nu_{\rm br}$. 

\subsection{Physical interpretation of thermal TDE variability} \label{subsec:physical_interp}

\begin{figure*}
    \centering
    \includegraphics[width=\linewidth]{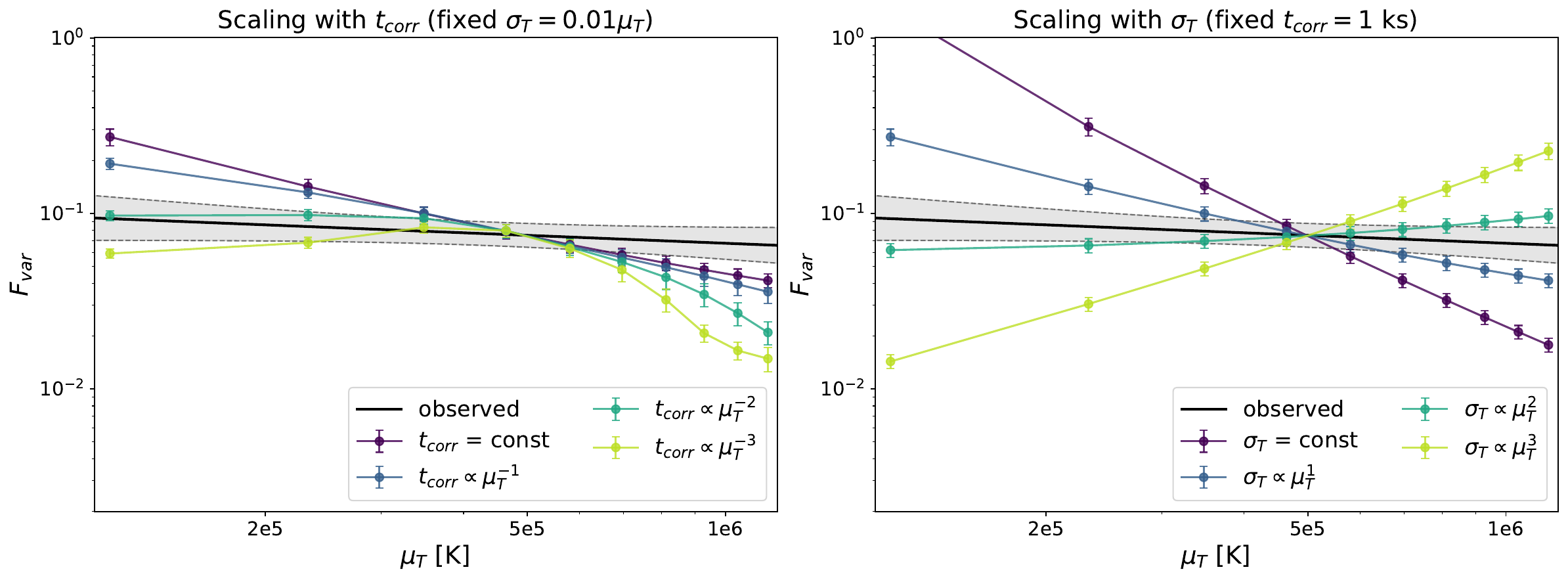}
    \caption{\textbf{Left:} The effect on observed $F_{\rm var}$ (at $>0.3$ keV) of trial $t_{\rm corr}\propto \mu_T^{\alpha}$ scalings. We tested values of $\alpha\in\{0,-1,-2,-3\}$. \textbf{Right:} same plot for $\sigma_T\propto \mu_T^{\alpha}$ scalings for $\alpha\in\{0,1,2,3\}$.}
    \label{fig:lc_sim}
\end{figure*}

\begin{figure}
    \centering
    \includegraphics[width=\linewidth]{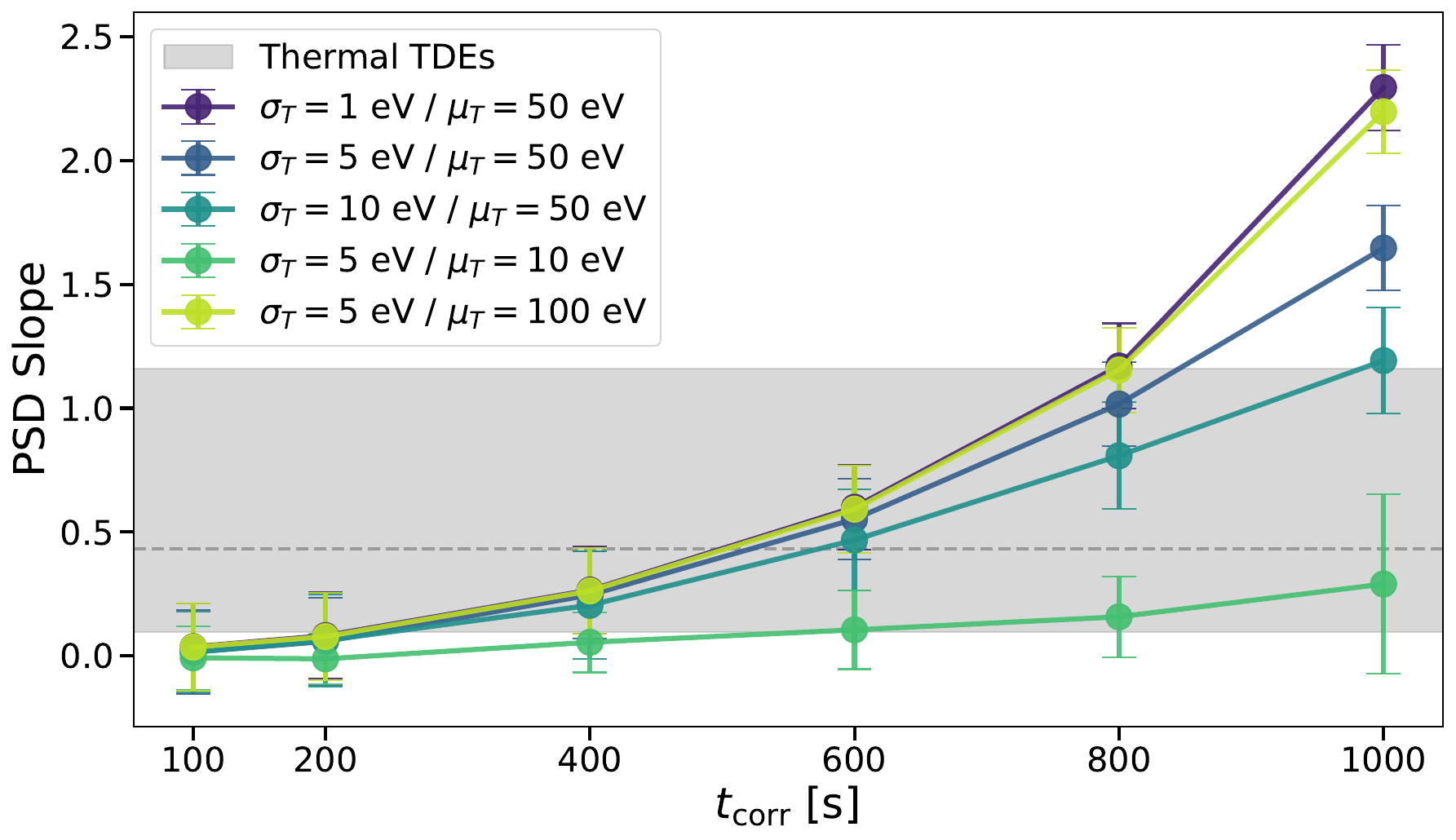}
    \caption{Mean and $1\sigma$ distributions of $\alpha_{\rm PSD}$ resulting from 50 realizations of stochastic light curves for various choices of $\sigma_T$/$\mu_T$, as a function of $t_{\rm corr}$. We shade the region of $\alpha_{\rm PSD}$ measured for the thermal TDE sample, finding that $t_{\rm corr}\lesssim 1$ ks is consistent with the population.}
    \label{fig:psd_slope_sim}
\end{figure}

We now turn our attention towards interpreting the results and their implications for the microphysical turbulent processes in TDE accretion disks, particularly those in pure-thermal states as the presence of only one continuum component makes them the simplest to model. Our observation that thermal disk-dominated TDEs exhibit systematically lower variability amplitudes, flatter PSDs, and stronger energy-dependent variability compared to TDEs with coronae can be understood through recent theoretical advances in accretion disk turbulence modeling, such as the numerical stochastic-viscosity hydrodynamic simulations of \cite{Mummery22}, \cite{Turner23}, and \cite{Mummery24b}. These simulations find that disk geometry---specifically the aspect ratio $H/R$ and associated turbulent coherence length---is a dominant factor controlling the observed variability amplitude at fixed observing energy. To intuitively understand why $H/R$ sets the variability, consider that there are on the order of $N_{\rm turb}\sim2\pi r/H$ independent turbulent cells in the disk; in geometrically thin disks, where $H/R \ll 1$, there are $N_{\rm turb}\gg 1$ cells contributing to the variability, whereas for thick disks with $H/R\sim1$ there are $N_{\rm turb}\sim2\pi$. When integrated over the entire emitting region, if there are many independent cells the variability in effect averages out (an example of the central limit theorem), resulting in higher variance for low $N_{\rm turb}$ and lower variance for $N_{\rm turb}\gg1$. Thus, if corona formation coincides with disk thickening, as might be expected for less radiatively efficient accretion flows, the accretion disk variability itself may also be suppressed. \cite{Turner23} found that doubling the coherence length from $H/R=0.1$ to 0.2 nearly doubles the temperature variance and produces factor $\sim$3 enhancements in luminosity variability, comparable to our observed difference between the medians of the thermal TDE ($F_{\rm var} \sim 8$\%) and corona TDE ($F_{\rm var}\sim 13$\%) populations.

The energy dependence of variability provides additional physical insight. \cite{Mummery22} showed that when disks are observed in the Wien tail, small temperature fluctuations are exponentially amplified into large luminosity variations, with the amplification factor increasing at higher photon energies. They demonstrated that even $\sim~1.5$\% temperature fluctuations can produce factor 2-5 luminosity variations when observed sufficiently deep in the Wien tail. This mechanism naturally explains why all 12 thermal TDEs with sufficient data quality to measure the Fourier-resolved spectrum (FRS) show a positive $\Gamma_{\rm FRS}$: the variability amplitude increases with energy because we probe deeper into the Wien tail. In sources with a corona, Comptonization redistributes photons to higher energies through a fundamentally different physical process that does not necessarily preserve the local temperature fluctuation signature, resulting in the more heterogeneous and generally flatter FRS trends we observe.

Naively extending this argument, one would infer that for a fixed observing band, lower-temperature sources should show systematically higher variability, as they are observed deeper in the Wien tail. However, we \textit{do not observe} this predicted anticorrelation between $F_{\rm var}$ and peak temperature ($T_p$) across the population of distinct sources. This result is not changed if we attempt to account for the different size scales of the X-ray emitting regions (which should roughly correlate with black hole mass) by measuring $F_\mathrm{var}$ at a frequency that scales with $R_p$. A resolution to this apparent contradiction would be found if the underlying variability properties (in effect the fractional variability in the temperature $\sigma_T/\mu_T$) systematically scaled with temperature in a manner which offsets the Wien tail enhancement. To test this hypothesis, we performed light curve simulations following the method of \cite{Mummery25}. We simulate stochastic temperature variability assuming that the disk temperature field is log-normally distributed \citep[as seen in simulations;][]{Turner23, Mummery24b} with mean $\mu_T$ and standard deviation $\sigma_T$ and that the temperature field has a fixed correlation time of $t_{\rm corr}$. This stochastic temperature field is then ``observed'' in the Wien tail, and the X-ray luminosity is computed as a function of time. 

In Fig.~\ref{fig:lc_sim} we test four different scaling laws for the variables $t_{\rm corr}$ (assuming an exponential form for the correlation function, \citealt{Mummery24b}) and $\sigma_T$ (the standard deviation of temperature fluctuations). We simulated 50 light curves for each choice of parameters, using a 20 second cadence and 50 ks duration to maintain similarity to our observational data, then computed the PSD and $F_{\rm var}$ exactly as in Section~\ref{sec:results}. We find that our observed lack of scaling can be reproduced if the amplitude of temperature fluctuations scales with the mean temperature as $\sigma_T \propto \mu_T^{1-2}$. This scaling suggests that hotter disk regions not only have higher mean temperatures, but also proportionally larger turbulent fluctuations, providing an important constraint on the microphysics of MHD turbulence in accretion flows.

With hindsight, this result is perhaps not surprising. If one turns to an MHD theory of thin accretion flows \citep{BalbPap99} then the temperature in the disk can be related to the surface density $\Sigma$ and the turbulent correlations in the velocity and magnetic field by 
\begin{equation}
    T^4 \propto \Sigma \left\langle \delta u^r \delta u^\phi - b^r b^\phi/\rho \right\rangle \equiv \Sigma W^{r\phi},
\end{equation}
where the first term in the middle equality denotes the Reynolds stress on the disk and the second the Maxwell stress\footnote{This expressions becomes usual $T^4 \propto \dot M$ when one assumes a steady state.}, and they are conventionally brought together into a single turbulent stress $W^{r\phi}$ \citep{BalbPap99}. The surface density of a TDE disk is reasonably well constrained from the stellar properties of the disrupted star $\Sigma \sim M_\star / r_T^2 \sim M_\star^{5/3} / R_\star^2 M_{\rm BH}^{2/3}$ and may not vary hugely across the TDE population, meaning that hotter TDE disks may well result from more vigorous turbulence. As the peak temperature only grows as the one quarter power of the correlations in the turbulent stress $\mu_T \sim (W^{r\phi})^{1/4}$, then one simply requires a stronger (and certainly plausible) correlation between temperature variability (which originates from turbulence) and the magnitude of the turbulent stress itself $\sigma_T \sim (W^{r\phi})^{\sim1/4-1/2}$ to reproduce the results found in this work. 

Using a similar method, we also examine constraints based on the observed PSD slopes ($\alpha_{\rm PSD}$) of the sample. In Fig.~\ref{fig:psd_slope_sim} we simulated light curves for several choices of $\sigma_T$/$\mu_T$, and observe the predicted increase in $\alpha_{\rm PSD}$ with $t_{\rm corr}$. Reproducing the thermal TDE values of $0.1-1.2$ requires $t_{\rm corr}\lesssim 1000$ seconds, with some scatter induced by the disk temperature variability. Compared to some characteristic timescales in the accretion flow (e.g. \citealt{Shakura73,Kara25}):
$$ t_{\rm dyn} = \frac{1}{\Omega} = 3\;\mathrm{min}\bigg(\frac{M_{\rm BH}}{10^6M_\odot}\bigg)\bigg(\frac{R}{10R_g}\bigg)^{3/2} $$
$$ t_{\rm therm} = \frac{1}{\alpha\Omega} = 27\;\mathrm{min}\bigg(\frac{0.1}{\alpha}\bigg)\bigg(\frac{M_{\rm BH}}{10^6M_\odot}\bigg)\bigg(\frac{R}{10R_g}\bigg)^{3/2}  $$
we see that the implied correlation timescales lie between the dynamical and thermal timescales.

This is a readily understandable result, which follows from the fact that we are observing these systems in their Wien tail and therefore are ``seeing'' emission from the very inner near-ISCO region. While turbulent temperature fluctuations sourced in the main body (far from ISCO) region have a hope of surviving for many orbital timescales, if they are sourced near to the ISCO it is expected that they will then quickly fall across the ISCO and rapidly spiral across the horizon on sub-dynamical timescales (see e.g., \citealt{MummeryStone24} for a discussion of this rapid near-ISCO evolution seen in GRMHD simulations). The correlation timescale we infer therefore likely represents the time it takes for a given turbulent fluctuations to cross the narrow near-ISCO-to-post-ISCO distance, before quickly disappearing once in this unstable regime. 

\section{Conclusion} \label{sec:conclusion}

We have systematically measured the properties of minutes-to-hours (1-10 kiloseconds) timescale variability observed in the soft X-ray (0.3-2 keV) light curves of TDEs using \XMM\ data. Our analysis extends Fourier timing methods developed in the AGN and XRB literature (Fig.~\ref{fig:lri_example}, Section~\ref{sec:methods}; see also \citealt{Vaughan03,Uttley14}) to the first such population study in TDEs. We find the following key results:
\begin{itemize} 
    \item We separately analyzed the populations of thermal disk-dominated TDEs and TDEs with coronae, also comparing both populations to similar-mass AGN \citep{Gonzales12,Liu16}. TDE coronae are significantly softer than their AGN counterparts, as previously noted in \cite{Guolo24}. We also find TDEs are, on average, as variable as AGN. TDEs with coronae are more variable than those with a pure-thermal spectrum, likely indicative of additional nonthermal fluctuations in the corona itself. 
    
    \item TDE power spectral densities (PSDs) are well-characterized by power laws with greater variability at lower frequencies, although their slopes are significantly shallower than red noise PSDs in AGN. Thermal TDEs show the flattest PSD slopes, consistent with a stochastic process of short correlation time ($t_{\rm corr} \lesssim 1000$s) between the dynamical ($\sim 180$s) and thermal ($\sim 1600$s) timescales at $10R_g$ for $10^6M_\odot$ black hole. On the other hand, TDEs with coronae show intermediate PSD slopes between thermal TDEs and similar-mass AGN, even in the 2-10 keV band. We propose this, along with the softer energy spectra of TDEs, is related to the scattering properties of seed photons in their higher-$\tau$, lower-electron temperature coronae (Fig. \ref{fig:schematic}).
    
    \item We tested for any scaling relations (and lack thereof) between various spectral-timing parameters, time since disruption, and black hole mass (Fig.~\ref{fig:rp_tp_fvar_rms_psd}, Table~\ref{tab:corr}). Notably, we do not find the tight inverse scaling of $F_{\rm var}$ with $M_{\rm BH}$ observed in AGN \citep{Ponti12}, though we find higher variability is associated with redder noise, larger disks, stronger coronae, and later-time disks.
    
    \item We also measured the energy-dependence of the variability and found that all thermal TDEs show increasing variability at higher energies. This indicates TDE X-ray variability is driven by disk temperature fluctuations, aligning with theoretical models that observing a blackbody-like spectrum in the Wien tail exponentially enhances small-scale temperature fluctuations \citep{Mummery24b,Mummery25}. We use these models to infer the amplitude of temperature fluctuations ($\sigma_T$) scales with the mean disk temperature ($\mu_T$) with an exponent between $\sigma_T\propto\mu_T^{1-2}$, which is an interesting constraint on the nature of microphysical turbulent processes in TDE disks.
    
    \item In sources with absorption features indicating potential outflows, we found disk temperature is positively correlated with line centroid and width, and inversely with line strength. These correlations may place constraints of the coupling between accretion properties and outflow ionization and/or velocity, though high-resolution data is needed to break degeneracies in the spectral modeling. In AT2022lri, the TDE with the highest SNR, there is a drop in the variability in the narrow band dominated by the emission line ($\sim 1$ keV), reminiscent of findings in AGN outflows \citep{Parker20}.
\end{itemize}

Our systematic characterization of TDE X-ray variability opens up several avenues for further study. Only a handful of sources thus far have sufficient-quality data over several epochs to enable long-term monitoring of the variability properties (Fig.~\ref{fig:evol}); longer-duration repeated exposures on individual X-ray bright TDEs are needed to better understand of how variability properties evolve alongside the long-term viscous expansion of the accretion disks. Moreover, very few sources have sufficient-quality data at hard X-rays ($\gtrsim 2$ keV), making it difficult to probe the corona-dominated bands in hard-state TDEs. This spectral-timing study represents a first step toward using the X-ray variability properties of TDEs as fundamental probes of the turbulence in rapidly-evolving accretion disks and the disk-corona coupling in newborn accretion flows. Our findings that the coronae formed in TDEs are fundamentally different from those in AGN in their spectral-timing properties is both an interesting question for further modeling and simulation work, and also suggests a novel method for variability-based selection of TDEs in X-ray surveys.

X-ray spectral-timing analysis of TDEs has traditionally been limited because variability studies require observation durations set by the intrinsic timescales of interest, not simply the time needed to accumulate sufficient photons for photometry or spectroscopy. Our new sample analysis is a first step towards opening the X-ray time-domain for TDEs, and we hope the results presented here motivate future observations of considerably longer duration ($\gtrsim$100 ks) to better understand the dynamic, turbulent disks and coronae formed in TDEs. Extensions of our work facilitated by ever-growing observational sample sizes from upcoming time-domain surveys such as the Vera C. Rubin Observatory's Legacy Survey of Space and Time \citep{LSST} and the wide-field X-ray telescope onboard the Einstein Probe mission \citep{Yuan22}, as well as pointed follow-up observations using next-generation X-ray mission concepts such as \textit{AXIS} \citep{Reynolds23} and \textit{NewAthena} \citep{Cruise25}, will provide a uniquely exciting opportunity to further our physical understanding of TDE accretion disks and X-ray coronae around supermassive black holes.

\section*{Acknowledgements}
We thank the anonymous referee for their comments which improved the manuscript. We thank Ehud Behar and Muryel Guolo for insightful conversations which informed this work. A.M. acknowledges support from the Ambrose Monell Foundation, the W.M. Keck Foundation and the John N. Bahcall Fellowship Fund at the Institute for Advanced Study.   R.A. was supported by NASA through the NASA Hubble Fellowship grant \#HST-HF2-51499.001-A awarded by the Space Telescope Science Institute, which is operated by the Association of Universities for Research in Astronomy, Incorporated, under NASA contract NAS5-26555. A.M. acknowledges support from the Ambrose Monell Foundation, the W.M. Keck Foundation and the John N. Bahcall Fellowship Fund at the Institute for Advanced Study. 

\software{Astropy \citep{Astropy}, \texttt{XSPEC} \citep{Arnaud96}, \texttt{emcee} \citep{Foreman-Mackey2013}, HEASoft \citep{HEASoft}, \textit{XMM-Newton} Science Analysis Software.}

\bibliography{refs}{}

@ARTICLE{Saxton2025,
       author = {{Saxton}, R.~D. and {Wevers}, T. and {van Velzen}, S. and {Alexander}, K. and {Liu}, Z. and {Mummery}, A. and {Giustini}, M. and {Miniutti}, G. and {Fuerst}, F. and {Kajava}, J.~J.~E. and {Read}, A.~M. and {Jonker}, P.~G. and {Rau}, A. and {Li}, D.-Y.},
        title = "{Rapid onset of a Comptonisation zone in the repeating tidal disruption event XMMSL2 J140446.9-251135}",
      journal = {\aap},
         year = 2025,
        month = dec,
       volume = {704},
          eid = {A165},
        pages = {A165},
          doi = {10.1051/0004-6361/202554193},
archivePrefix = {arXiv},
       eprint = {2510.02905},
 primaryClass = {astro-ph.HE},
       adsurl = {https://ui.adsabs.harvard.edu/abs/2025A&A...704A.165S}
}

@ARTICLE{Arevalo2006,
       author = {{Ar{\'e}valo}, P. and {Uttley}, P.},
        title = "{Investigating a fluctuating-accretion model for the spectral-timing properties of accreting black hole systems}",
      journal = {\mnras},
         year = 2006,
        month = apr,
       volume = {367},
       number = {2},
        pages = {801-814},
          doi = {10.1111/j.1365-2966.2006.09989.x},
archivePrefix = {arXiv},
       eprint = {astro-ph/0512394},
 primaryClass = {astro-ph},
       adsurl = {https://ui.adsabs.harvard.edu/abs/2006MNRAS.367..801A}
}

@ARTICLE{Giustini2019,
       author = {{Giustini}, Margherita and {Proga}, Daniel},
        title = "{A global view of the inner accretion and ejection flow around super massive black holes. Radiation-driven accretion disk winds in a physical context}",
      journal = {\aap},
         year = 2019,
        month = oct,
       volume = {630},
          eid = {A94},
        pages = {A94},
          doi = {10.1051/0004-6361/201833810},
archivePrefix = {arXiv},
       eprint = {1904.07341},
 primaryClass = {astro-ph.GA},
       adsurl = {https://ui.adsabs.harvard.edu/abs/2019A&A...630A..94G}
}

@ARTICLE{Ricci2020,
       author = {{Ricci}, C. and {Kara}, E. and {Loewenstein}, M. and {Trakhtenbrot}, B. and {Arcavi}, I. and {Remillard}, R. and {Fabian}, A.~C. and {Gendreau}, K.~C. and {Arzoumanian}, Z. and {Li}, R. and {Ho}, L.~C. and {MacLeod}, C.~L. and {Cackett}, E. and {Altamirano}, D. and {Gandhi}, P. and {Kosec}, P. and {Pasham}, D. and {Steiner}, J. and {Chan}, C.-H.},
        title = "{The Destruction and Recreation of the X-Ray Corona in a Changing-look Active Galactic Nucleus}",
      journal = {\apjl},
         year = 2020,
        month = jul,
       volume = {898},
       number = {1},
          eid = {L1},
        pages = {L1},
          doi = {10.3847/2041-8213/ab91a1},
archivePrefix = {arXiv},
       eprint = {2007.07275},
 primaryClass = {astro-ph.HE},
       adsurl = {https://ui.adsabs.harvard.edu/abs/2020ApJ...898L...1R}
}

@ARTICLE{Trakhtenbrot2019,
       author = {{Trakhtenbrot}, Benny and {Arcavi}, Iair and {MacLeod}, Chelsea L. and {Ricci}, Claudio and {Kara}, Erin and {Graham}, Melissa L. and {Stern}, Daniel and {Harrison}, Fiona A. and {Burke}, Jamison and {Hiramatsu}, Daichi and {Hosseinzadeh}, Griffin and {Howell}, D. Andrew and {Smartt}, Stephen J. and {Rest}, Armin and {Prieto}, Jose L. and {Shappee}, Benjamin J. and {Holoien}, Thomas W.-S. and {Bersier}, David and {Filippenko}, Alexei V. and {Brink}, Thomas G. and {Zheng}, WeiKang and {Li}, Ruancun and {Remillard}, Ronald A. and {Loewenstein}, Michael},
        title = "{1ES 1927+654: An AGN Caught Changing Look on a Timescale of Months}",
      journal = {\apj},
         year = 2019,
        month = sep,
       volume = {883},
       number = {1},
          eid = {94},
        pages = {94},
          doi = {10.3847/1538-4357/ab39e4},
archivePrefix = {arXiv},
       eprint = {1903.11084},
 primaryClass = {astro-ph.GA},
       adsurl = {https://ui.adsabs.harvard.edu/abs/2019ApJ...883...94T}
}

@ARTICLE{Ricci2021,
       author = {{Ricci}, C. and {Loewenstein}, M. and {Kara}, E. and {Remillard}, R. and {Trakhtenbrot}, B. and {Arcavi}, I. and {Gendreau}, K.~C. and {Arzoumanian}, Z. and {Fabian}, A.~C. and {Li}, R. and {Ho}, L.~C. and {MacLeod}, C.~L. and {Cackett}, E. and {Altamirano}, D. and {Gandhi}, P. and {Kosec}, P. and {Pasham}, D. and {Steiner}, J. and {Chan}, C.-H.},
        title = "{The 450 Day X-Ray Monitoring of the Changing-look AGN 1ES 1927+654}",
      journal = {\apjs},
         year = 2021,
        month = jul,
       volume = {255},
       number = {1},
          eid = {7},
        pages = {7},
          doi = {10.3847/1538-4365/abe94b},
archivePrefix = {arXiv},
       eprint = {2102.05666},
 primaryClass = {astro-ph.HE},
       adsurl = {https://ui.adsabs.harvard.edu/abs/2021ApJS..255....7R}
}

@ARTICLE{Masterson2022,
       author = {{Masterson}, Megan and {Kara}, Erin and {Ricci}, Claudio and {Garc{\'\i}a}, Javier A. and {Fabian}, Andrew C. and {Pinto}, Ciro and {Kosec}, Peter and {Remillard}, Ronald A. and {Loewenstein}, Michael and {Trakhtenbrot}, Benny and {Arcavi}, Iair},
        title = "{Evolution of a Relativistic Outflow and X-Ray Corona in the Extreme Changing-look AGN 1ES 1927+654}",
      journal = {\apj},
         year = 2022,
        month = jul,
       volume = {934},
       number = {1},
          eid = {35},
        pages = {35},
          doi = {10.3847/1538-4357/ac76c0},
archivePrefix = {arXiv},
       eprint = {2206.05140},
 primaryClass = {astro-ph.HE},
       adsurl = {https://ui.adsabs.harvard.edu/abs/2022ApJ...934...35M}
}

@ARTICLE{McHardy2007,
       author = {{McHardy}, I.~M. and {Ar{\'e}valo}, P. and {Uttley}, P. and {Papadakis}, I.~E. and {Summons}, D.~P. and {Brinkmann}, W. and {Page}, M.~J.},
        title = "{Discovery of multiple Lorentzian components in the X-ray timing properties of the Narrow Line Seyfert 1 Ark 564}",
      journal = {\mnras},
         year = 2007,
        month = dec,
       volume = {382},
       number = {3},
        pages = {985-994},
          doi = {10.1111/j.1365-2966.2007.12411.x},
archivePrefix = {arXiv},
       eprint = {0709.0262},
 primaryClass = {astro-ph},
       adsurl = {https://ui.adsabs.harvard.edu/abs/2007MNRAS.382..985M}
}

@ARTICLE{Kara2017,
       author = {{Kara}, E. and {Garc{\'\i}a}, J.~A. and {Lohfink}, A. and {Fabian}, A.~C. and {Reynolds}, C.~S. and {Tombesi}, F. and {Wilkins}, D.~R.},
        title = "{The high-Eddington NLS1 Ark 564 has the coolest corona}",
      journal = {\mnras},
         year = 2017,
        month = jul,
       volume = {468},
       number = {3},
        pages = {3489-3498},
          doi = {10.1093/mnras/stx792},
archivePrefix = {arXiv},
       eprint = {1703.09815},
 primaryClass = {astro-ph.HE},
       adsurl = {https://ui.adsabs.harvard.edu/abs/2017MNRAS.468.3489K}
}

@ARTICLE{Lewin2022,
       author = {{Lewin}, Collin and {Kara}, Erin and {Wilkins}, Dan and {Mastroserio}, Guglielmo and {Garc{\'\i}a}, Javier A. and {Zhang}, Rachel C. and {Alston}, William N. and {Connors}, Riley and {Dauser}, Thomas and {Fabian}, Andrew and {Ingram}, Adam and {Jiang}, Jiachen and {Lohfink}, Anne and {Lucchini}, Matteo and {Reynolds}, Christopher S. and {Tombesi}, Francesco and {Klis}, Michiel van der and {Wang}, Jingyi},
        title = "{X-Ray Reverberation Mapping of Ark 564 Using Gaussian Process Regression}",
      journal = {\apj},
         year = 2022,
        month = nov,
       volume = {939},
       number = {2},
          eid = {109},
        pages = {109},
          doi = {10.3847/1538-4357/ac978f},
archivePrefix = {arXiv},
       eprint = {2210.01810},
 primaryClass = {astro-ph.HE},
       adsurl = {https://ui.adsabs.harvard.edu/abs/2022ApJ...939..109L}
}

@ARTICLE{Uttley02,
       author = {{Uttley}, P. and {McHardy}, I.~M. and {Papadakis}, I.~E.},
        title = "{Measuring the broad-band power spectra of active galactic nuclei with RXTE}",
      journal = {\mnras},
         year = 2002,
        month = may,
       volume = {332},
       number = {1},
        pages = {231-250},
          doi = {10.1046/j.1365-8711.2002.05298.x},
archivePrefix = {arXiv},
       eprint = {astro-ph/0201134},
 primaryClass = {astro-ph},
       adsurl = {https://ui.adsabs.harvard.edu/abs/2002MNRAS.332..231U}
}

@ARTICLE{MummeryStone24,
       author = {{Mummery}, Andrew and {Stone}, James M.},
        title = "{The three-dimensional structure of black hole accretion flows within the plunging region}",
      journal = {\mnras},
         year = 2024,
        month = aug,
       volume = {532},
       number = {3},
        pages = {3395-3416},
          doi = {10.1093/mnras/stae1643},
archivePrefix = {arXiv},
       eprint = {2407.02164},
 primaryClass = {astro-ph.HE},
       adsurl = {https://ui.adsabs.harvard.edu/abs/2024MNRAS.532.3395M}
}

@ARTICLE{BalbPap99,
       author = {{Balbus}, Steven A. and {Papaloizou}, John C.~B.},
        title = "{On the Dynamical Foundations of {\ensuremath{\alpha}} Disks}",
      journal = {\apj},
         year = 1999,
        month = aug,
       volume = {521},
       number = {2},
        pages = {650-658},
          doi = {10.1086/307594},
archivePrefix = {arXiv},
       eprint = {astro-ph/9903035},
 primaryClass = {astro-ph},
       adsurl = {https://ui.adsabs.harvard.edu/abs/1999ApJ...521..650B}
}

@ARTICLE{Vaughan2005,
       author = {{Vaughan}, S.},
        title = "{A simple test for periodic signals in red noise}",
      journal = {\aap},
         year = 2005,
        month = feb,
       volume = {431},
        pages = {391-403},
          doi = {10.1051/0004-6361:20041453},
archivePrefix = {arXiv},
       eprint = {astro-ph/0412697},
 primaryClass = {astro-ph},
       adsurl = {https://ui.adsabs.harvard.edu/abs/2005A&A...431..391V}
}

@ARTICLE{Berger26,
       author = {{Berger}, Vera and {Kara}, Erin and {Chakraborty}, Joheen and {Masterson}, Megan and {Burdge}, Kevin},
        title = "{Disk-to-Corona State Transition and Extreme X-ray Variability in the Tidal Disruption Event AT2019teq}",
      journal = {arXiv e-prints},
         year = 2026,
        month = jan,
          eid = {arXiv:2601.04311},
        pages = {arXiv:2601.04311},
          doi = {10.48550/arXiv.2601.04311},
archivePrefix = {arXiv},
       eprint = {2601.04311},
 primaryClass = {astro-ph.HE},
       adsurl = {https://ui.adsabs.harvard.edu/abs/2026arXiv260104311B}
}

@ARTICLE{Miniutti19,
       author = {{Miniutti}, G. and {Saxton}, R.~D. and {Giustini}, M. and {Alexander}, K.~D. and {Fender}, R.~P. and {Heywood}, I. and {Monageng}, I. and {Coriat}, M. and {Tzioumis}, A.~K. and {Read}, A.~M. and {Knigge}, C. and {Gandhi}, P. and {Pretorius}, M.~L. and {Ag{\'\i}s-Gonz{\'a}lez}, B.},
        title = "{Nine-hour X-ray quasi-periodic eruptions from a low-mass black hole galactic nucleus}",
      journal = {\nat},
         year = 2019,
        month = sep,
       volume = {573},
       number = {7774},
        pages = {381-384},
          doi = {10.1038/s41586-019-1556-x},
archivePrefix = {arXiv},
       eprint = {1909.04693},
 primaryClass = {astro-ph.HE},
       adsurl = {https://ui.adsabs.harvard.edu/abs/2019Natur.573..381M}
}

@INPROCEEDINGS{Belloni16,
       author = {{Belloni}, Tomaso M. and {Motta}, Sara E.},
        title = "{Transient Black Hole Binaries}",
    booktitle = {Astrophysics of Black Holes: From Fundamental Aspects to Latest Developments},
         year = 2016,
       editor = {{Bambi}, Cosimo},
       series = {Astrophysics and Space Science Library},
       volume = {440},
        month = jan,
        pages = {61},
          doi = {10.1007/978-3-662-52859-4_2},
archivePrefix = {arXiv},
       eprint = {1603.07872},
 primaryClass = {astro-ph.HE},
       adsurl = {https://ui.adsabs.harvard.edu/abs/2016ASSL..440...61B}
}

@ARTICLE{Payne21,
       author = {{Payne}, Anna V. and {Shappee}, Benjamin J. and {Hinkle}, Jason T. and {Vallely}, Patrick J. and {Kochanek}, Christopher S. and {Holoien}, Thomas W. -S. and {Auchettl}, Katie and {Stanek}, K.~Z. and {Thompson}, Todd A. and {Neustadt}, Jack M.~M. and {Tucker}, Michael A. and {Armstrong}, James D. and {Brimacombe}, Joseph and {Cacella}, Paulo and {Cornect}, Robert and {Denneau}, Larry and {Fausnaugh}, Michael M. and {Flewelling}, Heather and {Grupe}, Dirk and {Heinze}, A.~N. and {Lopez}, Laura A. and {Monard}, Berto and {Prieto}, Jose L. and {Schneider}, Adam C. and {Sheppard}, Scott S. and {Tonry}, John L. and {Weiland}, Henry},
        title = "{ASASSN-14ko is a Periodic Nuclear Transient in ESO 253-G003}",
      journal = {\apj},
         year = 2021,
        month = apr,
       volume = {910},
       number = {2},
          eid = {125},
        pages = {125},
          doi = {10.3847/1538-4357/abe38d},
archivePrefix = {arXiv},
       eprint = {2009.03321},
 primaryClass = {astro-ph.HE},
       adsurl = {https://ui.adsabs.harvard.edu/abs/2021ApJ...910..125P}
}

@ARTICLE{Vaughan2010,
       author = {{Vaughan}, S.},
        title = "{A Bayesian test for periodic signals in red noise}",
      journal = {\mnras},
         year = 2010,
        month = feb,
       volume = {402},
       number = {1},
        pages = {307-320},
          doi = {10.1111/j.1365-2966.2009.15868.x},
archivePrefix = {arXiv},
       eprint = {0910.2706},
 primaryClass = {astro-ph.HE},
       adsurl = {https://ui.adsabs.harvard.edu/abs/2010MNRAS.402..307V}
}

@INPROCEEDINGS{Phinney89,
       author = {{Phinney}, E.~S.},
        title = "{Manifestations of a Massive Black Hole in the Galactic Center}",
    booktitle = {The Center of the Galaxy},
         year = 1989,
       editor = {{Morris}, Mark},
       series = {IAU Symposium},
       volume = {136},
        month = jan,
        pages = {543},
       adsurl = {https://ui.adsabs.harvard.edu/abs/1989IAUS..136..543P}
}

@ARTICLE{Rees88,
       author = {{Rees}, Martin J.},
        title = "{Tidal disruption of stars by black holes of {}10$^{6}$-{}10$^{8}$ solar masses in nearby galaxies}",
      journal = {\nat},
         year = 1988,
        month = jun,
       volume = {333},
       number = {6173},
        pages = {523-528},
          doi = {10.1038/333523a0},
       adsurl = {https://ui.adsabs.harvard.edu/abs/1988Natur.333..523R}
}

@ARTICLE{Foreman-Mackey2013,
       author = {{Foreman-Mackey}, Daniel and {Hogg}, David W. and {Lang}, Dustin and {Goodman}, Jonathan},
        title = "{emcee: The MCMC Hammer}",
      journal = {\pasp},
         year = 2013,
        month = mar,
       volume = {125},
       number = {925},
        pages = {306},
          doi = {10.1086/670067},
archivePrefix = {arXiv},
       eprint = {1202.3665},
 primaryClass = {astro-ph.IM},
       adsurl = {https://ui.adsabs.harvard.edu/abs/2013PASP..125..306F}
}

@ARTICLE{Steiner09,
       author = {{Steiner}, James F. and {Narayan}, Ramesh and {McClintock}, Jeffrey E. and {Ebisawa}, Ken},
        title = "{A Simple Comptonization Model}",
      journal = {\pasp},
         year = 2009,
        month = nov,
       volume = {121},
       number = {885},
        pages = {1279},
          doi = {10.1086/648535},
archivePrefix = {arXiv},
       eprint = {0810.1758},
 primaryClass = {astro-ph},
       adsurl = {https://ui.adsabs.harvard.edu/abs/2009PASP..121.1279S}
}

@ARTICLE{Mummery21b,
       author = {{Mummery}, Andrew},
        title = "{Tidal disruption event discs are larger than they seem: removing systematic biases in TDE X-ray spectral modelling}",
      journal = {\mnras},
         year = 2021,
        month = oct,
       volume = {507},
       number = {1},
        pages = {L24-L28},
          doi = {10.1093/mnrasl/slab088},
archivePrefix = {arXiv},
       eprint = {2108.10160},
 primaryClass = {astro-ph.HE},
       adsurl = {https://ui.adsabs.harvard.edu/abs/2021MNRAS.507L..24M}
}

@ARTICLE{Wang22,
       author = {{Wang}, Jingyi and {Kara}, Erin and {Lucchini}, Matteo and {Ingram}, Adam and {van der Klis}, Michiel and {Mastroserio}, Guglielmo and {Garc{\'\i}a}, Javier A. and {Dauser}, Thomas and {Connors}, Riley and {Fabian}, Andrew C. and {Steiner}, James F. and {Remillard}, Ron A. and {Cackett}, Edward M. and {Uttley}, Phil and {Altamirano}, Diego},
        title = "{The NICER ``Reverberation Machine'': A Systematic Study of Time Lags in Black Hole X-Ray Binaries}",
      journal = {\apj},
         year = 2022,
        month = may,
       volume = {930},
       number = {1},
          eid = {18},
        pages = {18},
          doi = {10.3847/1538-4357/ac6262},
archivePrefix = {arXiv},
       eprint = {2205.00928},
 primaryClass = {astro-ph.HE},
       adsurl = {https://ui.adsabs.harvard.edu/abs/2022ApJ...930...18W}
}

@ARTICLE{Kara16,
       author = {{Kara}, E. and {Alston}, W.~N. and {Fabian}, A.~C. and {Cackett}, E.~M. and {Uttley}, P. and {Reynolds}, C.~S. and {Zoghbi}, A.},
        title = "{A global look at X-ray time lags in Seyfert galaxies}",
      journal = {\mnras},
         year = 2016,
        month = oct,
       volume = {462},
       number = {1},
        pages = {511-531},
          doi = {10.1093/mnras/stw1695},
archivePrefix = {arXiv},
       eprint = {1605.02631},
 primaryClass = {astro-ph.HE},
       adsurl = {https://ui.adsabs.harvard.edu/abs/2016MNRAS.462..511K}
}

@ARTICLE{Barr86,
       author = {{Barr}, P. and {Mushotzky}, R.~F.},
        title = "{Limits of X-ray variability in active galactic nuclei.}",
      journal = {\nat},
         year = 1986,
        month = apr,
       volume = {320},
        pages = {421-423},
          doi = {10.1038/320421a0},
       adsurl = {https://ui.adsabs.harvard.edu/abs/1986Natur.320..421B}
}

@ARTICLE{Papadakis07,
       author = {{Papadakis}, I.~E. and {Ioannou}, Z. and {Kazanas}, D.},
        title = "{Fourier-Resolved Spectroscopy of Active Galactic Nuclei Using XMM-Newton Data. I. The 3-10 keV Band Results}",
      journal = {\apj},
         year = 2007,
        month = may,
       volume = {661},
       number = {1},
        pages = {38-51},
          doi = {10.1086/513307},
archivePrefix = {arXiv},
       eprint = {astro-ph/0701809},
 primaryClass = {astro-ph},
       adsurl = {https://ui.adsabs.harvard.edu/abs/2007ApJ...661...38P}
}

@ARTICLE{Konig24,
       author = {{K{\"o}nig}, Ole and {Mastroserio}, Guglielmo and {Dauser}, Thomas and {M{\'e}ndez}, Mariano and {Wang}, Jingyi and {Garc{\'\i}a}, Javier A. and {Steiner}, James F. and {Pottschmidt}, Katja and {Ballhausen}, Ralf and {Connors}, Riley M. and {Garc{\'\i}a}, Federico and {Grinberg}, Victoria and {Horn}, David and {Ingram}, Adam and {Kara}, Erin and {Kallman}, Timothy R. and {Lucchini}, Matteo and {Nathan}, Edward and {Nowak}, Michael A. and {Thalhammer}, Philipp and {van der Klis}, Michiel and {Wilms}, J{\"o}rn},
        title = "{Long term variability of Cygnus X-1. VIII. A spectral-timing look at low energies with NICER}",
      journal = {\aap},
         year = 2024,
        month = jul,
       volume = {687},
          eid = {A284},
        pages = {A284},
          doi = {10.1051/0004-6361/202449333},
archivePrefix = {arXiv},
       eprint = {2405.07754},
 primaryClass = {astro-ph.HE},
       adsurl = {https://ui.adsabs.harvard.edu/abs/2024A&A...687A.284K}
}

@ARTICLE{Revnivtsev99,
       author = {{Revnivtsev}, M. and {Gilfanov}, M. and {Churazov}, E.},
        title = "{The frequency resolved spectroscopy of CYG X-1: fast variability of the Fe K\_{\ensuremath{\alpha}} line}",
      journal = {\aap},
         year = 1999,
        month = jul,
       volume = {347},
        pages = {L23-L26},
          doi = {10.48550/arXiv.astro-ph/9906198},
archivePrefix = {arXiv},
       eprint = {astro-ph/9906198},
 primaryClass = {astro-ph},
       adsurl = {https://ui.adsabs.harvard.edu/abs/1999A&A...347L..23R}
}

@ARTICLE{Mummery22,
       author = {{Mummery}, Andrew and {Balbus}, Steven},
        title = "{The high-energy probability distribution of accretion disc luminosity fluctuations}",
      journal = {\mnras},
         year = 2022,
        month = dec,
       volume = {517},
       number = {3},
        pages = {3423-3431},
          doi = {10.1093/mnras/stac2844},
archivePrefix = {arXiv},
       eprint = {2210.01450},
 primaryClass = {astro-ph.HE},
       adsurl = {https://ui.adsabs.harvard.edu/abs/2022MNRAS.517.3423M}
}

@ARTICLE{Mummery24b,
       author = {{Mummery}, Andrew and {Turner}, Samuel G.~D.},
        title = "{The turbulent variability of accretion discs observed at high energies}",
      journal = {\mnras},
         year = 2024,
        month = jun,
       volume = {530},
       number = {4},
        pages = {4730-4746},
          doi = {10.1093/mnras/stae1014},
archivePrefix = {arXiv},
       eprint = {2404.09564},
 primaryClass = {astro-ph.HE},
       adsurl = {https://ui.adsabs.harvard.edu/abs/2024MNRAS.530.4730M}
}

@ARTICLE{Mummery21a,
       author = {{Mummery}, Andrew and {Balbus}, Steven A.},
        title = "{Hard X-ray emission from a Compton scattering corona in large black hole mass tidal disruption events}",
      journal = {\mnras},
         year = 2021,
        month = jul,
       volume = {504},
       number = {4},
        pages = {4730-4742},
          doi = {10.1093/mnras/stab1184},
archivePrefix = {arXiv},
       eprint = {2104.06195},
 primaryClass = {astro-ph.HE},
       adsurl = {https://ui.adsabs.harvard.edu/abs/2021MNRAS.504.4730M}
}

@ARTICLE{Payne23,
       author = {{Payne}, Anna V. and {Auchettl}, Katie and {Shappee}, Benjamin J. and {Kochanek}, Christopher S. and {Boyd}, Patricia T. and {Holoien}, Thomas W.-S. and {Fausnaugh}, Michael M. and {Ashall}, Chris and {Hinkle}, Jason T. and {Vallely}, Patrick J. and {Stanek}, K.~Z. and {Thompson}, Todd A.},
        title = "{Chandra, HST/STIS, NICER, Swift, and TESS Detail the Flare Evolution of the Repeating Nuclear Transient ASASSN -14ko}",
      journal = {\apj},
         year = 2023,
        month = jul,
       volume = {951},
       number = {2},
          eid = {134},
        pages = {134},
          doi = {10.3847/1538-4357/acd455},
archivePrefix = {arXiv},
       eprint = {2206.11278},
 primaryClass = {astro-ph.HE},
       adsurl = {https://ui.adsabs.harvard.edu/abs/2023ApJ...951..134P}
}

@ARTICLE{Mummery23,
       author = {{Mummery}, Andrew and {Wevers}, Thomas and {Saxton}, Richard and {Pasham}, Dheeraj},
        title = "{From X-rays to physical parameters: a comprehensive analysis of thermal tidal disruption event X-ray spectra}",
      journal = {\mnras},
         year = 2023,
        month = mar,
       volume = {519},
       number = {4},
        pages = {5828-5847},
          doi = {10.1093/mnras/stac3798},
archivePrefix = {arXiv},
       eprint = {2301.07419},
 primaryClass = {astro-ph.HE},
       adsurl = {https://ui.adsabs.harvard.edu/abs/2023MNRAS.519.5828M}
}

@ARTICLE{Dai18,
       author = {{Dai}, Lixin and {McKinney}, Jonathan C. and {Roth}, Nathaniel and {Ramirez-Ruiz}, Enrico and {Miller}, M. Coleman},
        title = "{A Unified Model for Tidal Disruption Events}",
      journal = {\apjl},
         year = 2018,
        month = jun,
       volume = {859},
       number = {2},
          eid = {L20},
        pages = {L20},
          doi = {10.3847/2041-8213/aab429},
archivePrefix = {arXiv},
       eprint = {1803.03265},
 primaryClass = {astro-ph.HE},
       adsurl = {https://ui.adsabs.harvard.edu/abs/2018ApJ...859L..20D}
}

@ARTICLE{Nardini15,
       author = {{Nardini}, E. and {Reeves}, J.~N. and {Gofford}, J. and {Harrison}, F.~A. and {Risaliti}, G. and {Braito}, V. and {Costa}, M.~T. and {Matzeu}, G.~A. and {Walton}, D.~J. and {Behar}, E. and {Boggs}, S.~E. and {Christensen}, F.~E. and {Craig}, W.~W. and {Hailey}, C.~J. and {Matt}, G. and {Miller}, J.~M. and {O'Brien}, P.~T. and {Stern}, D. and {Turner}, T.~J. and {Ward}, M.~J.},
        title = "{Black hole feedback in the luminous quasar PDS 456}",
      journal = {Science},
         year = 2015,
        month = feb,
       volume = {347},
       number = {6224},
        pages = {860-863},
          doi = {10.1126/science.1259202},
archivePrefix = {arXiv},
       eprint = {1502.06636},
 primaryClass = {astro-ph.HE},
       adsurl = {https://ui.adsabs.harvard.edu/abs/2015Sci...347..860N}
}

@ARTICLE{Vurm25,
       author = {{Vurm}, Indrek and {Linial}, Itai and {Metzger}, Brian D.},
        title = "{Radiation Transport Simulations of Quasiperiodic Eruptions from Star{\textendash}Disk Collisions}",
      journal = {\apj},
         year = 2025,
        month = apr,
       volume = {983},
       number = {1},
          eid = {40},
        pages = {40},
          doi = {10.3847/1538-4357/adb74d},
archivePrefix = {arXiv},
       eprint = {2410.05166},
 primaryClass = {astro-ph.HE},
       adsurl = {https://ui.adsabs.harvard.edu/abs/2025ApJ...983...40V}
}

@ARTICLE{Chakraborty25b,
       author = {{Chakraborty}, Joheen and {Kosec}, Peter and {Kara}, Erin and {Miniutti}, Giovanni and {Arcodia}, Riccardo and {Behar}, Ehud and {Giustini}, Margherita and {Hern{\'a}ndez-Garc{\'\i}a}, Lorena and {Masterson}, Megan and {Quintin}, Erwan and {Ricci}, Claudio and {S{\'a}nchez-S{\'a}ez}, Paula},
        title = "{Rapidly Varying Ionization Features in a Quasi-periodic Eruption: A Homologous Expansion Model for the Spectroscopic Evolution}",
      journal = {\apj},
         year = 2025,
        month = may,
       volume = {984},
       number = {2},
          eid = {124},
        pages = {124},
          doi = {10.3847/1538-4357/adb972},
archivePrefix = {arXiv},
       eprint = {2504.07167},
 primaryClass = {astro-ph.HE},
       adsurl = {https://ui.adsabs.harvard.edu/abs/2025ApJ...984..124C}
}

@ARTICLE{Guolo24,
       author = {{Guolo}, Muryel and {Gezari}, Suvi and {Yao}, Yuhan and {van Velzen}, Sjoert and {Hammerstein}, Erica and {Cenko}, S. Bradley and {Tokayer}, Yarone M.},
        title = "{A Systematic Analysis of the X-Ray Emission in Optically Selected Tidal Disruption Events: Observational Evidence for the Unification of the Optically and X-Ray-selected Populations}",
      journal = {\apj},
         year = 2024,
        month = may,
       volume = {966},
       number = {2},
          eid = {160},
        pages = {160},
          doi = {10.3847/1538-4357/ad2f9f},
archivePrefix = {arXiv},
       eprint = {2308.13019},
 primaryClass = {astro-ph.HE},
       adsurl = {https://ui.adsabs.harvard.edu/abs/2024ApJ...966..160G}
}

@ARTICLE{Turner23,
       author = {{Turner}, Samuel G.~D. and {Reynolds}, Christopher S.},
        title = "{A new 2D stochastic methodology for simulating variable accretion discs: propagating fluctuations and epicyclic motion}",
      journal = {\mnras},
         year = 2023,
        month = oct,
       volume = {525},
       number = {2},
        pages = {2287-2314},
          doi = {10.1093/mnras/stad2275},
archivePrefix = {arXiv},
       eprint = {2306.07199},
 primaryClass = {astro-ph.HE},
       adsurl = {https://ui.adsabs.harvard.edu/abs/2023MNRAS.525.2287T}
}

@ARTICLE{Sazonov21,
       author = {{Sazonov}, S. and {Gilfanov}, M. and {Medvedev}, P. and {Yao}, Y. and {Khorunzhev}, G. and {Semena}, A. and {Sunyaev}, R. and {Burenin}, R. and {Lyapin}, A. and {Meshcheryakov}, A. and {Uskov}, G. and {Zaznobin}, I. and {Postnov}, K.~A. and {Dodin}, A.~V. and {Belinski}, A.~A. and {Cherepashchuk}, A.~M. and {Eselevich}, M. and {Dodonov}, S.~N. and {Grokhovskaya}, A.~A. and {Kotov}, S.~S. and {Bikmaev}, I.~F. and {Zhuchkov}, R. Ya and {Gumerov}, R.~I. and {van Velzen}, S. and {Kulkarni}, S.},
        title = "{First tidal disruption events discovered by SRG/eROSITA: X-ray/optical properties and X-ray luminosity function at z < 0.6}",
      journal = {\mnras},
         year = 2021,
        month = dec,
       volume = {508},
       number = {3},
        pages = {3820-3847},
          doi = {10.1093/mnras/stab2843},
archivePrefix = {arXiv},
       eprint = {2108.02449},
 primaryClass = {astro-ph.HE},
       adsurl = {https://ui.adsabs.harvard.edu/abs/2021MNRAS.508.3820S}
}

@ARTICLE{Saxton20,
       author = {{Saxton}, R. and {Komossa}, S. and {Auchettl}, K. and {Jonker}, P.~G.},
        title = "{X-Ray Properties of TDEs}",
      journal = {\ssr},
         year = 2020,
        month = jul,
       volume = {216},
       number = {5},
          eid = {85},
        pages = {85},
          doi = {10.1007/s11214-020-00708-4},
       adsurl = {https://ui.adsabs.harvard.edu/abs/2020SSRv..216...85S}
}

@ARTICLE{Esquej06,
       author = {{Esquej}, P. and {Saxton}, R.~D. and {Freyberg}, M.~J. and {Read}, A.~M. and {Altieri}, B. and {Sanchez-Portal}, M. and {Hasinger}, G.},
        title = "{Candidate tidal disruption events from the XMM-Newton slew survey}",
      journal = {\aap},
         year = 2007,
        month = feb,
       volume = {462},
       number = {3},
        pages = {L49-L52},
          doi = {10.1051/0004-6361:20066072},
archivePrefix = {arXiv},
       eprint = {astro-ph/0612340},
 primaryClass = {astro-ph},
       adsurl = {https://ui.adsabs.harvard.edu/abs/2007A&A...462L..49E}
}

@ARTICLE{Parker20,
       author = {{Parker}, M.~L. and {Alston}, W.~N. and {Igo}, Z. and {Fabian}, A.~C.},
        title = "{Modelling X-ray RMS spectra I: intrinsically variable AGNs}",
      journal = {\mnras},
         year = 2020,
        month = feb,
       volume = {492},
       number = {1},
        pages = {1363-1369},
          doi = {10.1093/mnras/stz3470},
archivePrefix = {arXiv},
       eprint = {1910.12808},
 primaryClass = {astro-ph.HE},
       adsurl = {https://ui.adsabs.harvard.edu/abs/2020MNRAS.492.1363P}
}

@ARTICLE{Wen20,
       author = {{Wen}, Sixiang and {Jonker}, Peter G. and {Stone}, Nicholas C. and {Zabludoff}, Ann I. and {Psaltis}, Dimitrios},
        title = "{Continuum-fitting the X-Ray Spectra of Tidal Disruption Events}",
      journal = {\apj},
         year = 2020,
        month = jul,
       volume = {897},
       number = {1},
          eid = {80},
        pages = {80},
          doi = {10.3847/1538-4357/ab9817},
archivePrefix = {arXiv},
       eprint = {2003.12583},
 primaryClass = {astro-ph.HE},
       adsurl = {https://ui.adsabs.harvard.edu/abs/2020ApJ...897...80W}
}

@ARTICLE{Wevers21,
       author = {{Wevers}, T. and {Pasham}, D.~R. and {van Velzen}, S. and {Miller-Jones}, J.~C.~A. and {Uttley}, P. and {Gendreau}, K.~C. and {Remillard}, R. and {Arzoumanian}, Z. and {L{\"o}wenstein}, M. and {Chiti}, A.},
        title = "{Rapid Accretion State Transitions following the Tidal Disruption Event AT2018fyk}",
      journal = {\apj},
         year = 2021,
        month = may,
       volume = {912},
       number = {2},
          eid = {151},
        pages = {151},
          doi = {10.3847/1538-4357/abf5e2},
archivePrefix = {arXiv},
       eprint = {2101.04692},
 primaryClass = {astro-ph.HE},
       adsurl = {https://ui.adsabs.harvard.edu/abs/2021ApJ...912..151W}
}

@ARTICLE{Markowitz03,
       author = {{Markowitz}, A. and {Edelson}, R. and {Vaughan}, S.},
        title = "{Long-Term X-Ray Spectral Variability in Seyfert 1 Galaxies}",
      journal = {\apj},
         year = 2003,
        month = dec,
       volume = {598},
       number = {2},
        pages = {935-955},
          doi = {10.1086/379103},
archivePrefix = {arXiv},
       eprint = {astro-ph/0308312},
 primaryClass = {astro-ph},
       adsurl = {https://ui.adsabs.harvard.edu/abs/2003ApJ...598..935M}
}

@ARTICLE{Kara18,
       author = {{Kara}, E. and {Dai}, L. and {Reynolds}, C.~S. and {Kallman}, T.},
        title = "{Ultrafast outflow in tidal disruption event ASASSN-14li}",
      journal = {\mnras},
         year = 2018,
        month = mar,
       volume = {474},
       number = {3},
        pages = {3593-3598},
          doi = {10.1093/mnras/stx3004},
archivePrefix = {arXiv},
       eprint = {1711.06090},
 primaryClass = {astro-ph.HE},
       adsurl = {https://ui.adsabs.harvard.edu/abs/2018MNRAS.474.3593K}
}

@ARTICLE{Pasham24,
       author = {{Pasham}, Dheeraj R. and {Tombesi}, Francesco and {Sukov{\'a}}, Petra and {Zaja{\v{c}}ek}, Michal and {Rakshit}, Suvendu and {Coughlin}, Eric and {Kosec}, Peter and {Karas}, Vladim{\'\i}r and {Masterson}, Megan and {Mummery}, Andrew and {Holoien}, Thomas W. -S. and {Guolo}, Muryel and {Hinkle}, Jason and {Ripperda}, Bart and {Witzany}, Vojt{\v{e}}ch and {Shappee}, Ben and {Kara}, Erin and {Horesh}, Assaf and {van Velzen}, Sjoert and {Sfaradi}, Itai and {Kaplan}, David and {Burger}, Noam and {Murphy}, Tara and {Remillard}, Ronald and {Steiner}, James F. and {Wevers}, Thomas and {Arcodia}, Riccardo and {Buchner}, Johannes and {Merloni}, Andrea and {Malyali}, Adam and {Fabian}, Andy and {Fausnaugh}, Michael and {Daylan}, Tansu and {Altamirano}, Diego and {Payne}, Anna and {Ferraraa}, Elizabeth C.},
        title = "{A case for a binary black hole system revealed via quasi-periodic outflows}",
      journal = {Science Advances},
         year = 2024,
        month = mar,
       volume = {10},
       number = {13},
          eid = {eadj8898},
        pages = {eadj8898},
          doi = {10.1126/sciadv.adj8898},
archivePrefix = {arXiv},
       eprint = {2402.10140},
 primaryClass = {astro-ph.HE},
       adsurl = {https://ui.adsabs.harvard.edu/abs/2024SciA...10J8898P}
}

@ARTICLE{Miniutti13,
       author = {{Miniutti}, G. and {Saxton}, R.~D. and {Rodr{\'\i}guez-Pascual}, P.~M. and {Read}, A.~M. and {Esquej}, P. and {Colless}, M. and {Dobbie}, P. and {Spolaor}, M.},
        title = "{A high Eddington-ratio, true Seyfert 2 galaxy candidate: implications for broad-line region models}",
      journal = {\mnras},
         year = 2013,
        month = aug,
       volume = {433},
       number = {2},
        pages = {1764-1777},
          doi = {10.1093/mnras/stt850},
archivePrefix = {arXiv},
       eprint = {1305.3284},
 primaryClass = {astro-ph.CO},
       adsurl = {https://ui.adsabs.harvard.edu/abs/2013MNRAS.433.1764M}
}

@ARTICLE{Guolo25b,
       author = {{Guolo}, M. and {Mummery}, A. and {van Velzen}, S. and {Gezari}, S. and {Nicholl}, M. and {Yao}, Y. and {Karmen}, M. and {Ajay}, Y. and {Wevers}, T. and {LeBaron}, N. and {Chornock}, R.},
        title = "{Compact Accretion Disks in the Aftermath of Tidal Disruption Events: Parameter Inference from Joint X-ray Spectra and UV/Optical Photometry Fitting}",
      journal = {arXiv e-prints},
         year = 2025,
        month = oct,
          eid = {arXiv:2510.26774},
        pages = {arXiv:2510.26774},
          doi = {10.48550/arXiv.2510.26774},
archivePrefix = {arXiv},
       eprint = {2510.26774},
 primaryClass = {astro-ph.HE},
       adsurl = {https://ui.adsabs.harvard.edu/abs/2025arXiv251026774G}
}

@ARTICLE{Guolo25a,
       author = {{Guolo}, M. and {Mummery}, A. and {Ingram}, A. and {Nicholl}, M. and {Gezari}, S. and {Nathan}, E.},
        title = "{A Time-Dependent Solution for GSN 069 Disk Evolution: The Nature of 'Long-Lived' TDEs and Implications for QPE Models}",
      journal = {arXiv e-prints},
         year = 2025,
        month = apr,
          eid = {arXiv:2504.20148},
        pages = {arXiv:2504.20148},
          doi = {10.48550/arXiv.2504.20148},
archivePrefix = {arXiv},
       eprint = {2504.20148},
 primaryClass = {astro-ph.HE},
       adsurl = {https://ui.adsabs.harvard.edu/abs/2025arXiv250420148G}
}

@ARTICLE{Kosec25,
       author = {{Kosec}, P. and {Kara}, E. and {Brenneman}, L. and {Chakraborty}, J. and {Giustini}, M. and {Miniutti}, G. and {Pinto}, C. and {Rogantini}, D. and {Arcodia}, R. and {Middleton}, M. and {Sacchi}, A.},
        title = "{Detection of a Highly Ionized Outflow in the Quasiperiodically Erupting Source GSN 069}",
      journal = {\apj},
         year = 2025,
        month = jan,
       volume = {978},
       number = {1},
          eid = {10},
        pages = {10},
          doi = {10.3847/1538-4357/ad9249},
archivePrefix = {arXiv},
       eprint = {2406.17105},
 primaryClass = {astro-ph.HE},
       adsurl = {https://ui.adsabs.harvard.edu/abs/2025ApJ...978...10K}
}

@ARTICLE{Hammerstein23,
       author = {{Hammerstein}, Erica and {Cenko}, S. Bradley and {Gezari}, Suvi and {Veilleux}, Sylvain and {O'Connor}, Brendan and {van Velzen}, Sjoert and {Ward}, Charlotte and {Yao}, Yuhan and {Graham}, Matthew},
        title = "{Integral Field Spectroscopy of 13 Tidal Disruption Event Hosts from the Zwicky Transient Facility Survey}",
      journal = {\apj},
         year = 2023,
        month = nov,
       volume = {957},
       number = {2},
          eid = {86},
        pages = {86},
          doi = {10.3847/1538-4357/acfb84},
archivePrefix = {arXiv},
       eprint = {2307.15705},
 primaryClass = {astro-ph.CO},
       adsurl = {https://ui.adsabs.harvard.edu/abs/2023ApJ...957...86H}
}

@ARTICLE{Yao22,
       author = {{Yao}, Yuhan and {Lu}, Wenbin and {Guolo}, Muryel and {Pasham}, Dheeraj R. and {Gezari}, Suvi and {Gilfanov}, Marat and {Gendreau}, Keith C. and {Harrison}, Fiona and {Cenko}, S. Bradley and {Kulkarni}, S.~R. and {Miller}, Jon M. and {Walton}, Dominic J. and {Garc{\'\i}a}, Javier A. and {van Velzen}, Sjoert and {Alexander}, Kate D. and {Miller-Jones}, James C.~A. and {Nicholl}, Matt and {Hammerstein}, Erica and {Medvedev}, Pavel and {Stern}, Daniel and {Ravi}, Vikram and {Sunyaev}, R. and {Bloom}, Joshua S. and {Graham}, Matthew J. and {Kool}, Erik C. and {Mahabal}, Ashish A. and {Masci}, Frank J. and {Purdum}, Josiah and {Rusholme}, Ben and {Sharma}, Yashvi and {Smith}, Roger and {Sollerman}, Jesper},
        title = "{The Tidal Disruption Event AT2021ehb: Evidence of Relativistic Disk Reflection, and Rapid Evolution of the Disk-Corona System}",
      journal = {\apj},
         year = 2022,
        month = sep,
       volume = {937},
       number = {1},
          eid = {8},
        pages = {8},
          doi = {10.3847/1538-4357/ac898a},
archivePrefix = {arXiv},
       eprint = {2206.12713},
 primaryClass = {astro-ph.HE},
       adsurl = {https://ui.adsabs.harvard.edu/abs/2022ApJ...937....8Y}
}

@ARTICLE{Yao23,
       author = {{Yao}, Yuhan and {Ravi}, Vikram and {Gezari}, Suvi and {van Velzen}, Sjoert and {Lu}, Wenbin and {Schulze}, Steve and {Somalwar}, Jean J. and {Kulkarni}, S.~R. and {Hammerstein}, Erica and {Nicholl}, Matt and {Graham}, Matthew J. and {Perley}, Daniel A. and {Cenko}, S. Bradley and {Stein}, Robert and {Ricarte}, Angelo and {Chadayammuri}, Urmila and {Quataert}, Eliot and {Bellm}, Eric C. and {Bloom}, Joshua S. and {Dekany}, Richard and {Drake}, Andrew J. and {Groom}, Steven L. and {Mahabal}, Ashish A. and {Prince}, Thomas A. and {Riddle}, Reed and {Rusholme}, Ben and {Sharma}, Yashvi and {Sollerman}, Jesper and {Yan}, Lin},
        title = "{Tidal Disruption Event Demographics with the Zwicky Transient Facility: Volumetric Rates, Luminosity Function, and Implications for the Local Black Hole Mass Function}",
      journal = {\apjl},
         year = 2023,
        month = sep,
       volume = {955},
       number = {1},
          eid = {L6},
        pages = {L6},
          doi = {10.3847/2041-8213/acf216},
archivePrefix = {arXiv},
       eprint = {2303.06523},
 primaryClass = {astro-ph.HE},
       adsurl = {https://ui.adsabs.harvard.edu/abs/2023ApJ...955L...6Y}
}

@ARTICLE{Yao24,
       author = {{Yao}, Yuhan and {Guolo}, Muryel and {Tombesi}, Francesco and {Li}, Ruancun and {Gezari}, Suvi and {Garc{\'\i}a}, Javier A. and {Dai}, Lixin and {Chornock}, Ryan and {Lu}, Wenbin and {Kulkarni}, S.~R. and {Gendreau}, Keith C. and {Pasham}, Dheeraj R. and {Cenko}, S. Bradley and {Kara}, Erin and {Margutti}, Raffaella and {Ajay}, Yukta and {Wevers}, Thomas and {Kwan}, Tom M. and {Andreoni}, Igor and {Bloom}, Joshua S. and {Drake}, Andrew J. and {Graham}, Matthew J. and {Hammerstein}, Erica and {Laher}, Russ R. and {LeBaron}, Natalie and {Mahabal}, Ashish A. and {O'Connor}, Brendan and {Purdum}, Josiah and {Ravi}, Vikram and {Sears}, Huei and {Sharma}, Yashvi and {Smith}, Roger and {Sollerman}, Jesper and {Somalwar}, Jean J. and {Wold}, Avery},
        title = "{Subrelativistic Outflow and Hours-timescale Large-amplitude X-Ray Dips during Super-Eddington Accretion onto a Low-mass Massive Black Hole in the Tidal Disruption Event AT2022lri}",
      journal = {\apj},
         year = 2024,
        month = nov,
       volume = {976},
       number = {1},
          eid = {34},
        pages = {34},
          doi = {10.3847/1538-4357/ad7d93},
archivePrefix = {arXiv},
       eprint = {2405.11343},
 primaryClass = {astro-ph.HE},
       adsurl = {https://ui.adsabs.harvard.edu/abs/2024ApJ...976...34Y}
}

@ARTICLE{Liu25,
       author = {{Lin}, Zikun and {Wang}, Yanan and {Bu}, De-Fu and {Mao}, Junjie and {Liu}, Jifeng},
        title = "{Delayed Launch of Ultrafast Outflows in the Tidal Disruption Event AT2020afhd}",
      journal = {\apjl},
         year = 2025,
        month = aug,
       volume = {989},
       number = {1},
          eid = {L9},
        pages = {L9},
          doi = {10.3847/2041-8213/adf205},
archivePrefix = {arXiv},
       eprint = {2507.15482},
 primaryClass = {astro-ph.HE},
       adsurl = {https://ui.adsabs.harvard.edu/abs/2025ApJ...989L...9L}
}

@ARTICLE{Mummery24a,
       author = {{Mummery}, Andrew and {van Velzen}, Sjoert and {Nathan}, Edward and {Ingram}, Adam and {Hammerstein}, Erica and {Fraser-Taliente}, Ludovic and {Balbus}, Steven},
        title = "{Fundamental scaling relationships revealed in the optical light curves of tidal disruption events}",
      journal = {\mnras},
         year = 2024,
        month = jan,
       volume = {527},
       number = {2},
        pages = {2452-2489},
          doi = {10.1093/mnras/stad3001},
archivePrefix = {arXiv},
       eprint = {2308.08255},
 primaryClass = {astro-ph.HE},
       adsurl = {https://ui.adsabs.harvard.edu/abs/2024MNRAS.527.2452M}
}

@ARTICLE{Wevers20,
       author = {{Wevers}, Thomas},
        title = "{Fainter harder brighter softer: a correlation between {\ensuremath{\alpha}}$_{ox}$, X-ray spectral state, and Eddington ratio in tidal disruption events}",
      journal = {\mnras},
         year = 2020,
        month = sep,
       volume = {497},
       number = {1},
        pages = {L1-L6},
          doi = {10.1093/mnrasl/slaa097},
archivePrefix = {arXiv},
       eprint = {2006.06684},
 primaryClass = {astro-ph.HE},
       adsurl = {https://ui.adsabs.harvard.edu/abs/2020MNRAS.497L...1W}
}

@ARTICLE{Grotova25,
       author = {{Grotova}, I. and {Rau}, A. and {Baldini}, P. and {Goodwin}, A.~J. and {Liu}, Z. and {Merloni}, A. and {Salvato}, M. and {Anderson}, G.~E. and {Arcodia}, R. and {Buchner}, J. and {Krumpe}, M. and {Malyali}, A. and {Masterson}, M. and {Miller-Jones}, J.~C.~A. and {Nandra}, K. and {Shirley}, R.},
        title = "{The population of tidal disruption events discovered with eROSITA}",
      journal = {\aap},
         year = 2025,
        month = may,
       volume = {697},
          eid = {A159},
        pages = {A159},
          doi = {10.1051/0004-6361/202553669},
archivePrefix = {arXiv},
       eprint = {2504.08424},
 primaryClass = {astro-ph.HE},
       adsurl = {https://ui.adsabs.harvard.edu/abs/2025A&A...697A.159G}
}

@ARTICLE{Liu16,
       author = {{Liu}, Zhu and {Merloni}, Andrea and {Georgakakis}, Antonis and {Menzel}, Marie-Luise and {Buchner}, Johannes and {Nandra}, Kirpal and {Salvato}, Mara and {Shen}, Yue and {Brusa}, Marcella and {Streblyanska}, Alina},
        title = "{X-ray spectral properties of the AGN sample in the northern XMM-XXL field}",
      journal = {\mnras},
         year = 2016,
        month = jun,
       volume = {459},
       number = {2},
        pages = {1602-1625},
          doi = {10.1093/mnras/stw753},
archivePrefix = {arXiv},
       eprint = {1605.00207},
 primaryClass = {astro-ph.HE},
       adsurl = {https://ui.adsabs.harvard.edu/abs/2016MNRAS.459.1602L}
}

@ARTICLE{McHardy06,
       author = {{McHardy}, I.~M. and {Koerding}, E. and {Knigge}, C. and {Uttley}, P. and {Fender}, R.~P.},
        title = "{Active galactic nuclei as scaled-up Galactic black holes}",
      journal = {\nat},
         year = 2006,
        month = dec,
       volume = {444},
       number = {7120},
        pages = {730-732},
          doi = {10.1038/nature05389},
archivePrefix = {arXiv},
       eprint = {astro-ph/0612273},
 primaryClass = {astro-ph},
       adsurl = {https://ui.adsabs.harvard.edu/abs/2006Natur.444..730M}
}

@ARTICLE{Gezari21,
       author = {{Gezari}, Suvi},
        title = "{Tidal Disruption Events}",
      journal = {\araa},
         year = 2021,
        month = sep,
       volume = {59},
        pages = {21-58},
          doi = {10.1146/annurev-astro-111720-030029},
archivePrefix = {arXiv},
       eprint = {2104.14580},
 primaryClass = {astro-ph.HE},
       adsurl = {https://ui.adsabs.harvard.edu/abs/2021ARA&A..59...21G}
}

@ARTICLE{Mummery25,
       author = {{Mummery}, Andrew},
        title = "{The X-ray variability of tidal disruption events}",
      journal = {arXiv e-prints},
         year = 2025,
        month = may,
          eid = {arXiv:2505.09238},
        pages = {arXiv:2505.09238},
          doi = {10.48550/arXiv.2505.09238},
archivePrefix = {arXiv},
       eprint = {2505.09238},
 primaryClass = {astro-ph.HE},
       adsurl = {https://ui.adsabs.harvard.edu/abs/2025arXiv250509238M}
}

@ARTICLE{Shakura73,
       author = {{Shakura}, N.~I. and {Sunyaev}, R.~A.},
        title = "{Black holes in binary systems. Observational appearance.}",
      journal = {\aap},
         year = 1973,
        month = jan,
       volume = {24},
        pages = {337-355},
       adsurl = {https://ui.adsabs.harvard.edu/abs/1973A&A....24..337S}
}

@INPROCEEDINGS{Arnaud96,
       author = {{Arnaud}, K.~A.},
        title = "{XSPEC: The First Ten Years}",
    booktitle = {Astronomical Data Analysis Software and Systems V},
         year = 1996,
       editor = {{Jacoby}, George H. and {Barnes}, Jeannette},
       series = {Astronomical Society of the Pacific Conference Series},
       volume = {101},
        month = jan,
        pages = {17},
       adsurl = {https://ui.adsabs.harvard.edu/abs/1996ASPC..101...17A}
}

@software{HEASoft,
       author = {{Nasa High Energy Astrophysics Science Archive Research Center (Heasarc)}},
        title = "{HEAsoft: Unified Release of FTOOLS and XANADU}",
 howpublished = {Astrophysics Source Code Library, record ascl:1408.004},
         year = 2014,
        month = aug,
          eid = {ascl:1408.004},
archivePrefix = {ascl},
       eprint = {1408.004},
       adsurl = {https://ui.adsabs.harvard.edu/abs/2014ascl.soft08004N}
}

@ARTICLE{Astropy,
       author = {{Astropy Collaboration} and {Price-Whelan}, Adrian M. and {Lim}, Pey Lian and {Earl}, Nicholas and {Starkman}, Nathaniel and {Bradley}, Larry and {Shupe}, David L. and {Patil}, Aarya A. and {Corrales}, Lia and {Brasseur}, C.~E. and {N{\"o}the}, Maximilian and {Donath}, Axel and {Tollerud}, Erik and {Morris}, Brett M. and {Ginsburg}, Adam and {Vaher}, Eero and {Weaver}, Benjamin A. and {Tocknell}, James and {Jamieson}, William and {van Kerkwijk}, Marten H. and {Robitaille}, Thomas P. and {Merry}, Bruce and {Bachetti}, Matteo and {G{\"u}nther}, H. Moritz and {Aldcroft}, Thomas L. and {Alvarado-Montes}, Jaime A. and {Archibald}, Anne M. and {B{\'o}di}, Attila and {Bapat}, Shreyas and {Barentsen}, Geert and {Baz{\'a}n}, Juanjo and {Biswas}, Manish and {Boquien}, M{\'e}d{\'e}ric and {Burke}, D.~J. and {Cara}, Daria and {Cara}, Mihai and {Conroy}, Kyle E. and {Conseil}, Simon and {Craig}, Matthew W. and {Cross}, Robert M. and {Cruz}, Kelle L. and {D'Eugenio}, Francesco and {Dencheva}, Nadia and {Devillepoix}, Hadrien A.~R. and {Dietrich}, J{\"o}rg P. and {Eigenbrot}, Arthur Davis and {Erben}, Thomas and {Ferreira}, Leonardo and {Foreman-Mackey}, Daniel and {Fox}, Ryan and {Freij}, Nabil and {Garg}, Suyog and {Geda}, Robel and {Glattly}, Lauren and {Gondhalekar}, Yash and {Gordon}, Karl D. and {Grant}, David and {Greenfield}, Perry and {Groener}, Austen M. and {Guest}, Steve and {Gurovich}, Sebastian and {Handberg}, Rasmus and {Hart}, Akeem and {Hatfield-Dodds}, Zac and {Homeier}, Derek and {Hosseinzadeh}, Griffin and {Jenness}, Tim and {Jones}, Craig K. and {Joseph}, Prajwel and {Kalmbach}, J. Bryce and {Karamehmetoglu}, Emir and {Ka{\l}uszy{\'n}ski}, Miko{\l}aj and {Kelley}, Michael S.~P. and {Kern}, Nicholas and {Kerzendorf}, Wolfgang E. and {Koch}, Eric W. and {Kulumani}, Shankar and {Lee}, Antony and {Ly}, Chun and {Ma}, Zhiyuan and {MacBride}, Conor and {Maljaars}, Jakob M. and {Muna}, Demitri and {Murphy}, N.~A. and {Norman}, Henrik and {O'Steen}, Richard and {Oman}, Kyle A. and {Pacifici}, Camilla and {Pascual}, Sergio and {Pascual-Granado}, J. and {Patil}, Rohit R. and {Perren}, Gabriel I. and {Pickering}, Timothy E. and {Rastogi}, Tanuj and {Roulston}, Benjamin R. and {Ryan}, Daniel F. and {Rykoff}, Eli S. and {Sabater}, Jose and {Sakurikar}, Parikshit and {Salgado}, Jes{\'u}s and {Sanghi}, Aniket and {Saunders}, Nicholas and {Savchenko}, Volodymyr and {Schwardt}, Ludwig and {Seifert-Eckert}, Michael and {Shih}, Albert Y. and {Jain}, Anany Shrey and {Shukla}, Gyanendra and {Sick}, Jonathan and {Simpson}, Chris and {Singanamalla}, Sudheesh and {Singer}, Leo P. and {Singhal}, Jaladh and {Sinha}, Manodeep and {Sip{\H{o}}cz}, Brigitta M. and {Spitler}, Lee R. and {Stansby}, David and {Streicher}, Ole and {{\v{S}}umak}, Jani and {Swinbank}, John D. and {Taranu}, Dan S. and {Tewary}, Nikita and {Tremblay}, Grant R. and {de Val-Borro}, Miguel and {Van Kooten}, Samuel J. and {Vasovi{\'c}}, Zlatan and {Verma}, Shresth and {de Miranda Cardoso}, Jos{\'e} Vin{\'\i}cius and {Williams}, Peter K.~G. and {Wilson}, Tom J. and {Winkel}, Benjamin and {Wood-Vasey}, W.~M. and {Xue}, Rui and {Yoachim}, Peter and {Zhang}, Chen and {Zonca}, Andrea and {Astropy Project Contributors}},
        title = "{The Astropy Project: Sustaining and Growing a Community-oriented Open-source Project and the Latest Major Release (v5.0) of the Core Package}",
      journal = {\apj},
         year = 2022,
        month = aug,
       volume = {935},
       number = {2},
          eid = {167},
        pages = {167},
          doi = {10.3847/1538-4357/ac7c74},
archivePrefix = {arXiv},
       eprint = {2206.14220},
 primaryClass = {astro-ph.IM},
       adsurl = {https://ui.adsabs.harvard.edu/abs/2022ApJ...935..167A}
}

@ARTICLE{Panagiotou20,
       author = {{Panagiotou}, C. and {Papadakis}, I.~E. and {Kammoun}, E.~S. and {Dov{\v{c}}iak}, M.},
        title = "{Multiwavelength power-spectrum analysis of NGC 5548}",
      journal = {\mnras},
         year = 2020,
        month = dec,
       volume = {499},
       number = {2},
        pages = {1998-2006},
          doi = {10.1093/mnras/staa2920},
archivePrefix = {arXiv},
       eprint = {2009.09693},
 primaryClass = {astro-ph.GA},
       adsurl = {https://ui.adsabs.harvard.edu/abs/2020MNRAS.499.1998P}
}

@ARTICLE{LSST,
       author = {{Ivezi{\'c}}, {\v{Z}}eljko and {Kahn}, Steven M. and {Tyson}, J. Anthony and {Abel}, Bob and {Acosta}, Emily and {Allsman}, Robyn and {Alonso}, David and {AlSayyad}, Yusra and {Anderson}, Scott F. and {Andrew}, John and {Angel}, James Roger P. and {Angeli}, George Z. and {Ansari}, Reza and {Antilogus}, Pierre and {Araujo}, Constanza and {Armstrong}, Robert and {Arndt}, Kirk T. and {Astier}, Pierre and {Aubourg}, {\'E}ric and {Auza}, Nicole and {Axelrod}, Tim S. and {Bard}, Deborah J. and {Barr}, Jeff D. and {Barrau}, Aurelian and {Bartlett}, James G. and {Bauer}, Amanda E. and {Bauman}, Brian J. and {Baumont}, Sylvain and {Bechtol}, Ellen and {Bechtol}, Keith and {Becker}, Andrew C. and {Becla}, Jacek and {Beldica}, Cristina and {Bellavia}, Steve and {Bianco}, Federica B. and {Biswas}, Rahul and {Blanc}, Guillaume and {Blazek}, Jonathan and {Blandford}, Roger D. and {Bloom}, Josh S. and {Bogart}, Joanne and {Bond}, Tim W. and {Booth}, Michael T. and {Borgland}, Anders W. and {Borne}, Kirk and {Bosch}, James F. and {Boutigny}, Dominique and {Brackett}, Craig A. and {Bradshaw}, Andrew and {Brandt}, William Nielsen and {Brown}, Michael E. and {Bullock}, James S. and {Burchat}, Patricia and {Burke}, David L. and {Cagnoli}, Gianpietro and {Calabrese}, Daniel and {Callahan}, Shawn and {Callen}, Alice L. and {Carlin}, Jeffrey L. and {Carlson}, Erin L. and {Chandrasekharan}, Srinivasan and {Charles-Emerson}, Glenaver and {Chesley}, Steve and {Cheu}, Elliott C. and {Chiang}, Hsin-Fang and {Chiang}, James and {Chirino}, Carol and {Chow}, Derek and {Ciardi}, David R. and {Claver}, Charles F. and {Cohen-Tanugi}, Johann and {Cockrum}, Joseph J. and {Coles}, Rebecca and {Connolly}, Andrew J. and {Cook}, Kem H. and {Cooray}, Asantha and {Covey}, Kevin R. and {Cribbs}, Chris and {Cui}, Wei and {Cutri}, Roc and {Daly}, Philip N. and {Daniel}, Scott F. and {Daruich}, Felipe and {Daubard}, Guillaume and {Daues}, Greg and {Dawson}, William and {Delgado}, Francisco and {Dellapenna}, Alfred and {de Peyster}, Robert and {de Val-Borro}, Miguel and {Digel}, Seth W. and {Doherty}, Peter and {Dubois}, Richard and {Dubois-Felsmann}, Gregory P. and {Durech}, Josef and {Economou}, Frossie and {Eifler}, Tim and {Eracleous}, Michael and {Emmons}, Benjamin L. and {Fausti Neto}, Angelo and {Ferguson}, Henry and {Figueroa}, Enrique and {Fisher-Levine}, Merlin and {Focke}, Warren and {Foss}, Michael D. and {Frank}, James and {Freemon}, Michael D. and {Gangler}, Emmanuel and {Gawiser}, Eric and {Geary}, John C. and {Gee}, Perry and {Geha}, Marla and {Gessner}, Charles J.~B. and {Gibson}, Robert R. and {Gilmore}, D. Kirk and {Glanzman}, Thomas and {Glick}, William and {Goldina}, Tatiana and {Goldstein}, Daniel A. and {Goodenow}, Iain and {Graham}, Melissa L. and {Gressler}, William J. and {Gris}, Philippe and {Guy}, Leanne P. and {Guyonnet}, Augustin and {Haller}, Gunther and {Harris}, Ron and {Hascall}, Patrick A. and {Haupt}, Justine and {Hernandez}, Fabio and {Herrmann}, Sven and {Hileman}, Edward and {Hoblitt}, Joshua and {Hodgson}, John A. and {Hogan}, Craig and {Howard}, James D. and {Huang}, Dajun and {Huffer}, Michael E. and {Ingraham}, Patrick and {Innes}, Walter R. and {Jacoby}, Suzanne H. and {Jain}, Bhuvnesh and {Jammes}, Fabrice and {Jee}, M. James and {Jenness}, Tim and {Jernigan}, Garrett and {Jevremovi{\'c}}, Darko and {Johns}, Kenneth and {Johnson}, Anthony S. and {Johnson}, Margaret W.~G. and {Jones}, R. Lynne and {Juramy-Gilles}, Claire and {Juri{\'c}}, Mario and {Kalirai}, Jason S. and {Kallivayalil}, Nitya J. and {Kalmbach}, Bryce and {Kantor}, Jeffrey P. and {Karst}, Pierre and {Kasliwal}, Mansi M. and {Kelly}, Heather and {Kessler}, Richard and {Kinnison}, Veronica and {Kirkby}, David and {Knox}, Lloyd and {Kotov}, Ivan V. and {Krabbendam}, Victor L. and {Krughoff}, K. Simon and {Kub{\'a}nek}, Petr and {Kuczewski}, John and {Kulkarni}, Shri and {Ku}, John and {Kurita}, Nadine R. and {Lage}, Craig S. and {Lambert}, Ron and {Lange}, Travis and {Langton}, J. Brian and {Le Guillou}, Laurent and {Levine}, Deborah and {Liang}, Ming and {Lim}, Kian-Tat and {Lintott}, Chris J. and {Long}, Kevin E. and {Lopez}, Margaux and {Lotz}, Paul J. and {Lupton}, Robert H. and {Lust}, Nate B. and {MacArthur}, Lauren A. and {Mahabal}, Ashish and {Mandelbaum}, Rachel and {Markiewicz}, Thomas W. and {Marsh}, Darren S. and {Marshall}, Philip J. and {Marshall}, Stuart and {May}, Morgan and {McKercher}, Robert and {McQueen}, Michelle and {Meyers}, Joshua and {Migliore}, Myriam and {Miller}, Michelle and {Mills}, David J.},
        title = "{LSST: From Science Drivers to Reference Design and Anticipated Data Products}",
      journal = {\apj},
         year = 2019,
        month = mar,
       volume = {873},
       number = {2},
          eid = {111},
        pages = {111},
          doi = {10.3847/1538-4357/ab042c},
archivePrefix = {arXiv},
       eprint = {0805.2366},
 primaryClass = {astro-ph},
       adsurl = {https://ui.adsabs.harvard.edu/abs/2019ApJ...873..111I}
}

@INCOLLECTION{Yuan22,
       author = {{Yuan}, Weimin and {Zhang}, Chen and {Chen}, Yong and {Ling}, Zhixing},
        title = "{The Einstein Probe Mission}",
    booktitle = {Handbook of X-ray and Gamma-ray Astrophysics},
         year = 2022,
       editor = {{Bambi}, Cosimo and {Sangangelo}, Andrea},
          eid = {86},
        pages = {86},
          doi = {10.1007/978-981-16-4544-0_151-1},
       adsurl = {https://ui.adsabs.harvard.edu/abs/2022hxga.book...86Y}
}

@ARTICLE{Cruise25,
       author = {{Cruise}, Mike and {Guainazzi}, Matteo and {Aird}, James and {Carrera}, Francisco J. and {Costantini}, Elisa and {Corrales}, Lia and {Dauser}, Thomas and {Eckert}, Dominique and {Gastaldello}, Fabio and {Matsumoto}, Hironori and {Osten}, Rachel and {Petrucci}, Pierre-Olivier and {Porquet}, Delphine and {Pratt}, Gabriel W. and {Rea}, Nanda and {Reiprich}, Thomas H. and {Simionescu}, Aurora and {Spiga}, Daniele and {Troja}, Eleonora},
        title = "{The NewAthena mission concept in the context of the next decade of X-ray astronomy}",
      journal = {Nature Astronomy},
         year = 2025,
        month = jan,
       volume = {9},
        pages = {36-44},
          doi = {10.1038/s41550-024-02416-3},
archivePrefix = {arXiv},
       eprint = {2501.03100},
 primaryClass = {astro-ph.IM},
       adsurl = {https://ui.adsabs.harvard.edu/abs/2025NatAs...9...36C}
}

@INPROCEEDINGS{Reynolds23,
       author = {{Reynolds}, Christopher S. and {Kara}, Erin A. and {Mushotzky}, Richard F. and {Ptak}, Andrew and {Koss}, Michael J. and {Williams}, Brian J. and {Allen}, Steven W. and {Bauer}, Franz E. and {Bautz}, Marshall and {Bogadhee}, Arash and {Burdge}, Kevin B. and {Cappelluti}, Nico and {Cenko}, Brad and {Chartas}, George and {Chan}, Kai-Wing and {Corrales}, L{\'\i}a. and {Daylan}, Tansu and {Falcone}, Abraham D. and {Foord}, Adi and {Grant}, Catherine E. and {Habouzit}, M{\'e}lanie and {Haggard}, Daryl and {Herrmann}, Sven and {Hodges-Kluck}, Edmund and {Kargaltsev}, Oleg and {King}, George W. and {Kounkel}, Marina and {Lopez}, Laura A. and {Marchesi}, Stefano and {McDonald}, Michael and {Meyer}, Eileen and {Miller}, Eric D. and {Nynka}, Melania and {Okajima}, Takashi and {Pacucci}, Fabio and {Russell}, Helen R. and {Safi-Harb}, Samar and {Strassun}, Keivan G. and {Trindade Falc{\~a}o}, Anna and {Walker}, Stephen A. and {Wilms}, Joern and {Yukita}, Mihoko and {Zhang}, William W.},
        title = "{Overview of the advanced x-ray imaging satellite (AXIS)}",
    booktitle = {UV, X-Ray, and Gamma-Ray Space Instrumentation for Astronomy XXIII},
         year = 2023,
       editor = {{Siegmund}, Oswald H. and {Hoadley}, Keri},
       series = {Society of Photo-Optical Instrumentation Engineers (SPIE) Conference Series},
       volume = {12678},
        month = oct,
          eid = {126781E},
        pages = {126781E},
          doi = {10.1117/12.2677468},
archivePrefix = {arXiv},
       eprint = {2311.00780},
 primaryClass = {astro-ph.IM},
       adsurl = {https://ui.adsabs.harvard.edu/abs/2023SPIE12678E..1ER}
}

@ARTICLE{Auchettl17,
       author = {{Auchettl}, Katie and {Guillochon}, James and {Ramirez-Ruiz}, Enrico},
        title = "{New Physical Insights about Tidal Disruption Events from a Comprehensive Observational Inventory at X-Ray Wavelengths}",
      journal = {\apj},
         year = 2017,
        month = apr,
       volume = {838},
       number = {2},
          eid = {149},
        pages = {149},
          doi = {10.3847/1538-4357/aa633b},
archivePrefix = {arXiv},
       eprint = {1611.02291},
 primaryClass = {astro-ph.HE},
       adsurl = {https://ui.adsabs.harvard.edu/abs/2017ApJ...838..149A}
}

@ARTICLE{Gierlinski05,
       author = {{Gierli{\'n}ski}, Marek and {Zdziarski}, Andrzej A.},
        title = "{Patterns of energy-dependent variability from Comptonization}",
      journal = {\mnras},
         year = 2005,
        month = nov,
       volume = {363},
       number = {4},
        pages = {1349-1360},
          doi = {10.1111/j.1365-2966.2005.09527.x},
archivePrefix = {arXiv},
       eprint = {astro-ph/0506388},
 primaryClass = {astro-ph},
       adsurl = {https://ui.adsabs.harvard.edu/abs/2005MNRAS.363.1349G}
}

@ARTICLE{Cao23,
       author = {{Cao}, Z. and {Jonker}, P.~G. and {Wen}, S. and {Stone}, N.~C. and {Zabludoff}, A.~I.},
        title = "{The rapidly spinning intermediate-mass black hole 3XMM J150052.0+015452}",
      journal = {\mnras},
         year = 2023,
        month = feb,
       volume = {519},
       number = {2},
        pages = {2375-2390},
          doi = {10.1093/mnras/stac3539},
archivePrefix = {arXiv},
       eprint = {2211.16936},
 primaryClass = {astro-ph.HE},
       adsurl = {https://ui.adsabs.harvard.edu/abs/2023MNRAS.519.2375C}
}

@ARTICLE{Cao24,
       author = {{Cao}, Z. and {Jonker}, P.~G. and {Pasham}, D.~R. and {Wen}, S. and {Stone}, N.~C. and {Zabludoff}, A.~I.},
        title = "{Tidal Disruption Event AT2020ocn: Early Time X-Ray Flares Caused by a Possible Disk Alignment Process}",
      journal = {\apj},
         year = 2024,
        month = jul,
       volume = {970},
       number = {1},
          eid = {89},
        pages = {89},
          doi = {10.3847/1538-4357/ad496f},
archivePrefix = {arXiv},
       eprint = {2405.07642},
 primaryClass = {astro-ph.HE},
       adsurl = {https://ui.adsabs.harvard.edu/abs/2024ApJ...970...89C}
}

@ARTICLE{Kara25,
       author = {{Kara}, Erin and {Garc{\'\i}a}, Javier},
        title = "{Supermassive Black Holes in X-rays: From Standard Accretion to Extreme Transients}",
      journal = {arXiv e-prints},
         year = 2025,
        month = mar,
          eid = {arXiv:2503.22791},
        pages = {arXiv:2503.22791},
          doi = {10.48550/arXiv.2503.22791},
archivePrefix = {arXiv},
       eprint = {2503.22791},
 primaryClass = {astro-ph.HE},
       adsurl = {https://ui.adsabs.harvard.edu/abs/2025arXiv250322791K}
}

@ARTICLE{Arcodia21,
       author = {{Arcodia}, R. and {Ponti}, G. and {Merloni}, A. and {Nandra}, K.},
        title = "{Do stellar-mass and super-massive black holes have similar dining habits?}",
      journal = {\aap},
         year = 2020,
        month = jun,
       volume = {638},
          eid = {A100},
        pages = {A100},
          doi = {10.1051/0004-6361/202037969},
archivePrefix = {arXiv},
       eprint = {2004.07258},
 primaryClass = {astro-ph.HE},
       adsurl = {https://ui.adsabs.harvard.edu/abs/2020A&A...638A.100A}
}

@ARTICLE{Bambic24,
       author = {{Bambic}, Christopher J. and {Quataert}, Eliot and {Kunz}, Matthew W.},
        title = "{Local models of two-temperature accretion disc coronae - I. Structure, outflows, and energetics}",
      journal = {\mnras},
         year = 2024,
        month = jan,
       volume = {527},
       number = {2},
        pages = {2895-2918},
          doi = {10.1093/mnras/stad3261},
archivePrefix = {arXiv},
       eprint = {2304.06067},
 primaryClass = {astro-ph.HE},
       adsurl = {https://ui.adsabs.harvard.edu/abs/2024MNRAS.527.2895B}
}

@ARTICLE{Uttley14,
       author = {{Uttley}, P. and {Cackett}, E.~M. and {Fabian}, A.~C. and {Kara}, E. and {Wilkins}, D.~R.},
        title = "{X-ray reverberation around accreting black holes}",
      journal = {\aapr},
         year = 2014,
        month = aug,
       volume = {22},
          eid = {72},
        pages = {72},
          doi = {10.1007/s00159-014-0072-0},
archivePrefix = {arXiv},
       eprint = {1405.6575},
 primaryClass = {astro-ph.HE},
       adsurl = {https://ui.adsabs.harvard.edu/abs/2014A&ARv..22...72U}
}

@ARTICLE{Wevers24,
       author = {{Wevers}, T. and {Guolo}, M. and {Pasham}, D.~R. and {Coughlin}, E.~R. and {Tombesi}, F. and {Yao}, Y. and {Gezari}, S.},
        title = "{Delayed X-Ray Brightening Accompanied by Variable Ionized Absorption Following a Tidal Disruption Event}",
      journal = {\apj},
         year = 2024,
        month = mar,
       volume = {963},
       number = {1},
          eid = {75},
        pages = {75},
          doi = {10.3847/1538-4357/ad1878},
       adsurl = {https://ui.adsabs.harvard.edu/abs/2024ApJ...963...75W}
}

@ARTICLE{Kosec23,
       author = {{Kosec}, P. and {Pasham}, D. and {Kara}, E. and {Tombesi}, F.},
        title = "{Discovery of a Variable Multiphase Outflow in the X-Ray-emitting Tidal Disruption Event ASASSN-20qc}",
      journal = {\apj},
         year = 2023,
        month = sep,
       volume = {954},
       number = {2},
          eid = {170},
        pages = {170},
          doi = {10.3847/1538-4357/aced87},
archivePrefix = {arXiv},
       eprint = {2308.05250},
 primaryClass = {astro-ph.HE},
       adsurl = {https://ui.adsabs.harvard.edu/abs/2023ApJ...954..170K}
}

@ARTICLE{Hu22,
       author = {{Hu}, Jingwei and {Jin}, Chichuan and {Cheng}, Huaqing and {Yuan}, Weimin},
        title = "{A Systematic Study of the Short-term X-Ray Variability of Seyfert Galaxies. I. Diversity of the X-Ray rms Spectra}",
      journal = {\apj},
         year = 2022,
        month = sep,
       volume = {936},
       number = {2},
          eid = {105},
        pages = {105},
          doi = {10.3847/1538-4357/ac83ba},
archivePrefix = {arXiv},
       eprint = {2208.05921},
 primaryClass = {astro-ph.HE},
       adsurl = {https://ui.adsabs.harvard.edu/abs/2022ApJ...936..105H}
}

@ARTICLE{HI4PI,
       author = {{HI4PI Collaboration} and {Ben Bekhti}, N. and {Fl{\"o}er}, L. and {Keller}, R. and {Kerp}, J. and {Lenz}, D. and {Winkel}, B. and {Bailin}, J. and {Calabretta}, M.~R. and {Dedes}, L. and {Ford}, H.~A. and {Gibson}, B.~K. and {Haud}, U. and {Janowiecki}, S. and {Kalberla}, P.~M.~W. and {Lockman}, F.~J. and {McClure-Griffiths}, N.~M. and {Murphy}, T. and {Nakanishi}, H. and {Pisano}, D.~J. and {Staveley-Smith}, L.},
        title = "{HI4PI: A full-sky H I survey based on EBHIS and GASS}",
      journal = {\aap},
         year = 2016,
        month = oct,
       volume = {594},
          eid = {A116},
        pages = {A116},
          doi = {10.1051/0004-6361/201629178},
archivePrefix = {arXiv},
       eprint = {1610.06175},
 primaryClass = {astro-ph.GA},
       adsurl = {https://ui.adsabs.harvard.edu/abs/2016A&A...594A.116H}
}

@ARTICLE{Cash1979,
       author = {{Cash}, W.},
        title = "{Parameter estimation in astronomy through application of the likelihood ratio.}",
      journal = {\apj},
         year = 1979,
        month = mar,
       volume = {228},
        pages = {939-947},
          doi = {10.1086/156922},
       adsurl = {https://ui.adsabs.harvard.edu/abs/1979ApJ...228..939C}
}

@ARTICLE{Liu23,
       author = {{Liu}, Z. and {Malyali}, A. and {Krumpe}, M. and {Homan}, D. and {Goodwin}, A.~J. and {Grotova}, I. and {Kawka}, A. and {Rau}, A. and {Merloni}, A. and {Anderson}, G.~E. and {Miller-Jones}, J.~C.~A. and {Markowitz}, A.~G. and {Ciroi}, S. and {Di Mille}, F. and {Schramm}, M. and {Tang}, S. and {Buckley}, D.~A.~H. and {Gromadzki}, M. and {Jin}, C. and {Buchner}, J.},
        title = "{Deciphering the extreme X-ray variability of the nuclear transient eRASSt J045650.3{\ensuremath{-}}203750. A likely repeating partial tidal disruption event}",
      journal = {\aap},
         year = 2023,
        month = jan,
       volume = {669},
          eid = {A75},
        pages = {A75},
          doi = {10.1051/0004-6361/202244805},
archivePrefix = {arXiv},
       eprint = {2208.12452},
 primaryClass = {astro-ph.HE},
       adsurl = {https://ui.adsabs.harvard.edu/abs/2023A&A...669A..75L}
}

@ARTICLE{Ponti12,
       author = {{Ponti}, G. and {Papadakis}, I. and {Bianchi}, S. and {Guainazzi}, M. and {Matt}, G. and {Uttley}, P. and {Bonilla}, N.~F.},
        title = "{CAIXA: a catalogue of AGN in the XMM-Newton archive. III. Excess variance analysis}",
      journal = {\aap},
         year = 2012,
        month = jun,
       volume = {542},
          eid = {A83},
        pages = {A83},
          doi = {10.1051/0004-6361/201118326},
archivePrefix = {arXiv},
       eprint = {1112.2744},
 primaryClass = {astro-ph.HE},
       adsurl = {https://ui.adsabs.harvard.edu/abs/2012A&A...542A..83P}
}

@ARTICLE{Gonzales12,
       author = {{Gonz{\'a}lez-Mart{\'\i}n}, O. and {Vaughan}, S.},
        title = "{X-ray variability of 104 active galactic nuclei. XMM-Newton power-spectrum density profiles}",
      journal = {\aap},
         year = 2012,
        month = aug,
       volume = {544},
          eid = {A80},
        pages = {A80},
          doi = {10.1051/0004-6361/201219008},
archivePrefix = {arXiv},
       eprint = {1205.4255},
 primaryClass = {astro-ph.HE},
       adsurl = {https://ui.adsabs.harvard.edu/abs/2012A&A...544A..80G}
}

@ARTICLE{Vaughan03,
       author = {{Vaughan}, S. and {Edelson}, R. and {Warwick}, R.~S. and {Uttley}, P.},
        title = "{On characterizing the variability properties of X-ray light curves from active galaxies}",
      journal = {\mnras},
         year = 2003,
        month = nov,
       volume = {345},
       number = {4},
        pages = {1271-1284},
          doi = {10.1046/j.1365-2966.2003.07042.x},
archivePrefix = {arXiv},
       eprint = {astro-ph/0307420},
 primaryClass = {astro-ph},
       adsurl = {https://ui.adsabs.harvard.edu/abs/2003MNRAS.345.1271V}
}
\bibliographystyle{aasjournal}

\appendix
\counterwithin{figure}{section}
\counterwithin{table}{section}

\section{Supplemental figures and tables} \label{appendix:figs}

\begin{figure*}
    \includegraphics[width=\linewidth]{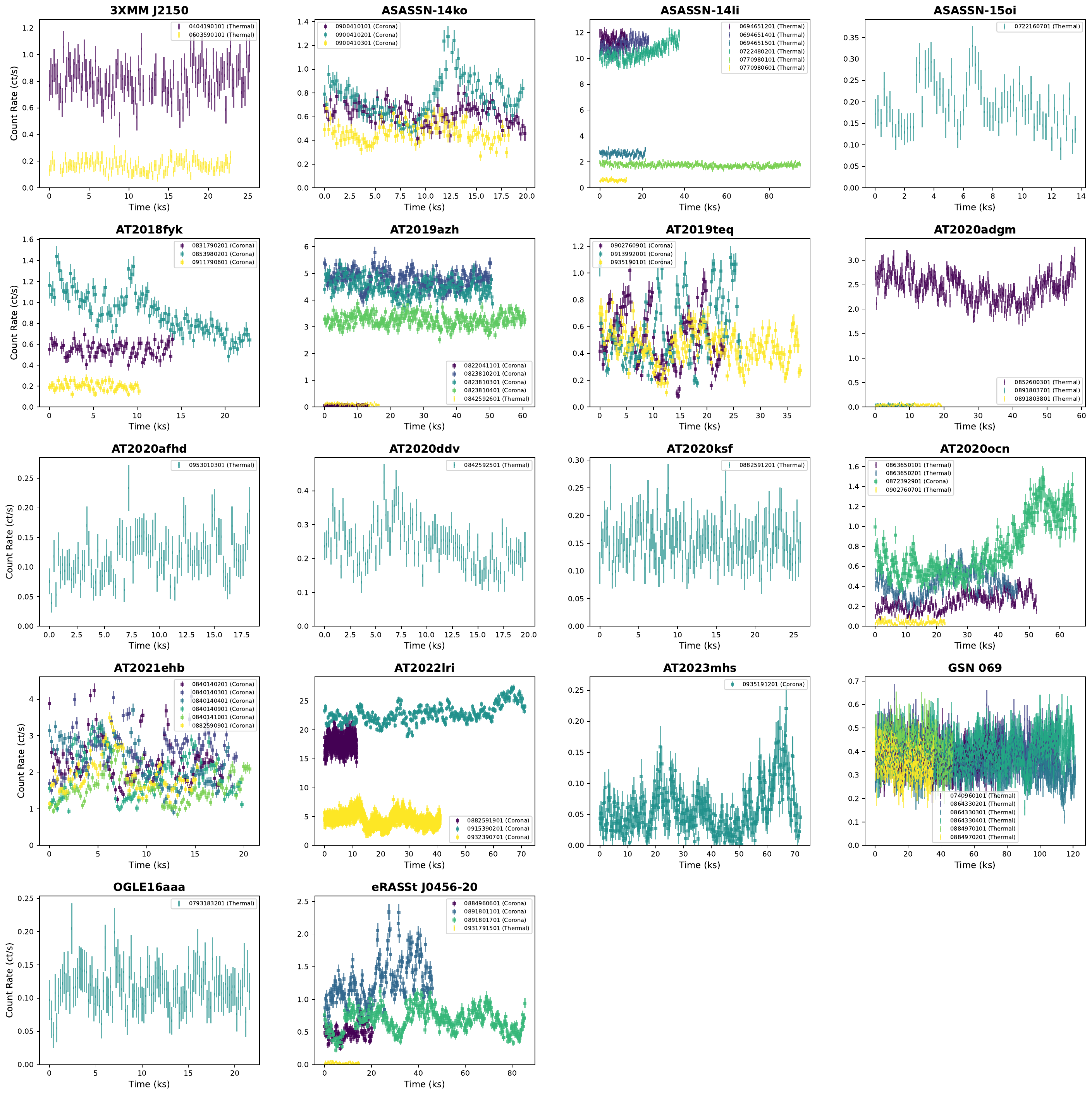}
    \caption{All \XMM\ 0.3-2 keV light curves for the 18 sources/54 observations in our sample, binned to 200 seconds for visual clarity. We retained only the longest continuous segments in each observation after filtering for background flare contamination. Observations with coronae are shown with square markers, while thermal observations are shown with vertical line markers.}
    \label{fig:lcs_all}
\end{figure*}

In Fig.~\ref{fig:lcs_all} we show all 0.3-2 keV \XMM\ light curves for all 18 sources/54 observations used in our analysis. We retained only the longest continuous segments in each observation after background filtering for background flare contamination, as Fourier timing techniques require uninterrupted baselines. We label the light curves by their OBSIDs and spectral type (thermal versus corona). Thermal observations are denoted by vertical line markers, while corona observations use square markers.

\begin{figure*}
    \includegraphics[width=\linewidth]{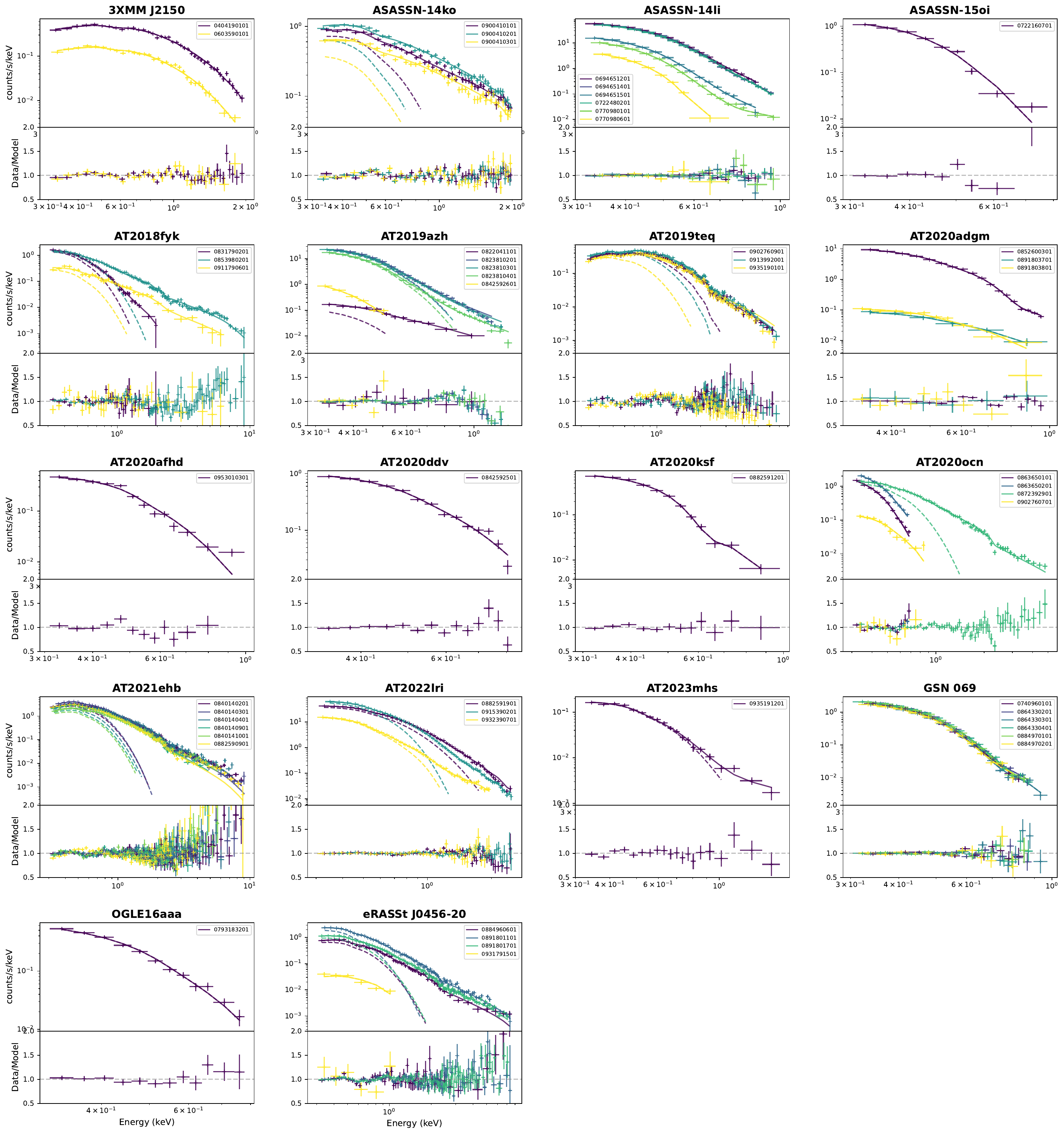}
    \caption{All \XMM\ spectra fit with \texttt{tbabs}$\times$\texttt{zashift}$\times$\texttt{simpl}$\times$\texttt{tdediscspec}. Dashed lines show the thermal component only, and solid lines show the combined thermal + corona continua.}
    \label{fig:xmm_spectra_all}
\end{figure*}

In Fig.~\ref{fig:xmm_spectra_all} we show all \XMM\ spectra, extending to the energy where the source component meets the background level. In the case of ASASSN-14ko only, we imposed a manual energy cutoff of 2 keV because the sources is blended with a background AGN; however, a \textit{Chandra} analysis was able to spatially resolve the sources and determine that the nucleus of ASASSN-14ko dominates the X-rays below $\lesssim 2$ keV \citep{Payne23}. All observations are fit with a spectral model \texttt{tbabs}$\times$\texttt{zashift}$\times$\texttt{simpl}$\times$\texttt{tdediscspec}. See Section~\ref{sec:methods} for further details.

\begin{figure*}
    \includegraphics[width=\linewidth]{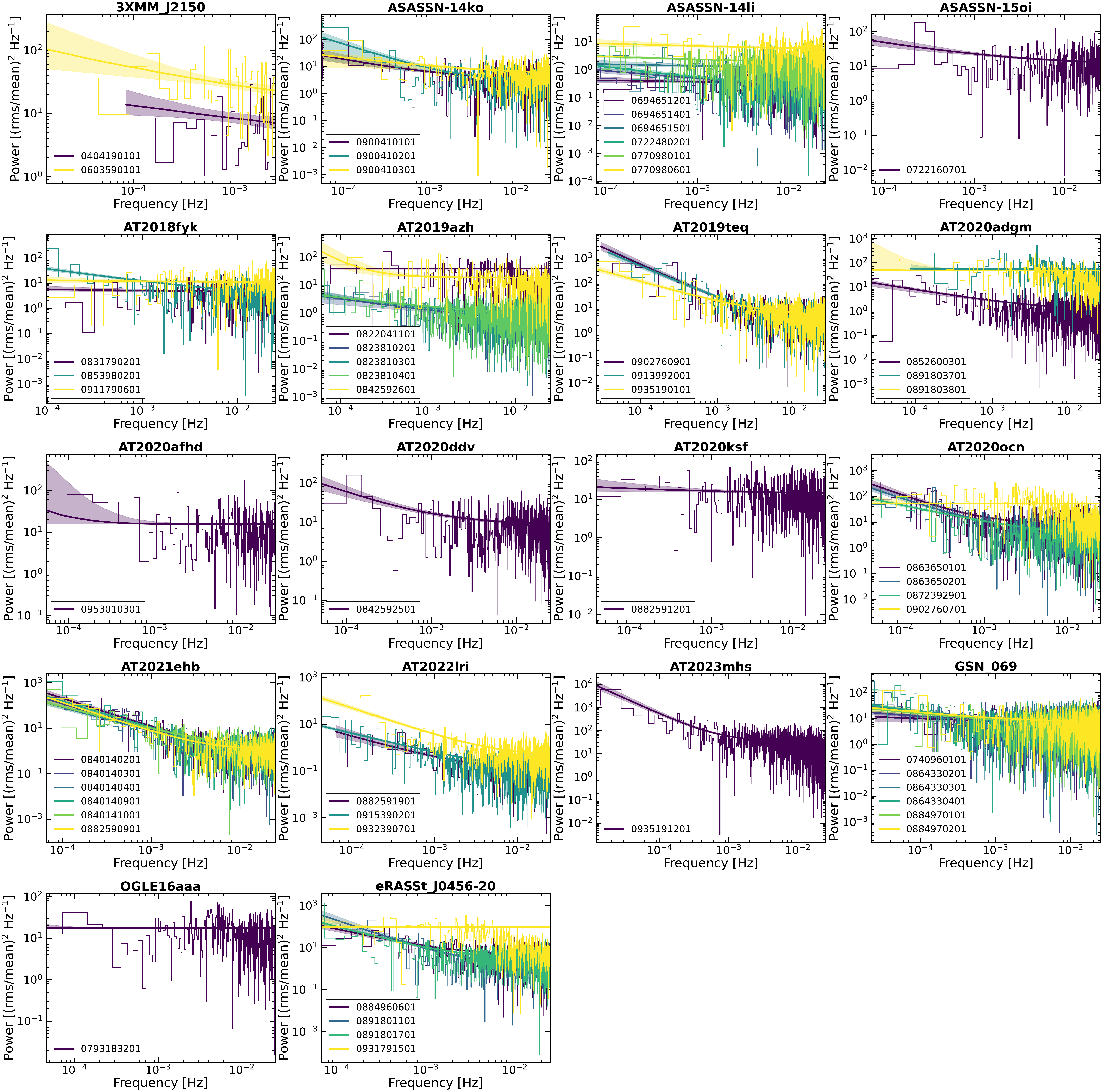}
    \caption{Soft X-ray (0.3-2 keV) power spectral densities (PSDs) fit with power-law + constant models and $1\sigma$ uncertainties.}
    \label{fig:psds_all}
\end{figure*}

\begin{figure*}
    \includegraphics[width=\linewidth]{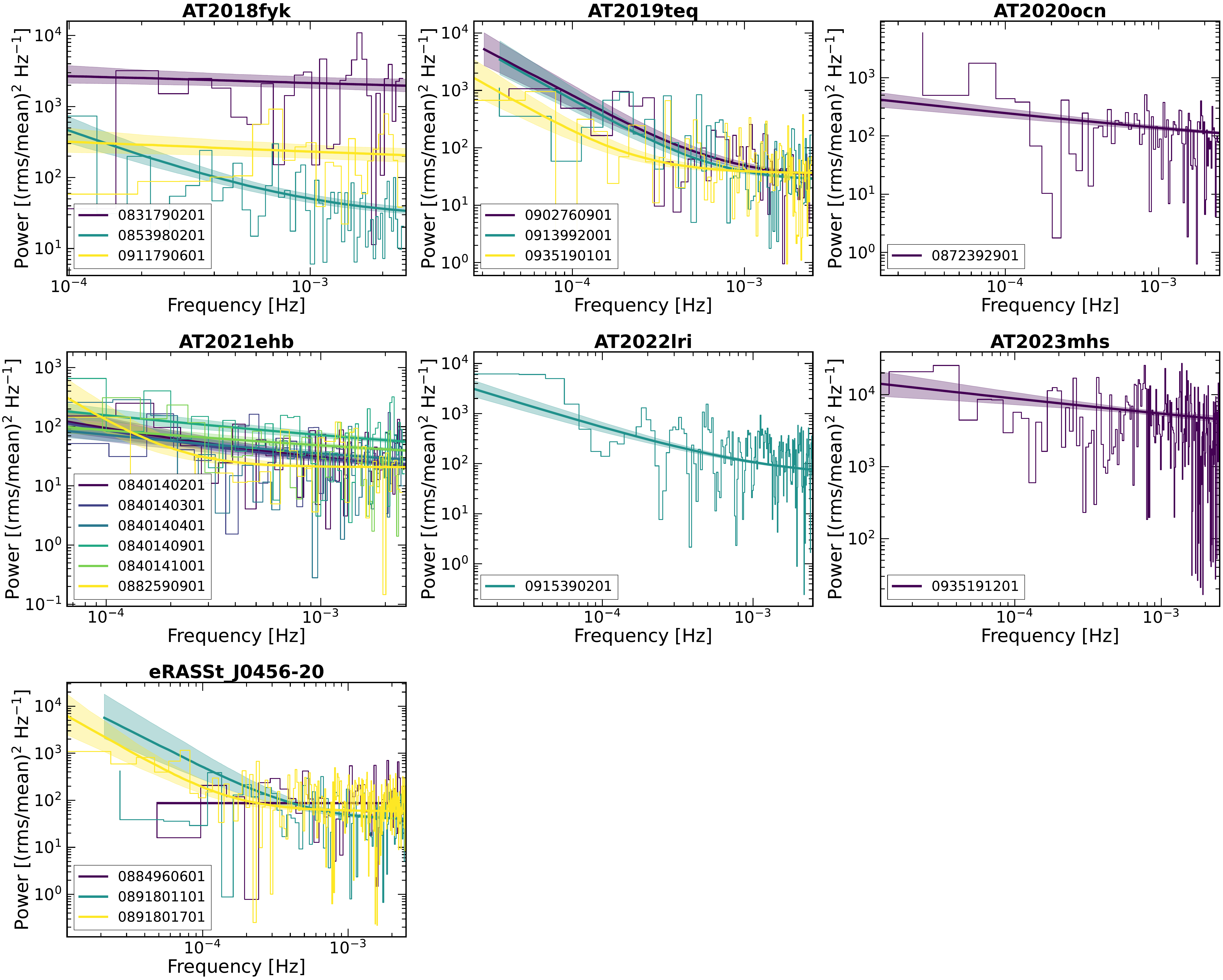}
    \caption{Hard X-ray (2-10 keV) power spectral densities (PSDs) fit with power-law + constant models and $1\sigma$ uncertainties. We exclude from this analysis ASASSN-14ko (due to the companion AGN dominating the 2-10 keV band) and AT2019azh (due to too few 2-10 keV counts for timing analysis).}
    \label{fig:psds_hard}
\end{figure*}

In Fig.~\ref{fig:psds_all} we show all 0.3-2 keV power spectral densities (PSDs) computed with Eq.~\ref{eq:psd} using 20-second light curve binning. We also show the power-law plus constant fits with $1\sigma$ uncertainties, which were performed as described in Section~\ref{subsec:psd}. Similarly, the 2-10 keV PSDs are shown in Fig.~\ref{fig:psds_hard}. These are computed and fit in the same manner as the soft-band PSDs, but with 200s time binning to increase signal-to-noise in the fainter hard band.

\begin{figure*}
    \includegraphics[width=\linewidth]{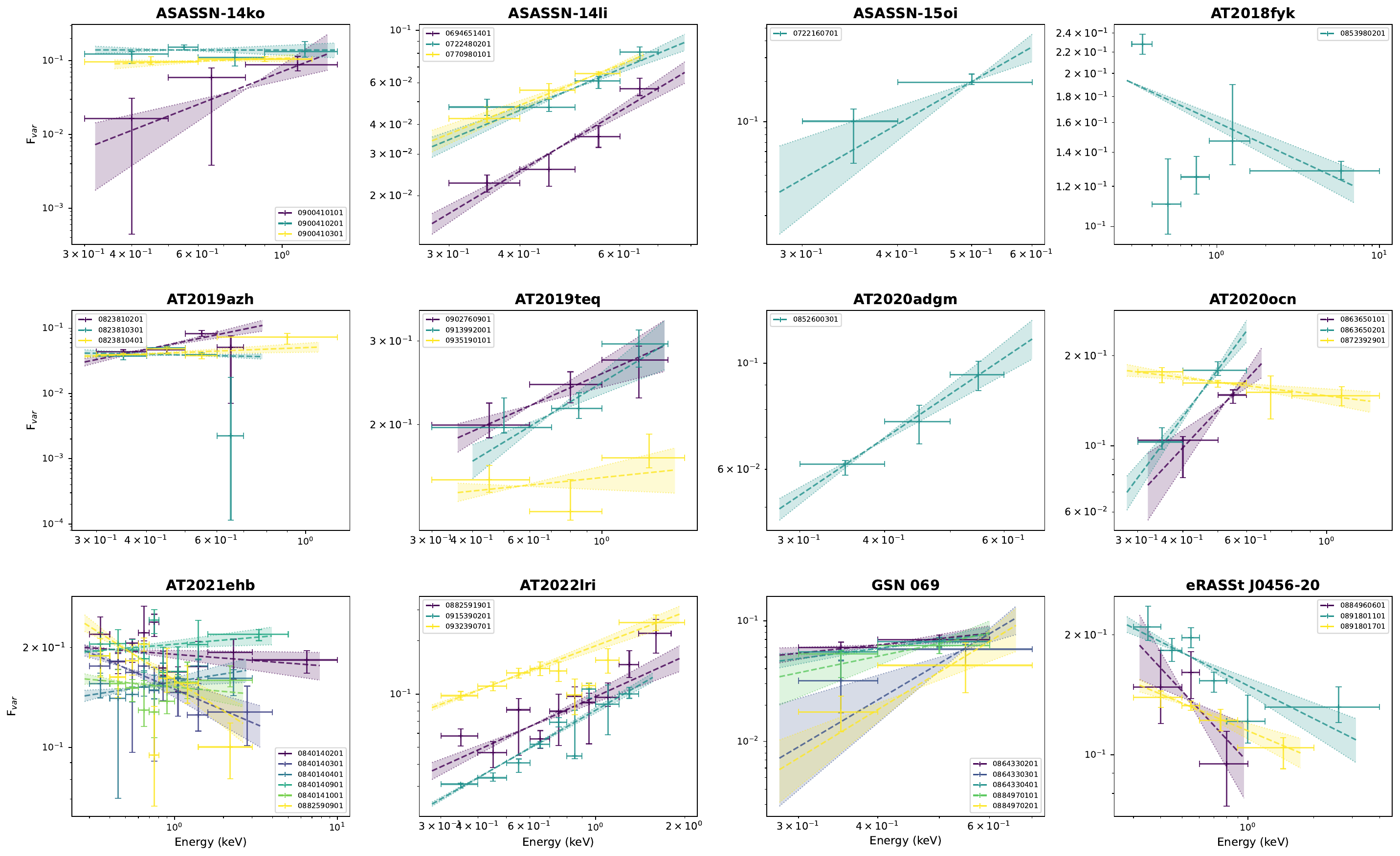}
    \caption{All Fourier-resolved spectra (FRS) fit with a single power-law model.}
    \label{fig:rms_spectra_all}
\end{figure*}

In Fig.~\ref{fig:rms_spectra_all} we show all Fourier-resolved spectra ($F_{\rm var}$ as a function of energy) along with power-law fits. To determine the FRS energy bin width, we initially split the light curves into energy-resolved bands with edges of 0.3/0.4/0.5/0.6/0.7/0.8/0.9/1/1.2/1.4/1.6/1.8/2/3/4/6/8/10 keV, then iteratively combined light curves starting from the lowest-energy bands until a minimum count rate of $\geq0.1$ cps was met. Once a band met this rate, we moved to the next band and repeated the procedure. We then separately computed the PSD for each light curve and measured $F_{\rm var}$ via Eq.~\ref{eq:fvar}, resulting in the FRS shown.

In Table~\ref{tab:tde_properties} we give the spectral and timing properties for each observation used in this work.

\clearpage
\pagestyle{empty}
\footnotesize
\setlength{\tabcolsep}{2pt}
\begin{sidewaystable*}
\centering
\tiny
\setlength{\tabcolsep}{3pt}
\caption{X-ray spectral and timing properties of the TDE sample}
\label{tab:tde_properties}
\begin{tabular}{llccccccccccccc}
\hline\hline
Source & ObsID & MJD & $T_p$ & $R_p$ & FracSctr & $\Gamma$ & C-stat/dof & $F_{\rm var}$ & RMS $\alpha$ & PSD $\alpha$ & $\log L_{\rm soft}$ & Line E & $\sigma$ & Strength \\
 & & & ($10^5$ K) & ($10^{12}$ cm) & & & & & & & (erg/s) & (keV) & (keV) & (keV) \\
\hline
\multirow{2}{*}{3XMM J2150} & 0404190101 & 53860.5 & $26.4_{-1.1}^{+1.1}$ & $0.160_{-0.034}^{+0.005}$ & --- & --- & 45/35 & --- & --- & --- & $42.859_{-0.004}^{+0.004}$ & --- & --- & --- \\
 & 0603590101 & 54989.3 & $21.8_{-7.9}^{+0.9}$ & $0.10_{-0.02}^{+0.11}$ & --- & --- & 40/30 & --- & --- & --- & $42.156_{-0.009}^{+0.016}$ & --- & --- & --- \\
\hline
\multirow{3}{*}{ASASSN-14ko} & 0900410101 & 60164.9 & $7.61_{-0.03}^{+1.15}$ & $0.299_{-0.005}^{+0.089}$ & $0.24_{-0.04}^{+0.05}$ & $1.98_{-0.04}^{+0.21}$ & 41/37 & $0.070_{-0.009}^{+0.010}$ & $2.0_{-0.8}^{+1.4}$ & $0.7 \pm 0.2$ & $42.533 \pm 0.005$ & --- & --- & --- \\
 & 0900410201 & 60170.7 & $6.07_{-0.06}^{+0.46}$ & $0.664_{-0.005}^{+0.211}$ & $0.36_{-0.05}^{+0.04}$ & $2.29_{-0.03}^{+0.08}$ & 57/37 & $0.12_{-0.06}^{+0.02}$ & $-0.0_{-0.2}^{+0.2}$ & $1.0 \pm 0.4$ & $42.621_{-0.005}^{+0.006}$ & --- & --- & --- \\
 & 0900410301 & 60171.8 & $6.5_{-0.3}^{+0.5}$ & $0.40_{-0.07}^{+0.07}$ & $0.25_{-0.04}^{+0.04}$ & $2.20_{-0.16}^{+0.06}$ & 64/37 & $0.09_{-0.03}^{+0.01}$ & $0.2_{-0.2}^{+0.2}$ & $0.4 \pm 0.2$ & $42.410_{-0.007}^{+0.008}$ & --- & --- & --- \\
\hline
\multirow{6}{*}{ASASSN-14li} & 0694651201 & 56998.0 & $4.91_{-0.01}^{+0.12}$ & $0.57_{-0.02}^{+0.03}$ & --- & --- & 21/8 & --- & --- & --- & $43.295_{-0.002}^{+0.002}$ & $0.727_{-0.006}^{+0.007}$ & $0.101_{-0.006}^{+0.006}$ & $0.17_{-0.01}^{+0.02}$ \\
 & 0722480201 & 56999.5 & $4.779_{-0.001}^{+0.086}$ & $1.93_{-0.04}^{+0.05}$ & --- & --- & 11/10 & $0.022_{-0.002}^{+0.002}$ & $1.0_{-0.2}^{+0.2}$ & $0.49 \pm 0.07$ & $43.273_{-0.001}^{+0.002}$ & $0.742_{-0.004}^{+0.006}$ & $0.113_{-0.003}^{+0.005}$ & $0.187_{-0.004}^{+0.012}$ \\
 & 0694651401 & 57023.9 & $4.78_{-0.01}^{+0.13}$ & $0.452_{-0.008}^{+0.023}$ & --- & --- & 24/10 & $0.020 \pm 0.002$ & $1.4_{-0.2}^{+0.2}$ & $0.37 \pm 0.10$ & $43.282_{-0.002}^{+0.002}$ & $0.744_{-0.005}^{+0.007}$ & $0.092_{-0.004}^{+0.006}$ & $0.180_{-0.007}^{+0.017}$ \\
 & 0694651501 & 57213.2 & $4.520_{-0.003}^{+0.023}$ & $0.7530_{-0.0138}^{+0.0008}$ & --- & --- & 16/8 & --- & --- & --- & $42.700_{-0.004}^{+0.004}$ & $0.7226_{-0.0003}^{+0.0040}$ & $0.1608_{-0.0058}^{+0.0004}$ & $0.678_{-0.005}^{+0.006}$ \\
 & 0770980101 & 57366.5 & $5.3_{-0.2}^{+0.2}$ & $0.41_{-0.03}^{+0.06}$ & --- & --- & 17/11 & $0.035 \pm 0.003$ & $1.0_{-0.2}^{+0.2}$ & $0.18 \pm 0.05$ & $42.540_{-0.003}^{+0.003}$ & $0.739_{-0.003}^{+0.012}$ & $0.168_{-0.005}^{+0.008}$ & $1.12_{-0.09}^{+0.09}$ \\
 & 0770980601 & 57544.0 & $3.68_{-0.08}^{+0.29}$ & $0.7_{-0.1}^{+0.1}$ & --- & --- & 5/3 & --- & --- & --- & $42.10_{-0.01}^{+0.01}$ & $0.67_{-0.01}^{+0.08}$ & $0.09_{-0.01}^{+0.05}$ & $0.68_{-0.04}^{+0.79}$ \\
\hline
\multirow{1}{*}{ASASSN-15oi} & 0722160701 & 57482.7 & $3.28_{-0.05}^{+0.45}$ & $3.8_{-1.1}^{+1.8}$ & --- & --- & 19/8 & $0.13_{-0.02}^{+0.02}$ & $3.3_{-1.4}^{+1.2}$ & $0.5 \pm 0.1$ & $42.27_{-0.01}^{+0.01}$ & --- & --- & --- \\
\hline
\multirow{3}{*}{AT2018fyk} & 0831790201 & 58461.8 & $7.5_{-2.5}^{+0.8}$ & $0.30_{-0.06}^{+0.79}$ & $0.28_{-0.08}^{+0.05}$ & $>4.4$ & 43/36 & --- & --- & --- & $42.803_{-0.006}^{+0.041}$ & --- & --- & --- \\
 & 0853980201 & 58783.4 & $9.5_{-0.2}^{+1.8}$ & $0.29_{-0.03}^{+0.05}$ & $0.22_{-0.04}^{+0.01}$ & $2.12_{-0.07}^{+0.06}$ & 194/153 & $0.116_{-0.007}^{+0.009}$ & $-0.15 \pm 0.02$ & $0.54 \pm 0.04$ & $42.944_{-0.005}^{+0.004}$ & --- & --- & --- \\
 & 0911790601 & 59720.0 & $7.6_{-1.0}^{+0.9}$ & $0.32_{-0.02}^{+0.27}$ & $0.30_{-0.09}^{+0.12}$ & $2.21_{-0.09}^{+0.25}$ & 86/96 & --- & --- & --- & $42.35_{-0.01}^{+0.01}$ & --- & --- & --- \\
\hline
\multirow{5}{*}{AT2019azh} & 0822041101 & 58579.8 & $2.9_{-0.3}^{+0.5}$ & $0.6_{-0.2}^{+0.4}$ & $0.14_{-0.07}^{+0.05}$ & $4.3_{-0.6}^{+0.2}$ & 17/14 & --- & --- & --- & $40.95_{-0.02}^{+0.03}$ & --- & --- & --- \\
 & 0823810201 & 58782.6 & $3.723_{-0.008}^{+0.073}$ & $1.27_{-0.03}^{+0.03}$ & $0.0212_{-0.0027}^{+0.0002}$ & $>4.9$ & 73/16 & $0.035_{-0.006}^{+0.005}$ & $1.3_{-0.3}^{+0.3}$ & $0.5 \pm 0.2$ & $42.977_{-0.002}^{+0.002}$ & --- & --- & --- \\
 & 0823810301 & 58784.6 & $3.79_{-0.06}^{+0.08}$ & $0.99_{-0.02}^{+0.05}$ & $0.0187_{-0.0012}^{+0.0004}$ & $>4.9$ & 93/19 & $0.038_{-0.003}^{+0.003}$ & $-0.1_{-0.2}^{+0.2}$ & $0.46 \pm 0.06$ & $42.950_{-0.002}^{+0.001}$ & --- & --- & --- \\
 & 0823810401 & 58788.6 & $3.52_{-0.02}^{+0.02}$ & $2.10_{-0.09}^{+0.16}$ & $0.0136_{-0.0024}^{+0.0008}$ & $>4.9$ & 82/21 & $0.037 \pm 0.002$ & $0.2_{-0.2}^{+0.2}$ & $0.42 \pm 0.04$ & $42.833_{-0.002}^{+0.002}$ & --- & --- & --- \\
 & 0842592601 & 58971.9 & $2.806_{-0.268}^{+0.010}$ & $2.4_{-0.9}^{+0.4}$ & --- & --- & 11/1 & $0.08_{-0.03}^{+0.04}$ & --- & $1.7 \pm 0.6$ & $41.42_{-0.02}^{+0.01}$ & --- & --- & --- \\
\hline
\multirow{3}{*}{AT2019teq} & 0902760901 & 59830.5 & $29.8_{-3.1}^{+1.1}$ & $0.10_{-0.01}^{+0.01}$ & $0.52_{-0.02}^{+0.06}$ & $2.61_{-0.07}^{+0.08}$ & 158/117 & $0.29_{-0.03}^{+0.04}$ & $0.3_{-0.1}^{+0.1}$ & $1.4 \pm 0.1$ & $43.051_{-0.004}^{+0.004}$ & --- & --- & --- \\
 & 0913992001 & 59902.3 & $21.9_{-0.6}^{+0.9}$ & $0.30_{-0.05}^{+0.02}$ & $0.942_{-0.061}^{+0.002}$ & $2.84_{-0.04}^{+0.04}$ & 133/123 & $0.27_{-0.02}^{+0.03}$ & $0.4_{-0.1}^{+0.1}$ & $1.30 \pm 0.08$ & $43.117_{-0.005}^{+0.006}$ & --- & --- & --- \\
 & 0935190101 & 60430.7 & $20.7_{-0.3}^{+1.2}$ & $0.099_{-0.008}^{+0.008}$ & $>0.9$ & $2.94_{-0.05}^{+0.06}$ & 125/120 & $0.17_{-0.01}^{+0.01}$ & $0.07_{-0.10}^{+0.10}$ & $0.88 \pm 0.06$ & $43.011_{-0.004}^{+0.007}$ & --- & --- & --- \\
\hline
\multirow{3}{*}{AT2020adgm} & 0852600301 & 59287.4 & $6.32_{-0.03}^{+0.18}$ & $0.9513_{-0.0953}^{+0.0008}$ & --- & --- & 73/10 & $0.052_{-0.005}^{+0.003}$ & $1.0_{-0.2}^{+0.2}$ & $0.6 \pm 0.2$ & $43.439_{-0.003}^{+0.002}$ & $0.873_{-0.005}^{+0.004}$ & $0.067_{-0.004}^{+0.012}$ & $0.149_{-0.001}^{+0.019}$ \\
 & 0891803701 & 59552.3 & $12.0_{-2.8}^{+1.7}$ & $0.15_{-0.07}^{+0.27}$ & --- & --- & 15/12 & --- & --- & --- & $41.47_{-0.04}^{+0.05}$ & --- & --- & --- \\
 & 0891803801 & 59556.3 & $7.7_{-0.6}^{+1.1}$ & $0.2_{-0.1}^{+0.1}$ & --- & --- & 13/13 & --- & --- & --- & $41.54_{-0.02}^{+0.02}$ & --- & --- & --- \\
\hline
\multirow{1}{*}{AT2020afhd} & 0953010301 & 60525.6 & $6.7_{-0.9}^{+0.2}$ & $0.30_{-0.09}^{+0.05}$ & --- & --- & 34/14 & --- & --- & --- & $41.47_{-0.02}^{+0.02}$ & --- & --- & --- \\
\hline
\multirow{1}{*}{AT2020ddv} & 0842592501 & 58967.9 & $7.02_{-0.01}^{+0.83}$ & $1.0_{-0.2}^{+0.2}$ & --- & --- & 15/9 & $0.12_{-0.01}^{+0.02}$ & --- & $0.8 \pm 0.2$ & $43.34_{-0.01}^{+0.01}$ & --- & --- & --- \\
\hline
\multirow{1}{*}{AT2020ksf} & 0882591201 & 59725.0 & $5.9_{-0.2}^{+0.3}$ & $0.9_{-0.2}^{+0.1}$ & --- & --- & 5/11 & --- & --- & --- & $42.70 \pm 0.01$ & $0.70_{-0.02}^{+0.03}$ & $0.05_{-0.02}^{+0.04}$ & $0.16_{-0.03}^{+0.10}$ \\
\hline
\multirow{4}{*}{AT2020ocn} & 0863650101 & 59048.1 & $4.3_{-0.3}^{+0.7}$ & $1.2_{-0.1}^{+0.6}$ & --- & --- & 25/6 & $0.14_{-0.01}^{+0.01}$ & $1.3_{-0.5}^{+0.5}$ & $1.2 \pm 0.2$ & $42.712_{-0.005}^{+0.006}$ & --- & --- & --- \\
 & 0863650201 & 59051.8 & $6.01_{-0.81}^{+0.03}$ & $1.2_{-0.3}^{+0.8}$ & --- & --- & 14/5 & $0.125_{-0.010}^{+0.010}$ & $1.6_{-0.3}^{+0.3}$ & $1.1 \pm 0.1$ & $42.920_{-0.007}^{+0.007}$ & --- & --- & --- \\
 & 0872392901 & 59349.7 & $9.5_{-0.1}^{+1.3}$ & $0.228_{-0.039}^{+0.003}$ & $0.35_{-0.05}^{+0.01}$ & $2.66_{-0.08}^{+0.04}$ & 114/96 & $0.119_{-0.006}^{+0.007}$ & $-0.15_{-0.09}^{+0.08}$ & $0.69 \pm 0.04$ & $43.057_{-0.003}^{+0.003}$ & --- & --- & --- \\
 & 0902760701 & 59713.0 & $8.9_{-1.6}^{+0.4}$ & $0.41_{-0.12}^{+0.08}$ & --- & --- & 16/10 & --- & --- & --- & $41.78_{-0.02}^{+0.02}$ & --- & --- & --- \\
\hline
\multirow{6}{*}{AT2021ehb} & 0882590901 & 59604.2 & $12.2_{-1.0}^{+0.3}$ & $0.124_{-0.007}^{+0.032}$ & $0.14_{-0.02}^{+0.02}$ & $2.69_{-0.11}^{+0.07}$ & 173/140 & $0.14_{-0.01}^{+0.02}$ & $-0.31_{-0.09}^{+0.08}$ & $1.3 \pm 0.1$ & $42.245_{-0.004}^{+0.004}$ & --- & --- & --- \\
 & 0840140201 & 59606.2 & $0.68_{-0.03}^{+0.01}$ & $25.1_{-1.3}^{+4.1}$ & $0.54_{-0.11}^{+0.08}$ & $3.08_{-0.01}^{+0.01}$ & 305/145 & $0.19_{-0.02}^{+0.02}$ & $-0.04_{-0.04}^{+0.03}$ & $1.27 \pm 0.08$ & $42.367 \pm 0.003$ & --- & --- & --- \\
 & 0840140301 & 59608.2 & $8.68_{-0.05}^{+0.64}$ & $0.159_{-0.024}^{+0.006}$ & $0.237_{-0.020}^{+0.009}$ & $2.92_{-0.05}^{+0.04}$ & 150/146 & $0.15_{-0.01}^{+0.02}$ & $-0.21_{-0.08}^{+0.07}$ & $1.14 \pm 0.07$ & $42.403 \pm 0.003$ & --- & --- & --- \\
 & 0840140401 & 59610.2 & $12.1_{-1.2}^{+0.5}$ & $0.12_{-0.01}^{+0.01}$ & $0.28_{-0.01}^{+0.04}$ & $2.68_{-0.03}^{+0.05}$ & 141/133 & $0.13_{-0.01}^{+0.01}$ & $0.08_{-0.05}^{+0.05}$ & $1.08 \pm 0.07$ & $42.344_{-0.003}^{+0.003}$ & --- & --- & --- \\
 & 0840140901 & 59612.2 & $9.5_{-0.4}^{+0.2}$ & $0.154_{-0.016}^{+0.002}$ & $0.29_{-0.03}^{+0.02}$ & $3.15_{-0.07}^{+0.04}$ & 120/94 & $0.16_{-0.01}^{+0.01}$ & $0.04 \pm 0.03$ & $1.18 \pm 0.07$ & $42.271_{-0.003}^{+0.003}$ & --- & --- & --- \\
 & 0840141001 & 59616.1 & $7.8_{-0.9}^{+0.3}$ & $0.79_{-0.12}^{+0.06}$ & $0.36_{-0.18}^{+0.03}$ & $2.77_{-0.32}^{+0.09}$ & 159/110 & $0.13_{-0.01}^{+0.01}$ & $-0.04_{-0.06}^{+0.06}$ & $1.01 \pm 0.07$ & $42.194_{-0.006}^{+0.014}$ & --- & --- & --- \\
\hline
\multirow{3}{*}{AT2022lri} & 0882591901 & 59939.3 & $12.45_{-0.22}^{+0.08}$ & $0.338_{-0.017}^{+0.009}$ & $0.26_{-0.01}^{+0.01}$ & $>5.0$ & 96/48 & $0.026_{-0.004}^{+0.005}$ & $0.8_{-0.1}^{+0.1}$ & $1.0 \pm 0.2$ & $43.767 \pm 0.001$ & --- & --- & --- \\
 & 0915390201 & 59949.9 & $7.851_{-0.002}^{+0.031}$ & $0.577_{-0.004}^{+0.004}$ & $0.1305_{-0.0015}^{+0.0008}$ & $>5.0$ & 57/45 & $0.032 \pm 0.001$ & $0.96_{-0.05}^{+0.05}$ & $0.85 \pm 0.03$ & $43.8911_{-0.0008}^{+0.0006}$ & $1.021_{-0.003}^{+0.003}$ & $0.1892_{-0.0018}^{+0.0007}$ & $-0.1358_{-0.0004}^{+0.0008}$ \\
 & 0932390701 & 60302.5 & $6.227_{-0.002}^{+0.062}$ & $0.7337_{-0.0032}^{+0.0001}$ & $0.0364_{-0.0018}^{+0.0007}$ & $3.57_{-0.03}^{+0.02}$ & 32/34 & $0.093_{-0.007}^{+0.007}$ & $0.63_{-0.08}^{+0.07}$ & $1.16 \pm 0.06$ & $43.211_{-0.002}^{+0.002}$ & $0.9920_{-0.0008}^{+0.0022}$ & $0.136_{-0.004}^{+0.008}$ & $-0.135_{-0.009}^{+0.003}$ \\
\hline
\multirow{1}{*}{AT2023mhs} & 0935191201 & 60493.8 & $7.9_{-0.1}^{+0.4}$ & $0.13_{-0.02}^{+0.02}$ & $0.021_{-0.005}^{+0.007}$ & $1.31_{-0.16}^{+0.06}$ & 26/28 & $0.28_{-0.02}^{+0.03}$ & --- & $1.35 \pm 0.10$ & $41.63_{-0.01}^{+0.01}$ & --- & --- & --- \\
\hline
\multirow{6}{*}{GSN 069} & 0740960101 & 56996.0 & $5.697_{-0.006}^{+0.478}$ & $0.279_{-0.009}^{+0.025}$ & --- & --- & 10/6 & --- & --- & --- & $41.688_{-0.005}^{+0.006}$ & $0.80_{-0.01}^{+0.06}$ & $0.195_{-0.006}^{+0.040}$ & $1.09_{-0.01}^{+0.51}$ \\
 & 0864330201 & 58997.9 & $4.64_{-0.01}^{+0.36}$ & $0.54_{-0.02}^{+0.04}$ & --- & --- & 9/7 & $0.062_{-0.003}^{+0.004}$ & $0.5_{-0.4}^{+0.4}$ & $0.34 \pm 0.04$ & $41.719_{-0.004}^{+0.005}$ & $0.717_{-0.004}^{+0.021}$ & $0.079_{-0.004}^{+0.022}$ & $0.183_{-0.002}^{+0.083}$ \\
 & 0864330301 & 59003.9 & $4.70_{-0.09}^{+0.41}$ & $0.40_{-0.01}^{+0.02}$ & --- & --- & 21/9 & $0.077_{-0.003}^{+0.004}$ & $3.1_{-1.6}^{+1.3}$ & $0.44 \pm 0.04$ & $41.658 \pm 0.003$ & $0.728_{-0.006}^{+0.018}$ & $0.078_{-0.005}^{+0.023}$ & $0.21_{-0.01}^{+0.08}$ \\
 & 0864330401 & 59013.8 & $5.3_{-0.1}^{+0.1}$ & $0.47_{-0.04}^{+0.06}$ & --- & --- & 14/8 & $0.077_{-0.004}^{+0.004}$ & $0.7_{-0.4}^{+0.4}$ & $0.48 \pm 0.04$ & $41.722 \pm 0.003$ & $0.77_{-0.02}^{+0.01}$ & $0.122_{-0.011}^{+0.009}$ & $0.34_{-0.05}^{+0.03}$ \\
 & 0884970101 & 59395.6 & $5.2_{-0.1}^{+0.4}$ & $0.278_{-0.009}^{+0.018}$ & --- & --- & 6/7 & $0.071 \pm 0.006$ & $1.1_{-0.8}^{+1.0}$ & $0.39 \pm 0.06$ & $41.726_{-0.005}^{+0.005}$ & $0.76_{-0.01}^{+0.03}$ & $0.120_{-0.007}^{+0.018}$ & $0.41_{-0.04}^{+0.09}$ \\
 & 0884970201 & 59551.2 & $4.49_{-0.10}^{+0.19}$ & $0.55_{-0.03}^{+0.05}$ & --- & --- & 10/5 & $0.074 \pm 0.008$ & $3.2_{-1.2}^{+1.1}$ & $0.39 \pm 0.08$ & $41.649_{-0.007}^{+0.005}$ & $0.70_{-0.01}^{+0.05}$ & $0.074_{-0.006}^{+0.034}$ & $0.16_{-0.03}^{+0.11}$ \\
\hline
\multirow{1}{*}{OGLE16aaa} & 0793183201 & 57722.6 & $6.7_{-0.7}^{+0.1}$ & $1.2_{-0.5}^{+0.3}$ & --- & --- & 7/8 & --- & --- & --- & $43.07_{-0.02}^{+0.01}$ & --- & --- & --- \\
\hline
\multirow{4}{*}{eRASSt J0456-20} & 0891801101 & 59447.2 & $11.8_{-0.5}^{+0.3}$ & $0.46_{-0.03}^{+0.03}$ & $0.22_{-0.01}^{+0.02}$ & $2.75_{-0.04}^{+0.05}$ & 197/132 & $0.17_{-0.02}^{+0.02}$ & $-0.28_{-0.07}^{+0.07}$ & $1.4 \pm 0.2$ & $43.359_{-0.002}^{+0.003}$ & --- & --- & --- \\
 & 0891801701 & 59480.6 & $14.1_{-0.5}^{+0.4}$ & $0.154_{-0.006}^{+0.011}$ & $0.227_{-0.016}^{+0.008}$ & $2.61_{-0.05}^{+0.01}$ & 192/131 & $0.133_{-0.009}^{+0.011}$ & $-0.23_{-0.07}^{+0.07}$ & $1.12 \pm 0.07$ & $43.088_{-0.002}^{+0.002}$ & --- & --- & --- \\
 & 0884960601 & 59657.0 & $0.980_{-0.001}^{+0.132}$ & $5.7_{-1.0}^{+1.1}$ & $0.66_{-0.21}^{+0.02}$ & $3.05_{-0.02}^{+0.04}$ & 150/121 & $0.13_{-0.01}^{+0.02}$ & $-0.6_{-0.4}^{+0.4}$ & $1.0 \pm 0.1$ & $42.950_{-0.005}^{+0.006}$ & --- & --- & --- \\
 & 0931791501 & 60198.9 & $15.0_{-1.3}^{+4.9}$ & $0.09_{-0.05}^{+0.04}$ & --- & --- & 26/17 & --- & --- & --- & $41.54_{-0.04}^{+0.03}$ & --- & --- & --- \\
\hline
\end{tabular}
\end{sidewaystable*}
\clearpage

\end{document}